\documentclass{aa}  

\usepackage{float}
\usepackage{graphicx}
\usepackage{txfonts}
\usepackage{natbib}
\bibpunct{(}{)}{,}{a}{}{,}  
\usepackage{pdflscape}
\usepackage{orcidlink}
\usepackage{hyperref}
\hypersetup{breaklinks=true,
            colorlinks,
            filecolor=blue,
            linkcolor=blue,
            citecolor=blue,
            urlcolor=blue,
            anchorcolor=blue} 
\newcommand\feh{\ensuremath{[\mathrm{Fe}/\mathrm{H}]~}}
\newcommand\afeh{\ensuremath{[\alpha/\mathrm{Fe}]~}}
\newcommand\dex{\ensuremath{~\mathrm{dex}}}

\usepackage{color}

\begin{document}

   \title{A comparative high-resolution spectroscopic analysis of in situ and accreted globular clusters}

   
   \authorrunning{Ceccarelli, E., et al.}
   
   \author{E. Ceccarelli
          \inst{1,2}
          \orcidlink{0009-0007-3793-9766}
          \and
          A. Mucciarelli
          \inst{2,1}
          \orcidlink{0000-0001-9158-8580}
          \and
          D. Massari
          \inst{1}
          \orcidlink{0000-0001-8892-4301}
          \and
          M. Bellazzini
          \inst{1}
          \orcidlink{0000-0001-8200-810X}
          \and
          T. Matsuno
          \inst{3,4}
          }

   \institute{INAF - Astrophysics and Space Science Observatory of 
              Bologna, Via Gobetti 93/3, 40129 Bologna, Italy\\ \email{edoardo.ceccarelli3@unibo.it}
         \and
             Department of Physics and Astronomy, University of 
             Bologna, Via Gobetti 93/2, 40129 Bologna, Italy 
         \and 
             Astronomisches Rechen-Institut, Zentrum für Astronomie der Universität Heidelberg, Mönchhofstraße 12-14, 69120 Heidelberg, Germany
         \and 
             Kapteyn Astronomical Institute, University of Groningen, Landleven 12, 9747 AD Groningen, The Netherlands
             }


 
  \abstract
  {Globular clusters (GCs) are extremely intriguing systems that help in reconstructing the assembly of the Milky Way via the characterisation of their chemo-chrono-dynamical properties. In this study, we use high-resolution spectroscopic archival data from UVES and UVES-FLAMES at the VLT to compare the chemistry of GCs dynamically tagged as either Galactic (NGC 6218, NGC 6522, and NGC 6626) or accreted from distinct merger events (NGC 362 and NGC 1261 from \textit{Gaia}-Sausage-Enceladus, and Ruprecht 106 from the Helmi Streams) in the metallicity regime where abundance patterns of field stars with different origin effectively separate $( -1.3 \le \feh \le -1.0 \dex)$. We find remarkable similarities in the abundances of the two \textit{Gaia}-Sausage-Enceladus GCs across all chemical elements. They both display depletion in the $\alpha$-elements (Mg, Si and Ca) and statistically significant differences in Zn and Eu compared to in situ GCs. Additionally, we confirm that Ruprecht 106 exhibits a completely different chemical makeup from the other target clusters, being underabundant in all chemical elements. This demonstrates that when high precision is achieved, the abundances of certain chemical elements can not only efficiently separate in situ from accreted GCs, but can also distinguish among GCs born in different progenitor galaxies. In the end, we investigate the possible origin of the chemical peculiarity of Ruprecht 106. Given that its abundances do not match the chemical patterns of the field stars associated with its most likely parent galaxy (i.e. the Helmi Streams), being depleted in the abundances of $\alpha$-elements in particular, we believe Ruprecht 106 to originate from a less massive galaxy compared to the progenitor of the Helmi Streams.}


   \maketitle
%

\section{Introduction} \label{introduction}

   The latest data releases of the ESA/\textit{Gaia} mission \citep{GC21,GC23} have instigated a profound transformation of our comprehension of the early chronicles of the Milky Way (MW). This enormous 6D phase space dataset has led to a clearer understanding of the roles played by various mergers in shaping the Galactic halo (see \citealt{helmi2020} for a review), as predicted by simulations in the $\Lambda$CDM cosmological framework \citep{white&frenk1991,moore1999,helmidezeeuw00,newton18}. The imprint of accretion events manifests dynamically as stellar streams \citep{helmi99,ibata2024}, substructures discernible in phase space \citep[e.g.][]{helmi2018, belokurov2018,myeong18,koppelman19}, and dwarf galaxies whose disruption is currently ongoing \citep{ibata94,majewski2003}. During such events, not only field stars but also globular clusters (GCs) can survive the merging processes, contributing to the formation of the present-day MW GC system \citep{brodie&starder2006,penarrubia2009,bellazzini2020,trujillo-gomez2021}. Leveraging the extremely precise kinematic data provided by the \textit{Gaia} mission, the orbits of MW GCs have been reconstructed with exceptional precision. Through the analysis of their orbital properties, MW GCs have been classified into distinct groups by several authors \citep{massari19,forbes2020,callingham2022,chen&gnedin2024}, distinguishing those accreted from disrupted dwarf galaxies, such as \textit{Gaia}-Sausage-Enceladus \citep[GSE, ][]{belokurov2018,helmi2018}, Sequoia \citep{myeong18}, and the Helmi Streams \citep{helmi99}, from those formed in situ. Complications may arise in the interpretation of these findings due to the complex and overlapping distribution of GCs in the spaces defined by their orbital parameters \citep{callingham2022}. This complexity is further exacerbated when accounting for the impact of a non-static potential in the computation of a GC orbit \citep{amarante2022,belokurov23,pagnini2023,chen&gnedin2024}. Nonetheless, it has been established that dynamically tagged groups of GCs lay on different sequences in the age--metallicity space \citep{kruijssen2019,massari19,myeong19}, confirming the dual nature of the MW GC system in this plane already found in the pre-\textit{Gaia} era \citep{marin-franch2009,forbes&bridges2010,leaman2013,vandenberg2013}. In particular, it has been demonstrated that older systems at fixed metallicity are more likely to be born in galaxies with a higher star formation efficiency ---such as the MW in its earliest phases--- compared to dwarf galaxies \citep[e.g.][]{massari2023}. 
   
   This intricate perspective of the assembly history of the MW can be further enriched by the addition of chemical abundance data obtained either from high-resolution spectra or large spectroscopic surveys \citep{venn2004,nissen&schuster2010,helmi2018,horta2020,minelli21,limberg22,malhan2022,naidu2022,horta23,monty23,ceccarelli2024}. 
   However, when looking at the distribution of GCs in chemical spaces, the differences between these two populations are subtle, complicating the task of interpreting the results. For instance, \citet{recioblanco2018} highlights a very low scatter in the \afeh ratio among GCs with $\feh < -0.8 \dex$ regardless of their origin. On a similar note, \citet{horta2020} employed the [Si/Fe] ratio from APOGEE DR17 \citep{apogee22} of 46 GCs to explore whether their distribution in chemical spaces reflects the kinematically defined classification by \citet{massari19}. Interestingly, \citet{horta2020} observed that the positions of both in situ and accreted subgroups in this chemical space align with those of their field-star counterparts. However, they were unable to effectively identify different behaviours from GCs that were brought in by separate progenitors, possibly because of the higher precision needed. Recently, \citet{belokurov&krastov2024} discriminated between Galactic and extragalactic origin for MW GCs using the [Al/Fe] ratios, building upon the findings of \citet{belokurov22}. However, caution should be exercised when using the abundances of light elements in GCs owing to the impact of multiple populations \citep{bastian&lardo18}. Finally, \citet{monty23,monty2023_2} showcased the efficacy of differential chemical analysis on GCs. Their works not only revealed chemical inhomogeneity at a precision level of 0.02 dex for all chemical elements in GCs, but also showed distinctions in the chemical compositions of two GCs, NGC 288 and NGC 362, associated with the same progenitor, GSE. Specifically, NGC 288 and NGC 362 exhibit dissimilarities in their chemical makeup, particularly in their neutron-capture elements. This difference could be interpreted either as stemming from chemical inhomogeneities in the progenitor or as a consequence of internal chemical evolution within the GCs. In light of all the mentioned complexities, which underline the challenges associated with interpreting results when comparing diverse studies, it becomes evident that maintaining homogeneity is imperative in conducting chemical analyses aimed at verifying the origin of GCs.
   
   In the present study, we derived fully homogeneous detailed elemental abundances of several species for six GCs with similar metallicity $( -1.3 \le \feh \le -1.0 \dex) $ that have been consistently associated with different progenitors (see below) according to all post-\textit{Gaia} investigations \citep{massari19, forbes2020, callingham2022, chen&gnedin2024}. These GCs lie in the metallicity range best suited to chemical tagging studies, where abundance trends of in situ and accreted objects start to significantly part ways \citep{nissen&schuster2010,helmi2018,matsuno22}. Three of the target clusters, namely NGC 362, NGC 1261, and Ruprecht 106 (Rup106 hereafter), lie on the younger branch of the Galactic GC age--metallicity relation \citep[AMR,][]{dotter2010,dotter2011}, which is interpreted as suggesting an accreted origin. Specifically, due to the characteristics of the spaces defined by the integrals of motion, they have been assigned to GSE (NGC 362 and NGC 1261) and the Helmi Streams (Rup106). Also, we selected three targets (NGC 6218, NGC 6522, and NGC 6626) that are generally considered to have formed in situ, as they fit the older branch of the AMR \citep{dotter2010,dotter2011,villanova17,kerber2018} and move on orbits compatible with either the MW disc (NGC 6218) or the MW bulge (NGC 6522 and NGC 6626). 
   The primary objective of the present work is to test precise chemical tagging as a tool to infer the origin of GCs. When complementing chrono-dynamical information, it is possible to use chemical tagging to clarify the differences among clusters with ambiguous associations. Furthermore, this approach enables determination of whether or not the progenitor galaxy and the associated GC candidates are chemically compatible. For this to be effective, we require a combination of high precision in abundance derivations, which can only be achieved using high-quality data, and homogeneity in the detailed analysis. This is why we reanalysed these six clusters, allowing us to compare them with each other with the utmost precision.

   The paper is structured as follows. In Sect. \ref{sec:observation} we list the observations that were collected to compile the spectroscopic dataset. In Sect. \ref{sec:abu} we describe the methodology employed to carry out the chemical analysis. In Sect. \ref{sec:abundances} we show the output of the analysis. In Sect. \ref{sec:literature} we compare our results with the literature. In Sect. \ref{sec:Rup106} we discuss the peculiarity of Ruprecht 106. In Sect. \ref{sec:conclusion} we summarise the main results of this work.   
\section{Observations}\label{sec:observation}

   \begin{figure*}
   \centering
   \begin{minipage}{0.35\textwidth}
        \centering
        \includegraphics[width=1.0\textwidth]{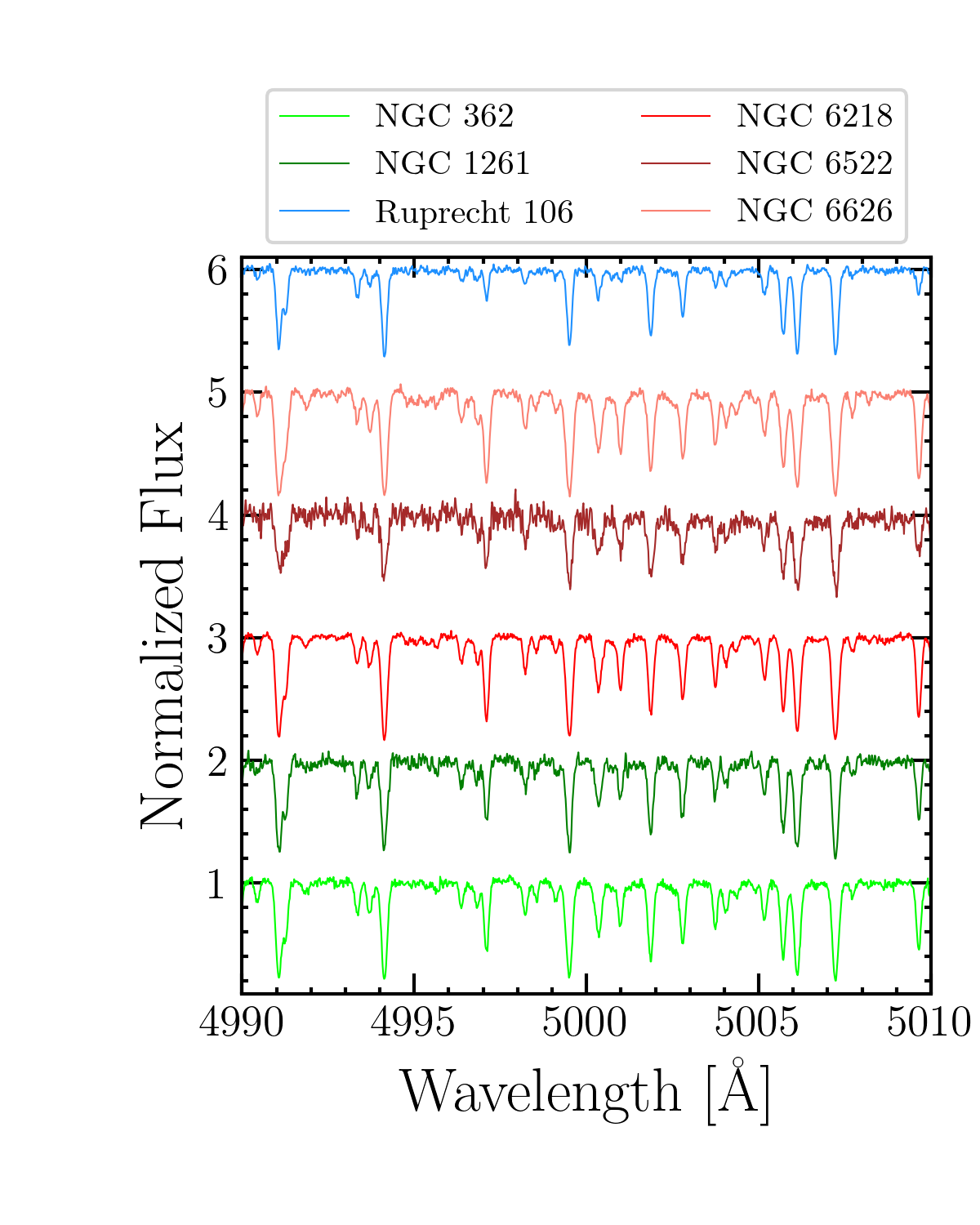} 
   \end{minipage}
   \begin{minipage}{0.35\textwidth}
        \centering
        \includegraphics[width=1.0\textwidth]{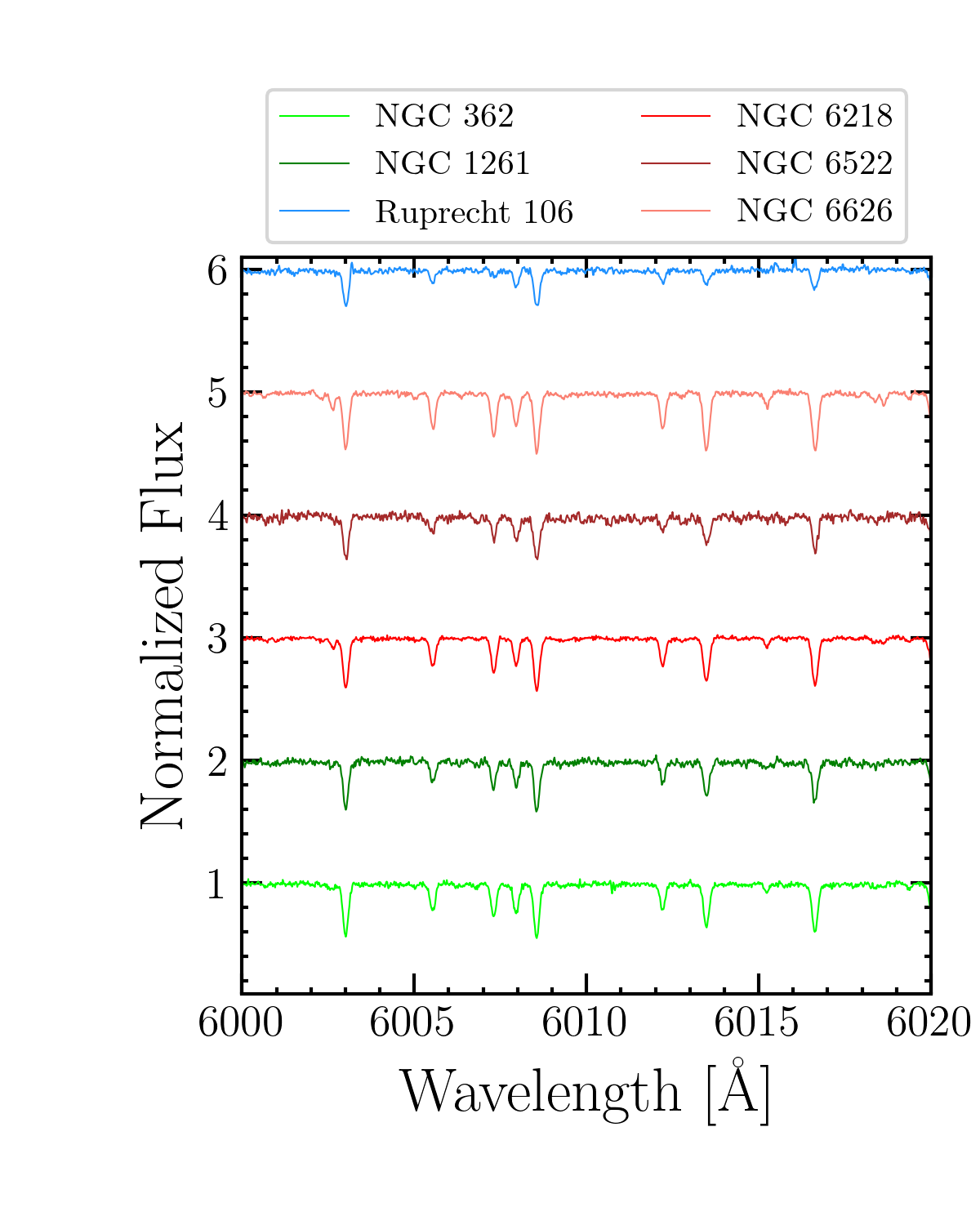}
   \end{minipage}   
   \caption{Spectra of stars from the six target clusters observed with UVES and UVES-FLAMES at the VLT at different wavelengths. The spectra have been vertically shifted for the sake of clarity.}
              \label{fig:spectra}%
    \end{figure*}   

   We retrieved our entire spectroscopic dataset from the ESO archive. The spectra of these stars were  all acquired using the multi-object spectrograph UVES-FLAMES \citep{pasquini02} or UVES \citep{dekker2000} mounted at the Very Large Telescope (VLT) of the European Southern Observatory. Stars were observed using the Red Arm 580 CD3 grating with a spectral coverage between 4800 and 6800 \r{A} and a spectral resolution of R=40,000. Typical signal-to-noise ratios for these spectra are in the range of 40 $\le$ S/N $\le$ 65 at 5000 \r{A} and 60 $\le$ S/N $\le$ 90 at 6000 \r{A}. The spectra were reduced using the dedicated ESO pipeline\footnote{\url{https://www.eso.org/sci/software/pipelines/}}, which includes bias subtraction, flat-fielding, wavelength calibration, spectral extraction, and order merging. For the stars observed with UVES at the VLT, the contribution of the sky was properly removed from each stellar spectrum during the pipeline reduction. For the stars observed with UVES-FLAMES at the VLT, the sky contribution was taken into account by acquiring the spectra of some nearby sky regions at the same time as the science targets and subtracting it from each individual exposure. In the end, single exposures of the same star were merged to obtain the final spectrum for each target.  Fig. \ref{fig:spectra} shows the spectra of six stars, one per target cluster, to showcase the quality of the dataset.
   
   As mentioned in Sect. \ref{introduction}, the sample of GCs studied here was selected based on their similar metallicity and dynamical association \citep[which is consistent among the different studies; e.g.][see below]{massari19,forbes2020,callingham2022}. The individual spectroscopic targets analysed in this work were selected as follows:
   \begin{itemize}
       \item NGC 362. This cluster has been categorised as accreted during the merger with GSE. The mean metallicity of NGC 362 is $\feh = -1.17 \pm 0.05 \dex$ \citep{carretta13}. We collected the spectra of 14 RGB stars observed under the ESO-VLT Programme 083.D-0208 (PI: E. Carretta).
       \item NGC 1261. This GC has also been associated with the GSE accretion event. The mean metallicity of this cluster is $\feh = -1.28 \pm 0.02 \dex$ \citep{marino21}. Our dataset includes the spectra of 12 RGB stars. Among them, 6 were observed under the ESO-VLT Programme 0101.D-0109 (PI: A. Marino), 4 under the ESO-VLT Programme 197.B-1074 (PI: G. F. Gilmore), and 2 under the ESO-VLT Programme 193.D-0232 (PI: F. R. Ferraro).
       \item NGC 6218. This system is an in situ GC located in the MW disc. NGC 6218 has a mean metallicity of $\feh = -1.33 \pm 0.08 \dex$ \citep{carretta09}. We include in our analysis 11 RGB stars observed under the ESO-VLT Programme 073.D-0211 (PI: E. Carretta). 
       \item NGC 6522. This ia an in situ cluster placed in the MW bulge. Its mean metallicity is $\feh = -1.16 \pm 0.05 \dex$ \citep{barbuy2021}. Our sample is composed of four RGB stars observed under the ESO-VLT Programme  097.D-0175 (PI: B. Barbuy).
       \item NGC 6626. This is also an in situ MW bulge GC with a mean metallicity of $\feh = -1.29 \pm 0.01 \dex$ \citep{villanova17}. We collected the spectra of 16 RGB stars observed under the ESO-VLT Programme  091.D-0535 (PI: C. Moni Bidin).       
       \item Ruprecht 106. Among the population of accreted GCs, one that has always attracted significant attention within the community is Rup106. Its orbit is typical of the Galactic halo \citep{frelijj2021} and due to its positioning in the spaces defined by the integrals of motion, this GC has been tentatively linked with the progenitor of the Helmi Streams. The chemical peculiarity of Rup106 has been revealed by several spectroscopic analyses \citep[e.g.][V13 hereafter]{brown1997,villanova13}, which find that all Rup106 target spectra are depleted in [$\alpha$/Fe] compared to GCs and field stars of the same metallicity, that is $\feh = -1.37 \pm 0.11 \dex$ \citep{lucertini23}. We collected the spectra of nine RGB stars that have been observed under the ESO-VLT Programme 069.D-0642 (PI: P. Fran\c{c}ois). 
   \end{itemize}

\section{Chemical analysis}\label{sec:abu}
\subsection{Stellar parameters}\label{sp}
%
\begin{table*}
\caption{Selected target information (extract).}          
\label{TabSP}      
\centering          
\begin{tabular}{ccccccc}  
\hline      
Cluster & Star ID & ID \textit{Gaia} DR3 & $T_{\mathrm{eff}}$ & log $g$ & $v_{\mathrm{t}}$ & \texttt{flag\_high\_ruwe} \\ 
 & & & (K) & (dex) & (km $\mathrm{s^{-1}}$) & \\ 
\hline 
NGC 362   	 & 	      1037	 & 	 4690839791797266688	 & 	4549	 & 	1.35	 & 	1.4	& 0  \\
NGC 362   	 & 	      1840	 & 	 4690841273563721728	 & 	4485	 & 	1.28	 & 	1.5	& 0	  \\
NGC 362   	 & 	      2683	 & 	 4690886834576765952	 & 	4142	 & 	0.69	 & 	1.6	& 0	  \\
NGC 362   	 & 	      3392	 & 	 4690888105887037184	 & 	4522	 & 	1.33	 & 	1.5	& 0	  \\
...      &    ...      & ...   & ...   & ...   & ... & ...      \\
\hline                  
\end{tabular}
\tablefoot{We list star ID from previous literature works, ID from \textit{Gaia} DR3, effective temperature, surface gravity, microturbulent velocity, and a flag to check the quality of the \textit{Gaia} photometry of the star (\texttt{flag\_high\_ruwe}=0 stands for \texttt{ruwe} $<1.4$). The entire table is available at the CDS.
}
\end{table*}
%
   \begin{figure*}
   \centering
   \begin{minipage}{0.35\textwidth}
        \centering
        \includegraphics[width=1.0\textwidth]{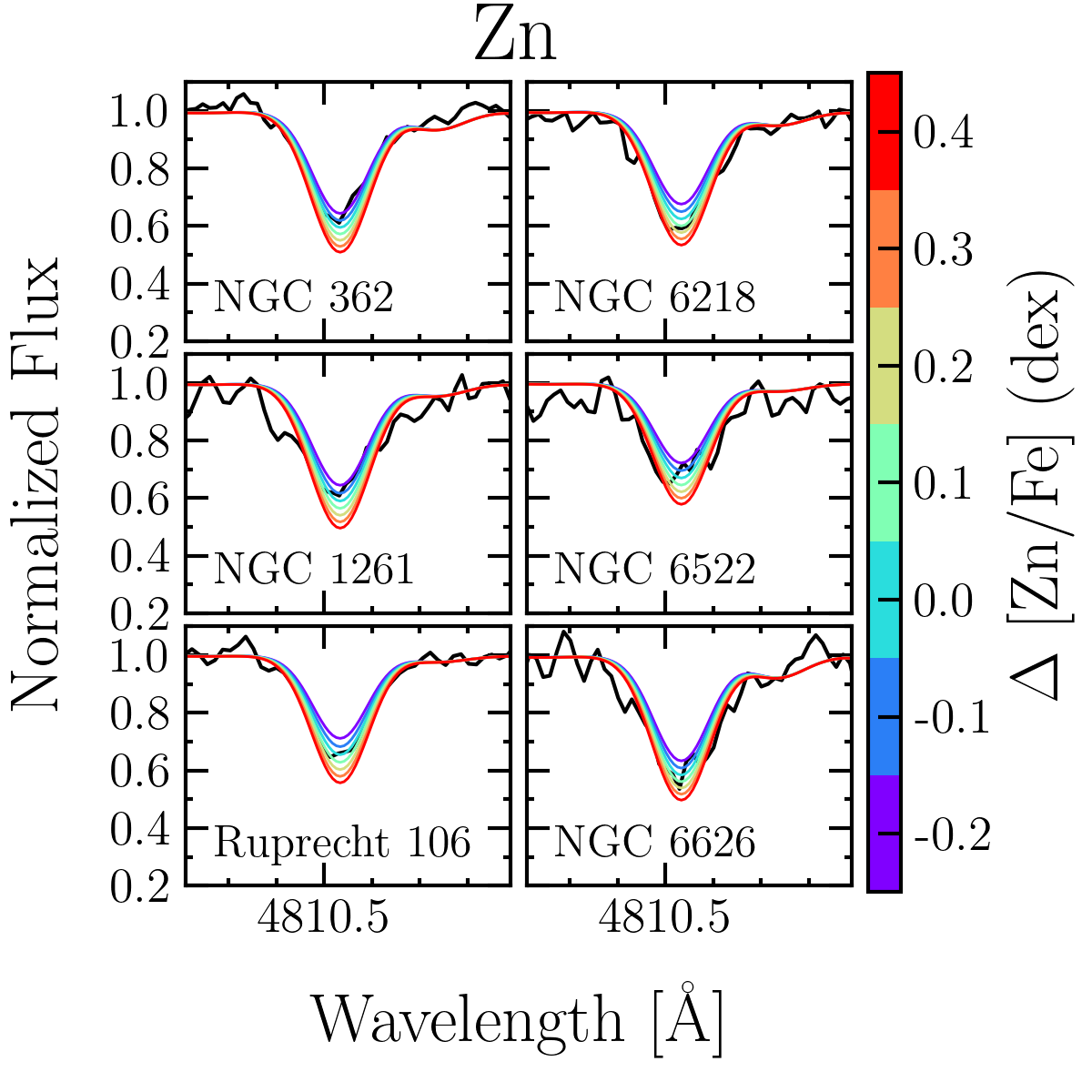} 
   \end{minipage}
   \begin{minipage}{0.35\textwidth}
        \centering
        \includegraphics[width=1.0\textwidth]{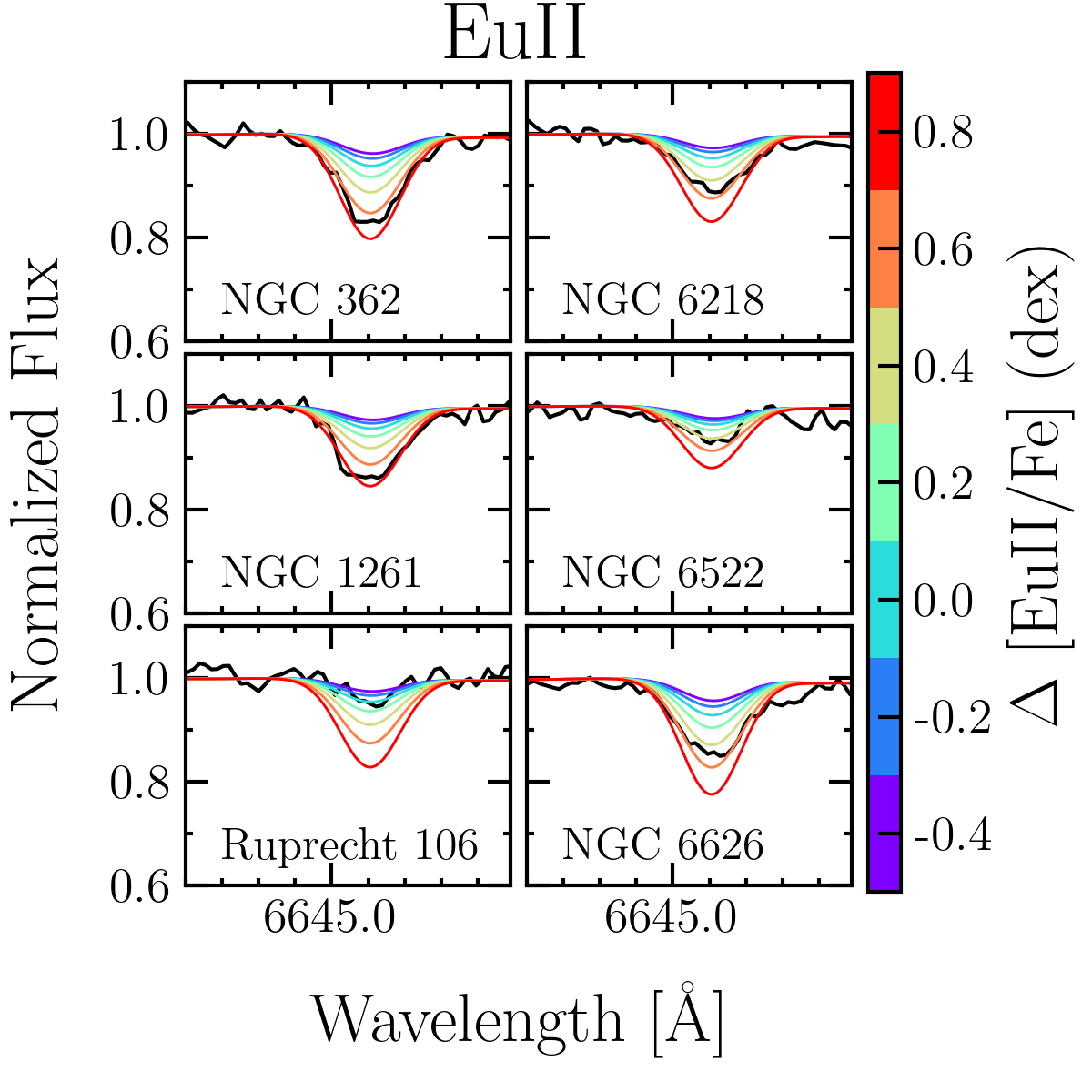}
   \end{minipage}   
   \caption{Comparison around the 4810.5 \r{A} Zn (left panel) and the 6645.1 \r{A} EuII (right panel) lines between the observed spectra and seven synthetic spectra computed assuming the derived atmospheric parameters for each star and varying the level of [Zn/Fe] and [EuII/Fe]. The variations are computed with respect to the abundances measured in Rup106.}
              \label{fig:fits}%
    \end{figure*}   
   We derived both the effective temperature ($T_{\mathrm{eff}}$) and the surface gravity (log $g$) exploiting the \textit{Gaia} Data Release 3 \citep[DR3,][]{GC23} photometric dataset. It is important to note that stars in GCs are located at significant distances and within crowded environments, which may result in a slight degradation of the quality of \textit{Gaia} photometry. Stars with \texttt{ruwe} $>1.4$ were flagged because their photometry might be of lower quality \citep{GC21}. 
   
    We determined the $T_{\mathrm{eff}}$  for all the targets of our sample using the empirically calibrated $\mathrm{(BP-RP)_{0}}$--$T_{\mathrm{eff}}$ relation provided by \citet{mucciarelli21}. The choice of using the $\mathrm{(BP-RP)}$ colour is guided by the fact that it is the most extended in wavelength among \textit{Gaia} colours, thus ensuring an optimal sampling of the spectral energy distribution. This approach also eliminates the need to rely on measurements from other photometric systems, which may have lower precision than \textit{Gaia}. We verified that using other \textit{Gaia} colours results in $T_{\mathrm{eff}}$ differences that are consistently smaller than 50 K, which is below the current uncertainties. The colour excess E(B-V) adopted for each GC is taken from \citet{harris2010}. Given the high extinction in the direction of the Galactic bulge, we corrected the \textit{Gaia} photometry of NGC 6522 and NGC 6626 for effects of differential reddening following the prescription described in \citet{milone2012}, and using the \citet{cardelli1989} extinction law. This analysis was performed on the catalogues provided by \citet{vasiliev&baumgardt2021}, focusing on stars with a probability membership of $> 90 \%$. The effects of the extinction on the observed colour $\mathrm{(BP-RP)}$ have been taken into account following the iterative prescription described in \citet{GC18_extinction}. As the colour--$T_{\mathrm{eff}}$ relation we employed is sensitive to the metallicity of the star, we assumed the \feh value listed in the \citet{harris2010} catalogue as representative of the GC under investigation to get an initial estimate of $T_{\mathrm{eff}}$. We note that estimating the $T_{\mathrm{eff}}$ using the derived \feh values leads to differences from the initial values consistently lower than 20 K. Internal errors in $T_{\mathrm{eff}}$ due to the uncertainties in photometric data, reddening, and the assumed colour--$T_{\mathrm{eff}}$ relation are in the range of 80 - 110 K. We  estimated surface gravities from the Stefan-Boltzmann relation using the photometric $T_{\mathrm{eff}}$ and a representative stellar mass of 0.8 $\mathrm{M_{\odot}}$, typical for RGB stars in old isochrones at this metallicity, the G-band bolometric corrections described by \citet{andrae18}, and the distance provided by \citet{baumgardt&vasiliev2021}. The uncertainties on the log $g$ were derived through the propagation of the errors on $T_{\mathrm{eff}}$, photometry, and distance, and they are always under $0.1 \dex$. In the end, we calculated microturbulent velocities ($v_{\mathrm{t}}$) by minimising the trend between iron abundances and reduced equivalent widths, defined as $\mathrm{log_{10}}$(EW/$\lambda$). To do so, we used $\ge 100$ iron lines per star on average. We assumed a conservative uncertainty of 0.2 km $\mathrm{s^{-1}}$. All the atmospheric parameters are listed in Table \ref{TabSP}.
   

\subsection{Deriving the abundances}

   The determination of chemical abundances of Mg, Si, Ca, TiI, TiII, Cr, Fe, Ni, and Zn was conducted by means of a comparative analysis between the observed equivalent widths (EWs) ---measured with the code \texttt{DAOSPEC} \citep{daospec} and using the automatic tool \texttt{4DAO} \citep{4dao}--- and theoretical line strengths. This analysis was executed using the code \texttt{GALA} \citep{gala}.
   
   We employed the spectral synthesis using the proprietary code \texttt{SALVADOR} to derive the chemical abundances for the species that have hyperfine or isotopic splitting transitions (ScII, V, Mn, Co, Cu, YII, BaII, LaII, and EuII). \texttt{SALVADOR} runs a $\chi^{2}$-minimisation between the observed line and a grid of suitable synthetic spectra computed on the fly by the code \texttt{SYNTHE} \citep{kurucz} varying only the abundance of the matching element and keeping the stellar parameters fixed. Synthetic spectra were computed using all the atomic and molecular transitions available in the Kurucz/Castelli\footnote{\url{http://wwwuser.oats.inaf.it/castelli/linelists.html}} database.

   We used \texttt{ATLAS9} \citep{kurucz} model atmospheres computed assuming plane-parallel geometry, hydrostatic and radiative equilibrium, and local thermodynamic equilibrium for all the chemical elements. We started from an $\alpha$-enhanced model atmosphere ([$\alpha$/Fe] = +0.4 dex) for all the target GCs except for Rup106, for which we used a solar-scaled chemical mixture according to the values of \afeh we derived.  Examples of fits around absorption lines of elements of interest (i.e. Zn and EuII; see Sects. \ref{sec:iron-peak} and \ref{sec:nc}) are displayed in Fig. \ref{fig:fits}. Different synthetic spectra are computed for each star by assuming the atmospheric parameters derived as described in Sect. \ref{sp}, and varying the abundances of Zn and EuII. Finally, we scale the abundance ratios to the solar values using the chemical composition described in \citet{grevesse1998}, as \texttt{ATLAS9} model atmospheres are computed based on this solar
   mixture \citep{castelli&kurucz2003}.

\subsection{Abundance uncertainties}

   During the determination of the uncertainties associated with abundance ratios, it is imperative to account for two principal sources of error: first, internal errors stemming from the measurement of the EW, and second, errors originating from the selection of atmospheric parameters.
   
   We estimated the uncertainties attributed to EW measurements as the dispersion observed around the mean of individual line measurements divided by the root mean square of the number of lines employed in the analysis.
   
   The internal error on chemical abundances derived with spectral synthesis were quantified using Monte Carlo simulations. To do so, we repeated the analysis on 1000 noisy synthetic spectra obtained adding Poissonian noise in order to mimic the S/N of the real spectra. The uncertainty is estimated as the standard deviation of the abundance distribution of the 1000 noisy synthetic spectra.
   
   Finally, we performed new calculations of the chemical abundances, considering also the uncertainties associated with the atmospheric parameters. This involved varying one stellar parameter at a time while keeping the others fixed. In the end, all the sources of error are added in quadrature. Given that our chemical abundances are expressed as abundance ratios ([X/Fe]), the uncertainties on the iron abundance \feh have been taken into account. We estimated the total uncertainty as the squared sum of these two components. Therefore, the final errors were calculated as:
   
   \begin{flalign}
      &\sigma_{\feh}  = 
      \sqrt{\frac{\sigma_{Fe}^{2}}{N_{Fe}}  +
      (\delta_{Fe}^{T_{\mathrm{eff}}})^{2}  +
      (\delta_{Fe}^{\mathrm{log} \; g})^{2} + 
      (\delta_{Fe}^{v_{\mathrm{t}}})^{2}}  ,\\ \nonumber
      & \\ \nonumber
      &\sigma_{[X/Fe]} = \\ 
      & \sqrt{\frac{\sigma_{X}^{2}}{N_{X}} + \frac{\sigma_{Fe}^{2}}{N_{Fe}} + (\delta_{X}^{T_{\mathrm{eff}}} - \delta_{Fe}^{T_{\mathrm{eff}}})^{2} + (\delta_{X}^{\mathrm{log} \; g} - \delta_{Fe}^{\mathrm{log} \; g})^{2} +(\delta_{X}^{v_{\mathrm{t}}} - \delta_{Fe}^{v_{\mathrm{t}}})^{2}},
   \end{flalign}
   where $\sigma_{X,Fe}$ is the dispersion around the mean of chemical abundances, $N_{X,Fe}$ is the number of used lines, and $\delta_{X,Fe}^{i}$ are the abundance differences obtained by varying the parameter $i$. 

\section{Abundances of target GCs}\label{sec:abundances}
   \begin{figure}
   \centering
   \includegraphics[width=.35\textwidth]{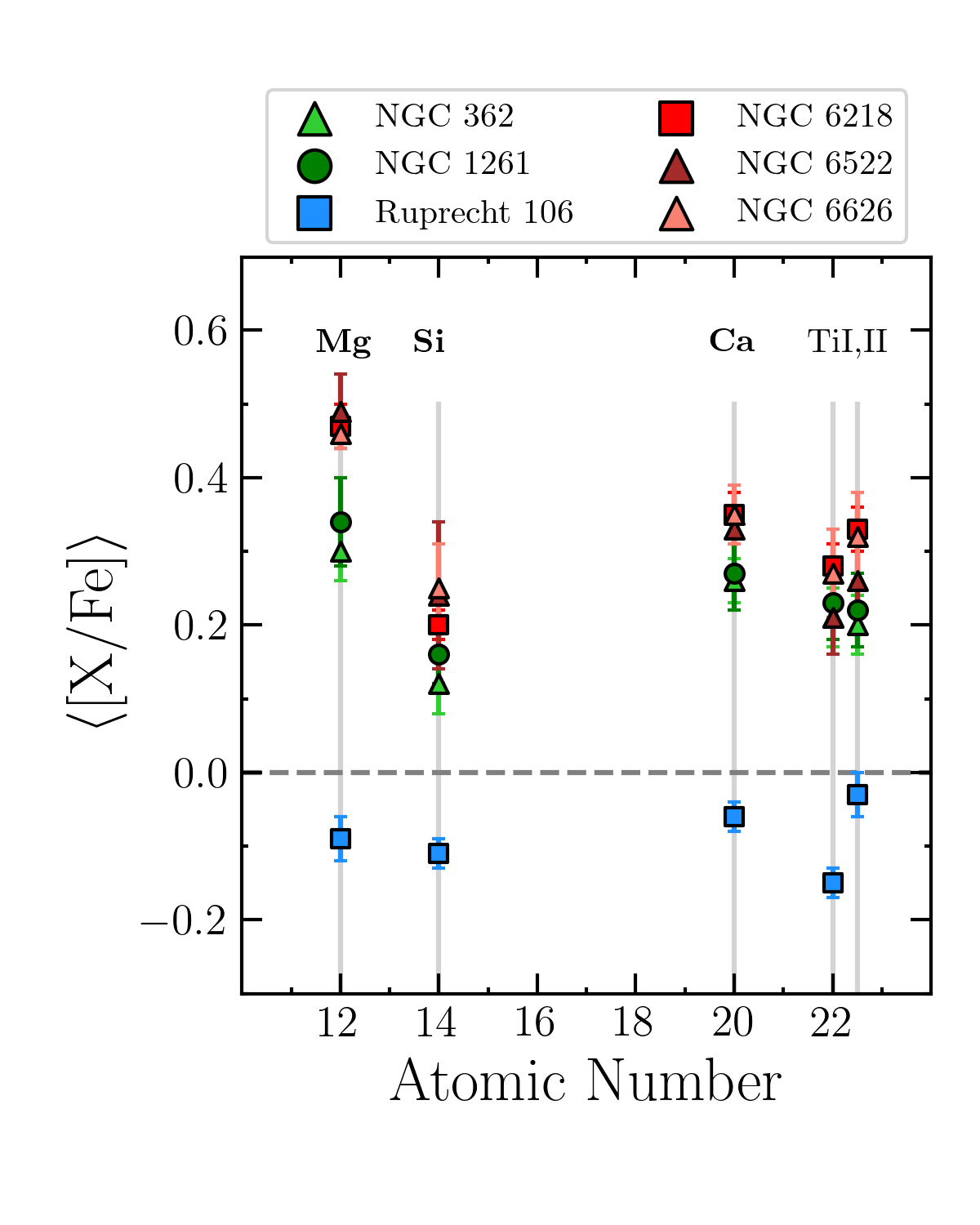}
   \caption{Comparison of mean abundance ratios of the $\alpha$-elements [Mg/Fe], [Si/Fe], [Ca/Fe], [TiI/Fe], and [TiII/Fe] for target GCs NGC 362 (green triangles), NGC 1261 (dark green circles), NGC 6218 (red squares), NGC 6522 (brown trianges), NGC 6626 (pink triangles), and Rup106 (blue squares). Error bars indicate the standard deviation.}
              \label{FigAlpha}%
    \end{figure}    

   \begin{figure}
   \centering
   \includegraphics[width=.35\textwidth]{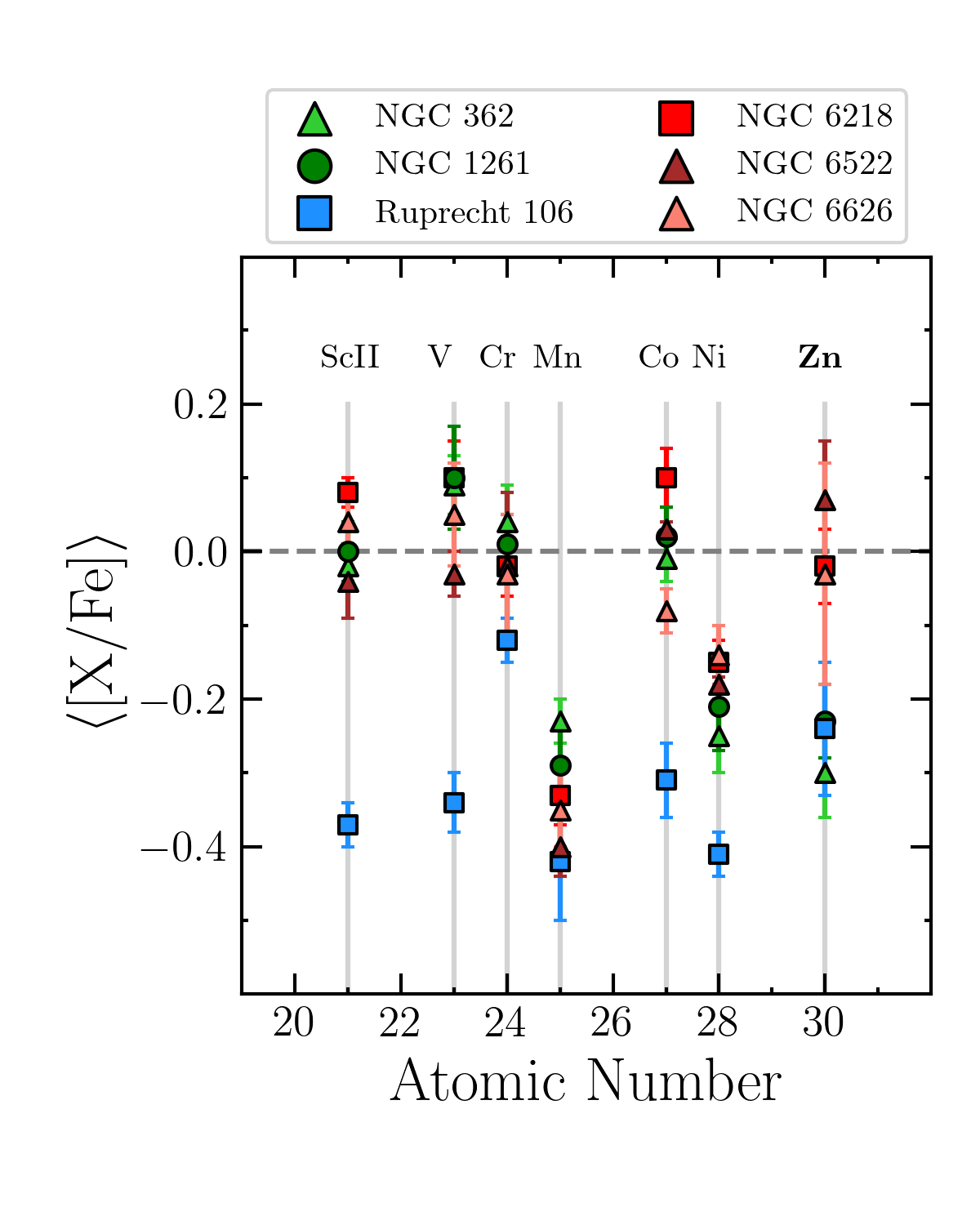}
   \includegraphics[width=.35\textwidth]{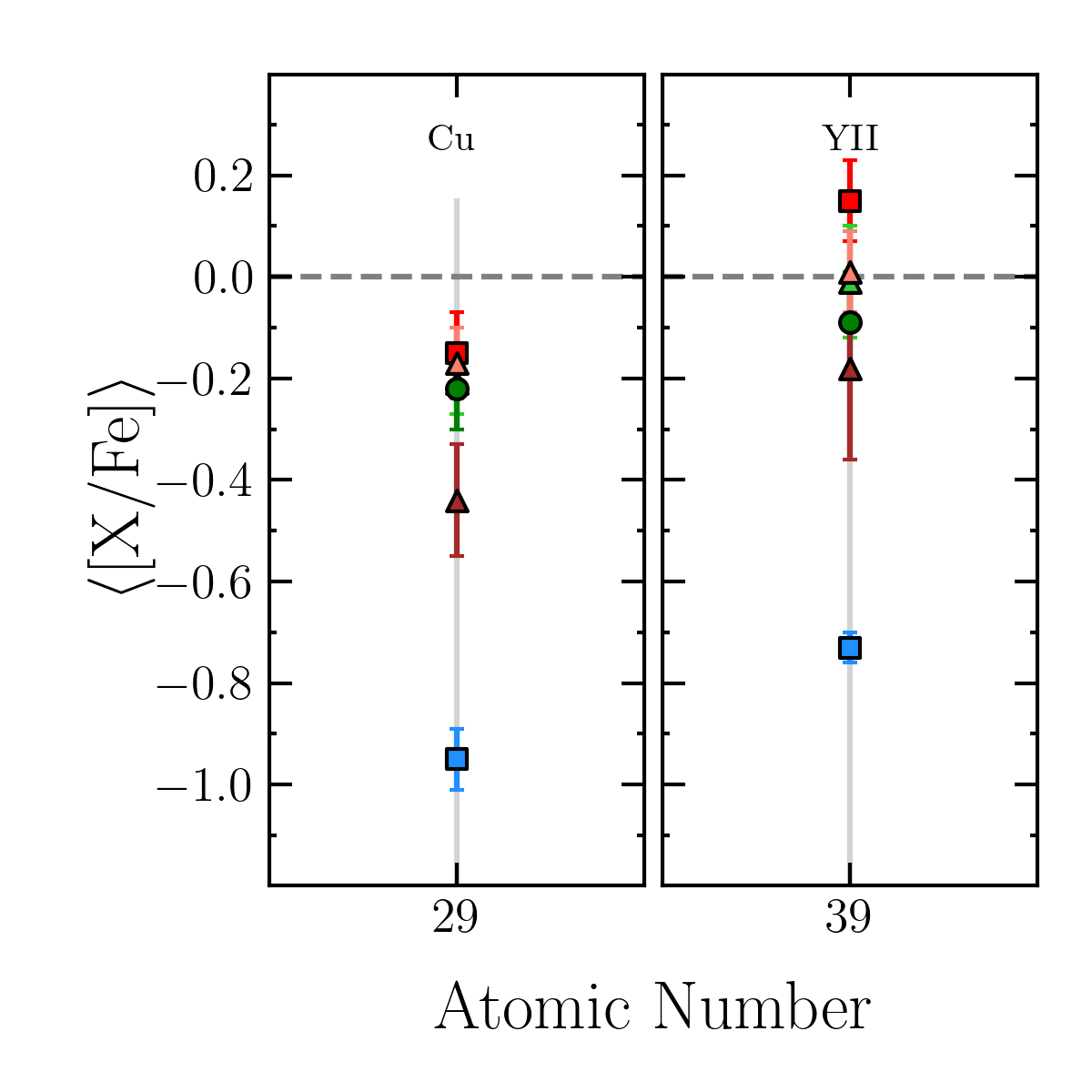}   
   \caption{Comparison of mean abundance ratios of the iron-peak elements [ScII/Fe], [V/Fe], [Cr/Fe], [Mn/Fe], [Co/Fe], [Ni/Fe], and [Zn/Fe] for target GCs (top panel). In the bottom panel we plot the two elements ([Cu/Fe] and [YII/Fe]) with the largest differences observed for Rup106. The colour coding is the same as in Fig. \ref{FigAlpha}. Error bars indicate the standard deviation.}
              \label{FigIronPeak}%
    \end{figure}   

   \begin{figure}
   \centering
   \includegraphics[width=.35\textwidth]{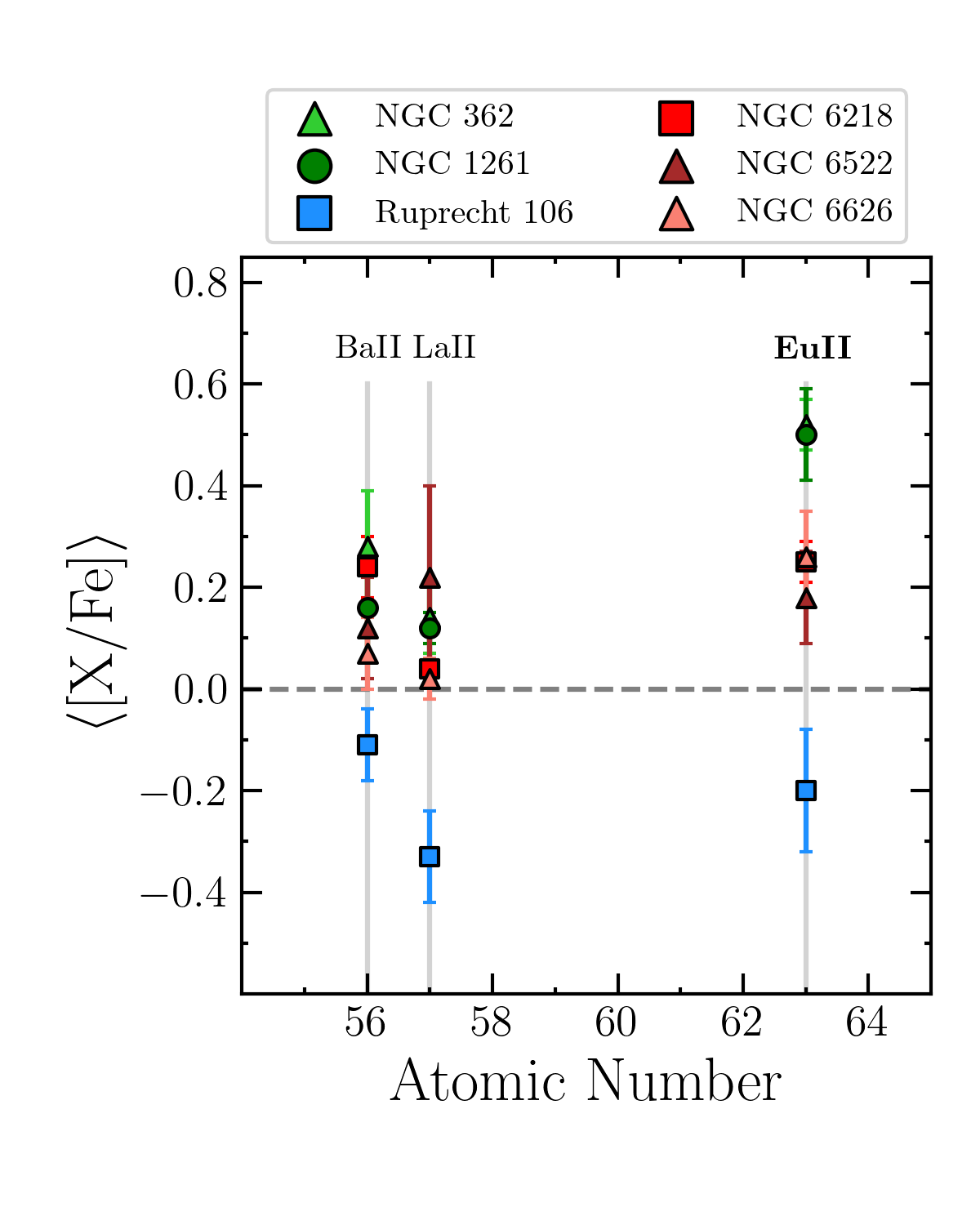}
   \caption{Comparison of mean abundance ratios of the neutron-capture elements [BaII/Fe], [LaII/Fe], and [EuII/Fe] for our target GCs. The colour coding is the same as in Fig. \ref{FigAlpha}. Error bars indicate the standard deviation.}
              \label{FigNeutronCapture}%
    \end{figure}    

   \begin{figure}
   \centering
   \includegraphics[width=.35\textwidth]{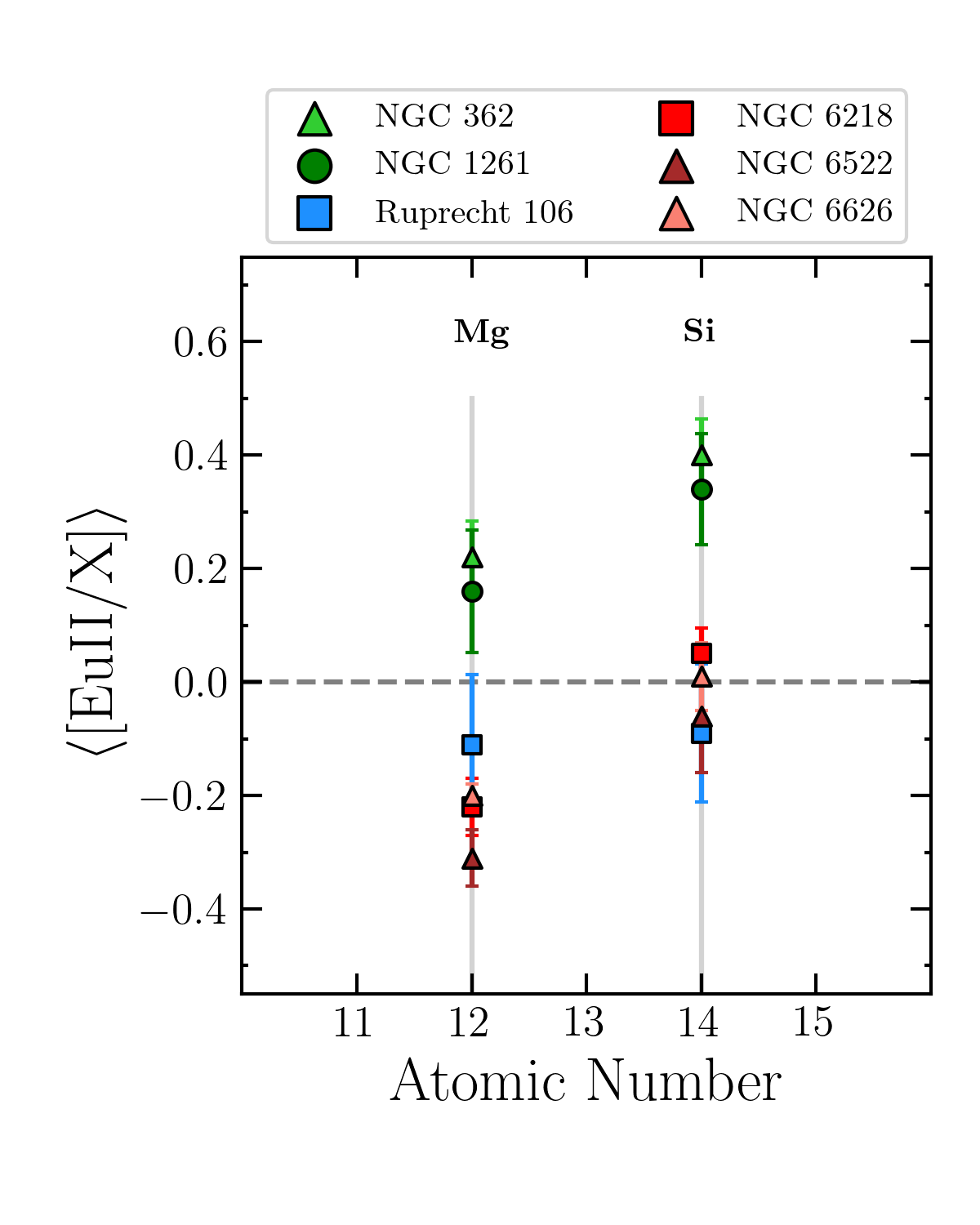}
   \includegraphics[width=.35\textwidth]{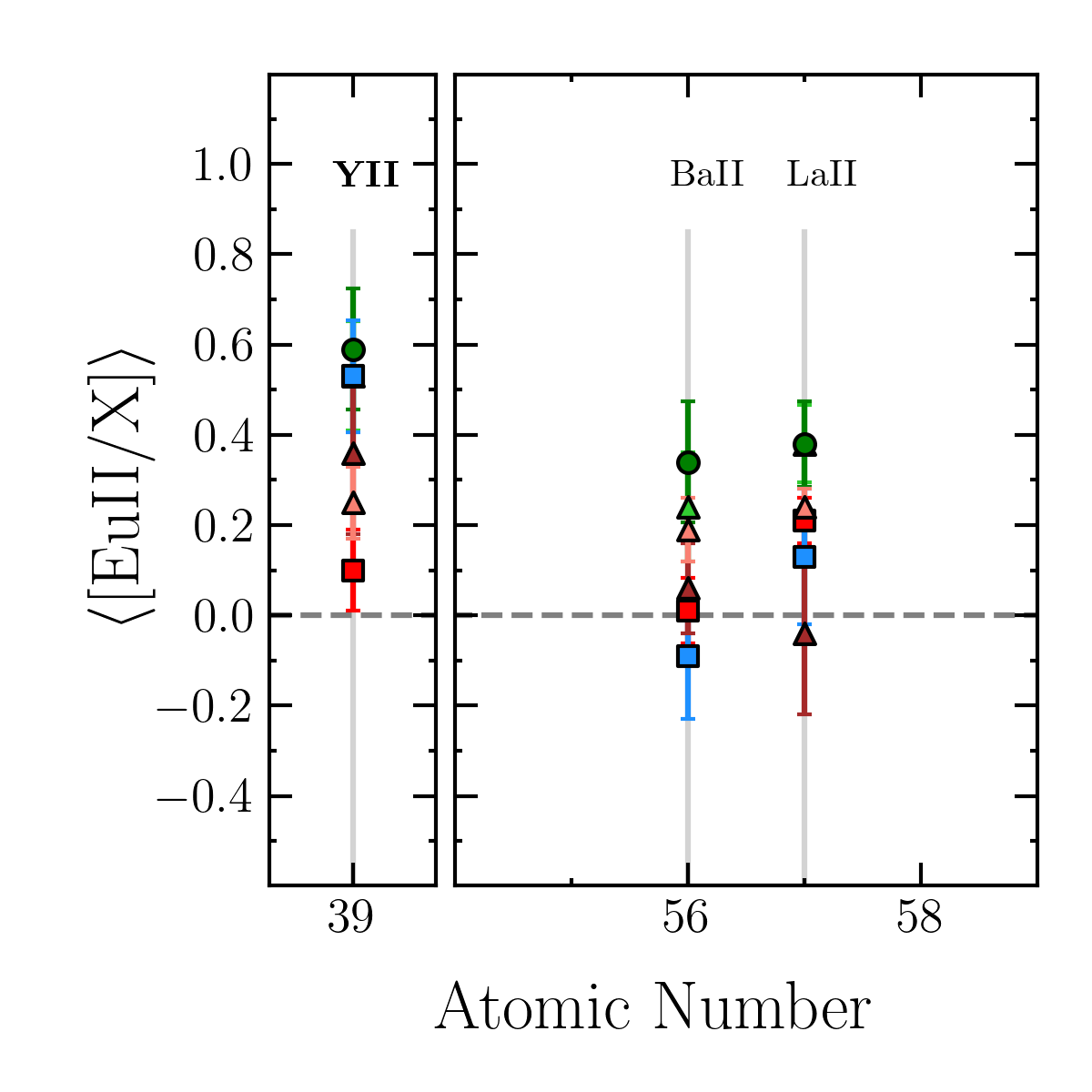}   
   \caption{Comparison of mean abundance ratios of EuII relative to other chemical elements for target GCs. The colour coding is the same as in Fig. \ref{FigAlpha}. Error bars indicate the standard deviation.}
              \label{FigEu}%
    \end{figure}    
\begin{table*}
  \caption{Chemical abundances for the $\alpha$-elements for the target stars (extract).}\label{TabAbuAlpha}
  \centering
  \begin{tabular}{cccccccc} 
   \hline             
Cluster & Star ID & [Fe/H] & [Mg/Fe] & [Si/Fe] & [Ca/Fe] & [TiI/Fe] & [TiII/Fe]  \\ 
 & & (dex) & (dex) & (dex) & (dex) & (dex) & (dex)  \\ 
\hline 
NGC 362 &                       1037     &      -1.04   $\pm$   0.09     &       0.28    $\pm$   0.04     &      0.10    $\pm$   0.10     &      0.27    $\pm$   0.05     &       0.20    $\pm$   0.09     &      0.21    $\pm$   0.12      \\
NGC 362 &                       1840     &      -1.04   $\pm$   0.09     &       0.25    $\pm$   0.04     &      0.08    $\pm$   0.10     &      0.22    $\pm$   0.06     &       0.18    $\pm$   0.10     &      0.17    $\pm$   0.11      \\
NGC 362 &                       2683     &      -1.12   $\pm$   0.06     &       0.36    $\pm$   0.05     &      0.20    $\pm$   0.11     &      0.30    $\pm$   0.08     &       0.24    $\pm$   0.15     &      0.21    $\pm$   0.11      \\
NGC 362 &                       3392     &      -1.07   $\pm$   0.09     &       0.28    $\pm$   0.03     &      0.11    $\pm$   0.10     &      0.24    $\pm$   0.05     &       0.17    $\pm$   0.09     &      0.22    $\pm$   0.12      \\
...      &   ...      & ...   & ...   & ...   & ...   & ...   & ...         \\
   \hline
  \end{tabular}
\tablefoot{The entire table is available at the CDS.}  
\end{table*}

\begin{table*}
  \caption{Chemical abundances for the iron-peak elements for the target stars (extract).}\label{TabAbu}
  \centering
  \begin{tabular}{ccccccc} 
   \hline             
Cluster & Star ID & [ScII/Fe] & [V/Fe] & [Cr/Fe] & [Mn/Fe]  \\ 
 & & (dex) & (dex) & (dex) & (dex)  \\ 
\hline 
NGC 362 &                       1037     &      -0.03   $\pm$   0.08     &       0.09    $\pm$   0.18     &      -0.01   $\pm$   0.09     &      -0.26   $\pm$   0.15     \\
NGC 362 &                       1840     &      -0.04   $\pm$   0.07     &       0.08    $\pm$   0.18     &      -0.04   $\pm$   0.09     &      -0.22   $\pm$   0.15     \\
NGC 362 &                       2683     &      0.01    $\pm$   0.07     &       0.13    $\pm$   0.18     &      0.12    $\pm$   0.12     &      -0.20   $\pm$   0.15     \\
NGC 362 &                       3392     &      -0.07   $\pm$   0.08     &       0.06    $\pm$   0.18     &      -0.01   $\pm$   0.08     &      -0.27   $\pm$   0.15     \\
...  &  ...    & ...   & ...   & ...   & ...                 \\
   \hline
Cluster & Star ID & [Co/Fe] & [Ni/Fe] & [Cu/Fe] & [Zn/Fe] \\ 
 & & (dex) & (dex) & (dex) & (dex)   \\ 
\hline 
NGC 362 &                       1037     &      0.01    $\pm$   0.13     &       -0.21   $\pm$   0.04     &      -0.26   $\pm$   0.18     &      -0.20   $\pm$   0.15      \\
NGC 362 &                       1840     &      0.03    $\pm$   0.13     &       -0.22   $\pm$   0.04     &      -0.27   $\pm$   0.18     &      -0.27   $\pm$   0.14      \\
NGC 362 &                       2683     &      -0.02   $\pm$   0.13     &       -0.31   $\pm$   0.04     &      -0.15   $\pm$   0.18     &      -0.39   $\pm$   0.13      \\
NGC 362 &                       3392     &      0.00    $\pm$   0.13     &       -0.14   $\pm$   0.04     &      -0.31   $\pm$   0.18     &      -0.24   $\pm$   0.14      \\
...  &  ...    & ...   & ...   & ...   & ...                  \\
   \hline   
  \end{tabular}
\tablefoot{The entire table is available at the CDS.}    
\end{table*}

\begin{table*}
  \caption{Chemical abundances for the neutron-capture elements for the target stars (extract).}\label{TabAbuNC}
  \centering
  \begin{tabular}{cccccc} 
   \hline             
Cluster & Star ID & [YII/Fe] & [BaII/Fe] & [LaII/Fe] & [EuII/Fe]  \\ 
 & & (dex) & (dex) & (dex) & (dex)   \\ 
\hline 
NGC 362 &                       1037     &      -0.01   $\pm$   0.15     &       0.27    $\pm$   0.17     &      0.15    $\pm$   0.05     &      0.51    $\pm$   0.05      \\
NGC 362 &                       1840     &      -0.05   $\pm$   0.15     &       0.24    $\pm$   0.17     &      0.13    $\pm$   0.05     &      0.50    $\pm$   0.04      \\
NGC 362 &                       2683     &      0.05    $\pm$   0.15     &       0.28    $\pm$   0.17     &      0.16    $\pm$   0.05     &      0.52    $\pm$   0.04      \\
NGC 362 &                       3392     &      -0.10   $\pm$   0.15     &       0.19    $\pm$   0.17     &      0.12    $\pm$   0.05     &      0.44    $\pm$   0.05      \\
...              & ...   & ...   & ...   & ...     & ...              \\
   \hline
  \end{tabular}
\tablefoot{The entire table is available at the CDS.}    
\end{table*}

\begin{table*}
  \caption{Mean abundances ratios for the six GCs analysed in this work.}\label{TabAbuMean}
  \centering
  \begin{tabular}{cccc} 
   \hline             
Element & NGC 362 & NGC 1261 & NGC 6218  \\ 
\hline 
$\langle$[Fe/H]$\rangle$              	          & 	-1.06	       $\pm$	0.01	   (0.03) & 	-1.14	       $\pm$	0.02	   (0.06) & 	-1.24	       $\pm$	0.01	      (0.02)	           \\
$\langle$[Mg/Fe]$\rangle$             	          & 	0.30	       $\pm$	0.01	   (0.04) & 	0.34	       $\pm$	0.02	   (0.06) & 	0.47	       $\pm$	0.01	      (0.03)	           \\
$\langle$[Si/Fe]$\rangle$             	          & 	0.12	       $\pm$	0.01	   (0.04) & 	0.16	       $\pm$	0.01	   (0.04) & 	0.20	       $\pm$	0.01	      (0.02)	           \\
$\langle$[Ca/Fe]$\rangle$             	          & 	0.26	       $\pm$	0.01	   (0.03) & 	0.27	       $\pm$	0.01	   (0.05) & 	0.35	       $\pm$	0.01	      (0.03)	           \\
$\langle$[ScII/Fe]$\rangle$           	          & 	-0.02	       $\pm$	0.01	   (0.03) & 	0.00	       $\pm$	0.01	   (0.04) & 	0.08	       $\pm$	0.01	      (0.02)	           \\
$\langle$[Ti/Fe]$\rangle$             	          & 	0.21	       $\pm$	0.01	   (0.04) & 	0.23	       $\pm$	0.01	   (0.05) & 	0.28	       $\pm$	0.01	      (0.03)	           \\
$\langle$[TiII/Fe]$\rangle$           	          & 	0.20	       $\pm$	0.01	   (0.04) & 	0.22	       $\pm$	0.01	   (0.05) & 	0.33	       $\pm$	0.01	      (0.03)	           \\
$\langle$[V/Fe]$\rangle$              	          & 	0.09	       $\pm$	0.01	   (0.04) & 	0.10	       $\pm$	0.02	   (0.07) & 	0.10	       $\pm$	0.02	      (0.05)	           \\
$\langle$[Cr/Fe]$\rangle$             	          & 	0.04	       $\pm$	0.01	   (0.05) & 	0.01	       $\pm$	0.01	   (0.04) & 	-0.02	       $\pm$	0.01	      (0.04)	           \\
$\langle$[Mn/Fe]$\rangle$             	          & 	-0.23	       $\pm$	0.01	   (0.03) & 	-0.29	       $\pm$	0.01	   (0.05) & 	-0.33	       $\pm$	0.01	      (0.04)	           \\
$\langle$[Co/Fe]$\rangle$             	          & 	-0.01	       $\pm$	0.01	   (0.03) & 	0.02	       $\pm$	0.01	   (0.04) & 	0.10	       $\pm$	0.01	      (0.04)	           \\
$\langle$[Ni/Fe]$\rangle$             	          & 	-0.25	       $\pm$	0.01	   (0.05) & 	-0.21	       $\pm$	0.02	   (0.06) & 	-0.15	       $\pm$	0.01	      (0.03)	           \\
$\langle$[Cu/Fe]$\rangle$             	          & 	-0.21	       $\pm$	0.02	   (0.06) & 	-0.22	       $\pm$	0.02	   (0.08) & 	-0.15	       $\pm$	0.02	      (0.08)	           \\
$\langle$[Zn/Fe]$\rangle$             	          & 	-0.30	       $\pm$	0.02	   (0.06) & 	-0.23	       $\pm$	0.01	   (0.05) & 	-0.02	       $\pm$	0.02	      (0.05)	           \\
$\langle$[YII/Fe]$\rangle$            	          & 	-0.01	       $\pm$	0.03	   (0.11) & 	-0.09	       $\pm$	0.03	    (0.10) & 	0.15	       $\pm$	0.02	      (0.08)	           \\
$\langle$[BaII/Fe]$\rangle$           	          & 	0.28	       $\pm$	0.03	   (0.11) & 	0.16	       $\pm$	0.03	    (0.10) & 	0.24	       $\pm$	0.02	      (0.06)	           \\
$\langle$[LaII/Fe]$\rangle$           	          & 	0.14	       $\pm$	0.02	   (0.07) & 	0.12	       $\pm$	0.01	   (0.03) & 	0.04	       $\pm$	0.01	      (0.03)	           \\
$\langle$[EuII/Fe]$\rangle$           	          & 	0.52	       $\pm$	0.01	   (0.05) & 	0.50	       $\pm$	0.02	   (0.09) & 	0.25	       $\pm$	0.01	      (0.04)	           \\
   \hline             
Element & NGC 6522 & NGC 6626 & Ruprecht 106  \\ 
\hline 
$\langle$[Fe/H]$\rangle$               	          & 	-1.07	       $\pm$	0.01	   (0.05) & 	-1.11	       $\pm$	0.01	   (0.06) & 	-1.30	       $\pm$	0.01	      (0.05)	           \\
$\langle$[Mg/Fe]$\rangle$              	          & 	0.49	       $\pm$	0.03	   (0.05) & 	0.46	       $\pm$	0.01	   (0.02) & 	-0.09	       $\pm$	0.01	      (0.03)	           \\
$\langle$[Si/Fe]$\rangle$              	          & 	0.24	       $\pm$	0.05	    (0.10) & 	0.25	       $\pm$	0.02	   (0.06) & 	-0.11	       $\pm$	0.01	      (0.02)	           \\
$\langle$[Ca/Fe]$\rangle$              	          & 	0.33	       $\pm$	0.01	   (0.02) & 	0.35	       $\pm$	0.01	   (0.04) & 	-0.06	       $\pm$	0.01	      (0.02)	           \\
$\langle$[ScII/Fe]$\rangle$            	          & 	-0.04	       $\pm$	0.02	   (0.05) & 	0.04	       $\pm$	0.01	   (0.04) & 	-0.37	       $\pm$	0.01	      (0.03)	           \\
$\langle$[Ti/Fe]$\rangle$              	          & 	0.21	       $\pm$	0.03	   (0.05) & 	0.27	       $\pm$	0.01	   (0.06) & 	-0.15	       $\pm$	0.01	      (0.02)	           \\
$\langle$[TiII/Fe]$\rangle$            	          & 	0.26	       $\pm$	0.03	   (0.07) & 	0.32	       $\pm$	0.01	   (0.06) & 	-0.03	       $\pm$	0.01	      (0.03)	           \\
$\langle$[V/Fe]$\rangle$               	          & 	-0.03	       $\pm$	0.01	   (0.03) & 	0.05	       $\pm$	0.02	   (0.07) & 	-0.34	       $\pm$	0.01	      (0.04)	           \\
$\langle$[Cr/Fe]$\rangle$              	          & 	-0.02	       $\pm$	0.05	    (0.10) & 	-0.03	       $\pm$	0.02	   (0.08) & 	-0.12	       $\pm$	0.01	      (0.03)	           \\
$\langle$[Mn/Fe]$\rangle$              	          & 	-0.40	       $\pm$	0.02	   (0.04) & 	-0.35	       $\pm$	0.01	   (0.05) & 	-0.42	       $\pm$	0.03	      (0.08)	           \\
$\langle$[Co/Fe]$\rangle$              	          & 	0.03	       $\pm$	0.01	   (0.01) & 	-0.08	       $\pm$	0.01	   (0.03) & 	-0.31	       $\pm$	0.02	      (0.05)	           \\
$\langle$[Ni/Fe]$\rangle$              	          & 	-0.18	       $\pm$	0.01	   (0.01) & 	-0.14	       $\pm$	0.01	   (0.04) & 	-0.41	       $\pm$	0.01	      (0.03)	           \\
$\langle$[Cu/Fe]$\rangle$              	          & 	-0.44	       $\pm$	0.05	   (0.11) & 	-0.17	       $\pm$	0.02	   (0.07) & 	-0.95	       $\pm$	0.02	      (0.06)	           \\
$\langle$[Zn/Fe]$\rangle$              	          & 	0.07	       $\pm$	0.04	   (0.08) &     -0.03	       $\pm$	0.04	    (0.15) & 	-0.24	       $\pm$	0.03	      (0.09)	           \\
$\langle$[YII/Fe]$\rangle$             	          & 	-0.18	       $\pm$	0.09	   (0.18) & 	0.01	       $\pm$	0.02	   (0.08) & 	-0.73	       $\pm$	0.01	      (0.03)	           \\
$\langle$[BaII/Fe]$\rangle$            	          & 	0.12	       $\pm$	0.05	    (0.10) & 	0.07	       $\pm$	0.02	   (0.07) & 	-0.11	       $\pm$	0.02	      (0.07)	           \\
$\langle$[LaII/Fe]$\rangle$            	          & 	0.22	       $\pm$	0.10	   (0.18) & 	0.02	       $\pm$	0.01	   (0.04) & 	-0.33	       $\pm$	0.03	      (0.09)	           \\
$\langle$[EuII/Fe]$\rangle$            	          & 	0.18	       $\pm$	0.06	   (0.09) & 	0.26	       $\pm$	0.02	   (0.09) & 	-0.20	       $\pm$	0.04	      (0.12)	           \\

   \hline
  \end{tabular}
\tablefoot{The standard deviation is reported in parenthesis.}     
\end{table*}

    In this section, we discuss the abundances of $\alpha$-, iron-peak, and neutron-capture elements of our target GCs. We exercise caution when interpreting the abundances of certain chemical elements that may be influenced by the presence of multiple populations within GCs \citep[e.g. O, Na, Mg, and Al; see ][]{bastian&lardo18}. Among these species, we focus on Mg, which exhibits minimal dispersion within five out of the six GCs in our sample. This outcome aligns with expectations, particularly considering that most of the targeted systems belong to the low-mass and high-metallicity regime of the MW GC system, where the efficiency of the MgAl burning channel diminishes \citep{ventura2013,dellagli2018,deimer2024}. Indeed, we find spread in the distribution of [Mg/Fe] only in the most massive among our target cluster, NGC 6626. As we are only concerned with the pristine chemical composition of the gas that formed the cluster, we derived the mean value of [Mg/Fe] using only first-generation stars. The results obtained from the homogeneous chemical analysis of 66 RGB stars of the sample of six GCs are listed in Tables \ref{TabAbuAlpha} - \ref{TabAbuNC}. The mean abundances for each GC are available in Table \ref{TabAbuMean}. The average abundance ratios of the target GCs are illustrated in Figs. \ref{FigAlpha} - \ref{FigNeutronCapture}. 

\subsection{$\alpha$-elements}

    Fig. \ref{FigAlpha} shows the mean values of the $\alpha$-elements abundances. These chemical elements are predominantly synthesised in massive stars and subsequently dispersed into the interstellar medium through core-collapse supernovae \citep[CC-SNe,][]{kobayashi2009,romano2010,kobayashi2020} events. Additionally, there is a minor but discernible contribution from Type Ia supernovae (SNe Ia), which is particularly notable for Ca and Ti, as discussed in \citet{kobayashi2020}. 
    
    First of all, we find that the two GCs associated to GSE (green filled symbols) show identical abundances, within the uncertainties, in all the analysed $\alpha$-elements. Second, we show that the mean abundances of Mg, Si, and Ca of the three in situ GCs, namely NGC 6218, NGC 6522, and NGC 6626 (red, brown and pink filled symbols, respectively), are fully compatible among each other. For Ti, this consistency is less evident. Third, we find that NGC 362 and NGC 1261 are $\alpha$-depleted compared to the in situ GCs. They display differences in the $\alpha$-element at the $1\sigma$ level in Mg and Ca when comparing the average abundances and the respective standard deviations (see Appendix A.1 for a comparison with literature). Finally, we derive subsolar abundances in the $\alpha$-elements of Rup106 (blue filled squares), specifically with \afeh $\sim - 0.1 \dex$, which is in good agreement with the literature \citep[][V13; see Appendix A.2 for a comparison with the results of V13]{brown1997}. This is, on average, $0.3 - 0.4 \dex$ lower than the other five target clusters. 

    These results guide us as to the degree to which we can use $\alpha$-elements to distinguish the origin of GCs. The observed trend is a good match to our expectations based on the cluster chrono-dynamical associations. Indeed, GSE clusters NGC 1261 and NGC 362 share the same $\alpha$-element abundances, as expected for clusters accreted from the same progenitor, with lower \afeh values compared to NGC 6218, NGC 6522, and NGC 6626. Indeed, the in situ GCs consistently show higher \afeh abundances, as predicted for systems born in the MW at this metallicity, where the contribution of SNIa to the gas chemical enrichment was not yet significant. Finally, Rup106 has an even lower \afeh than the GSE GCs, and can be distinguished from them in this respect; this indicates an even more different birth environment, probably characterised by very low star formation efficiency.

\subsection{Iron-peak elements} \label{sec:iron-peak}

    Iron-peak elements primarily originate in Type II CC-SNe and hypernovae (HNe), with some contribution also from SNe Ia. Specifically, elements such as Sc, Cu, and Zn are predominantly synthesised by massive stars, while for V and Co the contribution of SNe Ia is not negligible. On the other hand, Cr, Mn, and Ni are primarily produced by SNe Ia \citep{romano2010,kobayashi2020}. 
    
    The results shown in Fig. \ref{FigIronPeak} indicate that most of the iron-peak elements do not provide a means to effectively and clearly discriminate between in situ and GSE clusters. 
    Indeed, our target GCs display coherent values in all the chemical spaces with the exception, once again, of Rup106, which is underabundant in ScII, V, Mn, Co, and Ni. In the bottom panel of Fig. \ref{FigIronPeak}, we highlight the two elements where Rup106 is most depleted compared to the other systems. Specifically, Cu is the most underabundant among the chemical elements derived in this work ([Cu/Fe] $= -0.95 \dex$ for Rup106). Such low values of [Cu/Fe] can be reproduced in stellar systems with a chemical evolutionary model assuming very inefficient star formation \citep[star formation rate $< 5 \times 10^{-4} \; \mathrm{M_{\odot}} \; \mathrm{yr^{-1}}$,][]{mucciarelli2021NatAs}. 
    Thus, the pronounced depletion in all of these elements might indicate that Rup106 was born in an environment where the contribution of massive stars was extremely low.
    
    Among the iron-peak elements, the only one to demonstrate a statistically significant difference between accreted and in situ GCs is Zn: all the in situ GCs are overabundant with respect to GSE GCs and Rup106 at the 1$\sigma$ level. In passing, we note that NGC 6626 shows a significant spread in [Zn/Fe], as already found in \citet{villanova17}.  Zn therefore appears to be a good tracer of the in situ or accreted origin of GCs, but it is not sensitive enough to help distinguish among different accreted progenitors. Such a conclusion was also drawn by \citet{minelli21}, even though these authors used a sample of GCs at higher metallicity. Indeed, high [Zn/Fe] ratios are expected for in situ GCs, and the rationale behind this expectation lies in the fact that the primary contributors to Zn production are hypernovae, which are linked to stars with masses of around $25 - 30 \; \mathrm{M_{\odot}}$ \citep{romano2010, kobayashi2020}. Consequently, galaxies with lower star formation rates should manifest lower [Zn/Fe] ratios due to the reduced impact of massive stars \citep{yan2020}.

\subsection{Neutron-capture elements}\label{sec:nc}

    Slow ($s$-) neutron-capture elements are mainly synthesised in asymptotic giant branch (AGB) stars. In particular, light $s$-process elements (e.g. Y) are produced to a large extent by intermediate-mass AGB stars. Conversely, heavy $s$-process elements (e.g. Ba, La) are mainly produced in AGB stars with masses lower than $4 \; \mathrm{M_{\odot}}$ \citep{kobayashi2020}. Rapid ($r$-) neutron-capture elements (e.g. Eu) are synthesised during a broad spectrum of events, such as special types of CC-SNe \citep[rCCSNe, e.g. magnetorotational SNe, collapsars;][]{mosta2018,siegel2019,kobayashi2020} and neutron-star mergers \citep[NSMs,][]{lattimer1974}.

    First of all, we observe that the in situ NGC 6218 is enhanced ($\ge 0.2 \dex$) in [YII/Fe] compared to the other two in situ GCs and two GSE GCs. Additionally, among the neutron-capture elements, this is the chemical space in which Rup106 is the most depleted (see the bottom panel of Fig. \ref{FigIronPeak}). This finding is particularly intriguing, as Cu, the other element where Rup106 is notably depleted, is also produced by massive stars through weak $s$-processes, with a small contribution also by AGB stars \citep{kobayashi2009,kobayashi2020}. Thus, if the lack of weak $s$-process products is responsible for the low derived Cu abundance, light neutron-capture elements are also expected to be similarly depleted, as is observed for Y.
    
    When examining the abundances of the other neutron-capture elements (see Fig. \ref{FigNeutronCapture}), the most notable distinctions between various GCs arise in the \textit{r}-process element Eu. The two GSE GCs exhibit values that are $\sim 0.3 \dex$ higher compared to in situ GCs. The enhancement in \textit{r}-process of NGC 362 and NGC 1261 was also highlighted by \citet{monty2023_2} and by \citet{koch-hansen2021}, respectively, and resembles a behaviour seen in both GSE field stars at this metallicity \citep{nissen&schuster2011,fishlock2017,aguado21,matsuno2021,naidu22} and surviving dwarf galaxies, such as the Large Magellanic Cloud, Sculptor, and Fornax \citep{tolstoy09,letarte2010,vanderswaelmen2013,lemasle2014}. This might be explained by a joint effect of the different star formation efficiency in the progenitor galaxy and the impact of delayed $r$-process sources (i.e. NSMs), with minimum delay times of the order of $10$ - $1000$ Myr in producing Eu \citep{cescutti2015,siegel2019,naidu22,ou2024}. Once again, Rup106 is $0.3 - 0.6 \dex$ underabundant in all neutron-capture elements, indicating a different birth environment. 
    
    Recently, the [EuII/$\alpha$] ratio has been proposed as a strong discriminant between accreted and in situ GCs due to its effectiveness in tracing different timescales of star formation \citep{monty2024,ou2024}. As shown in the top panel of Fig. \ref{FigEu}, the observed differences between the two GSE GCs and the in situ GCs are consistent with literature results, while we do not find signs of a consistent enhancement in [EuII/$\alpha$] for Rup106. Thus, we recommend caution when using this diagnostic, as it might be effective for individual progenitors like GSE, but not in general for birth environments spanning a wide range of formation and evolutionary properties. Moreover, as demonstrated by the [EuII/BaII] and the [EuII/LaII] ratios (see the bottom panel of Fig. \ref{FigEu}), the in situ GCs and Rup106 show signs of higher efficiency in the production of heavy \textit{s-}process elements with respect to the \textit{r-}process elements when compared to the GSE GCs. We highlight that the accreted systems considered here can be either \textit{r-}process dominated (GSE) or \textit{s-}process dominated (Rup106). Moreover, we find a difference of $\ge 0.2 \dex$ in [EuII/YII] between the three accreted clusters and the three in situ systems. Interestingly, the [EuII/YII] ratio is similar in all accreted systems, regardless of the \textit{r-/s-} process dominance. This comparison may suggest that the applicability of [EuII/YII] as a tracer of the accreted or in situ origin is more general than [EuII/BaII] and [EuII/LaII]. Thus, this abundance ratio can play an important role in discriminating between in situ and accreted GCs at this metallicity, but, as for Zn, we point out that this diagnostic might lack the required sensitivity to discriminate between independent accreted progenitors.

\section{Comparison with field stars from the likely progenitor system}\label{sec:literature}

   \begin{figure*}
   \centering
   \begin{minipage}{0.33\textwidth}
        \centering
        \includegraphics[width=1.0\textwidth]{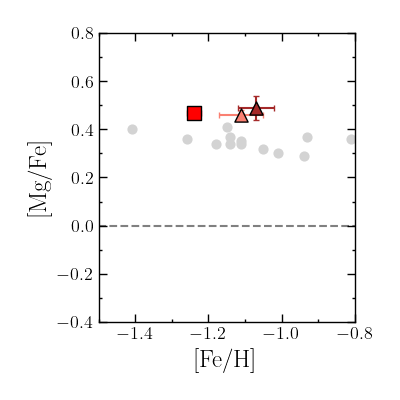} 
        \includegraphics[width=1.0\textwidth]{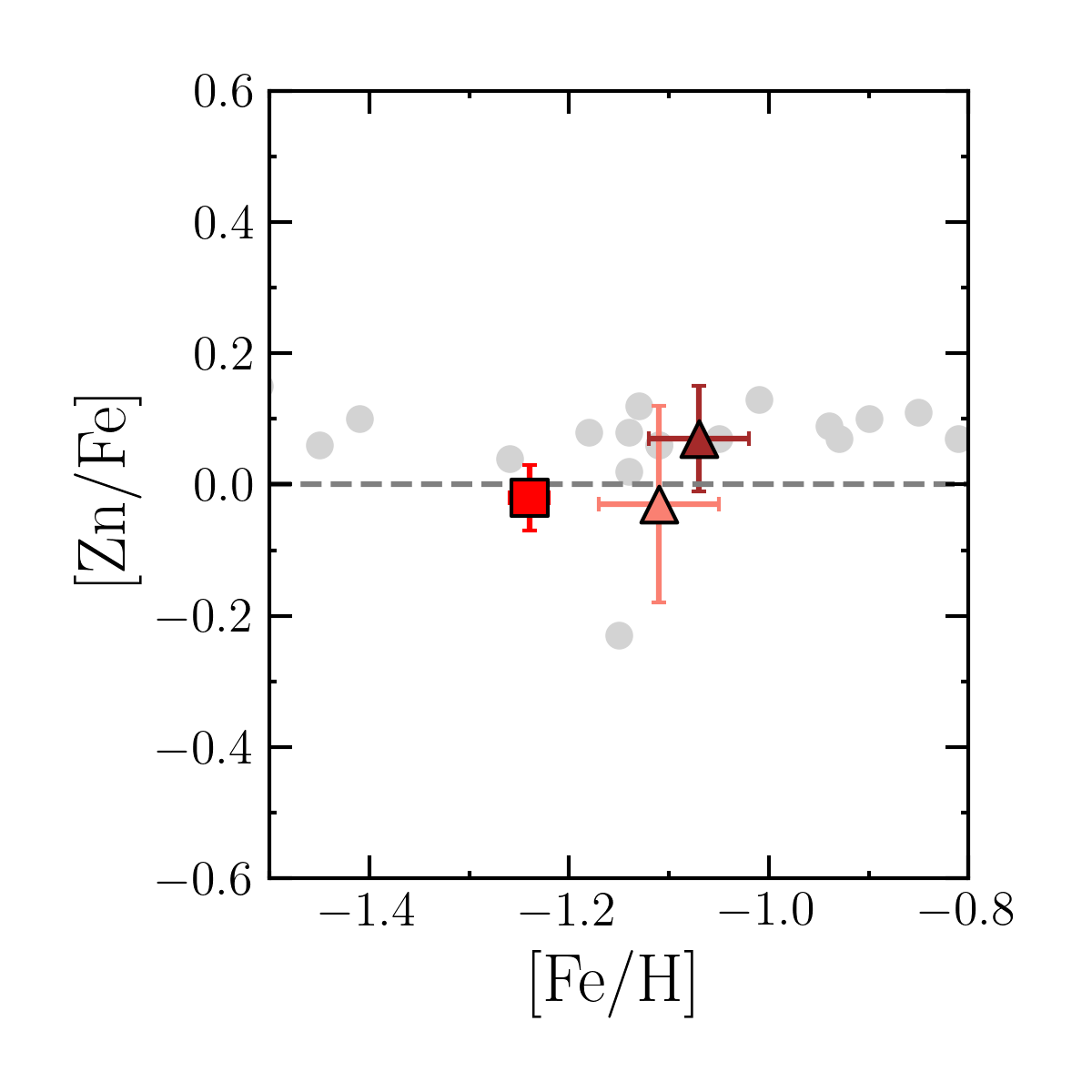}
   \end{minipage}\hfill
   \begin{minipage}{0.33\textwidth}
        \centering
        \includegraphics[width=1.0\textwidth]{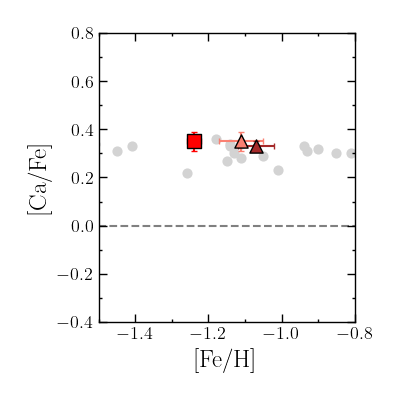}
        \includegraphics[width=1.0\textwidth]{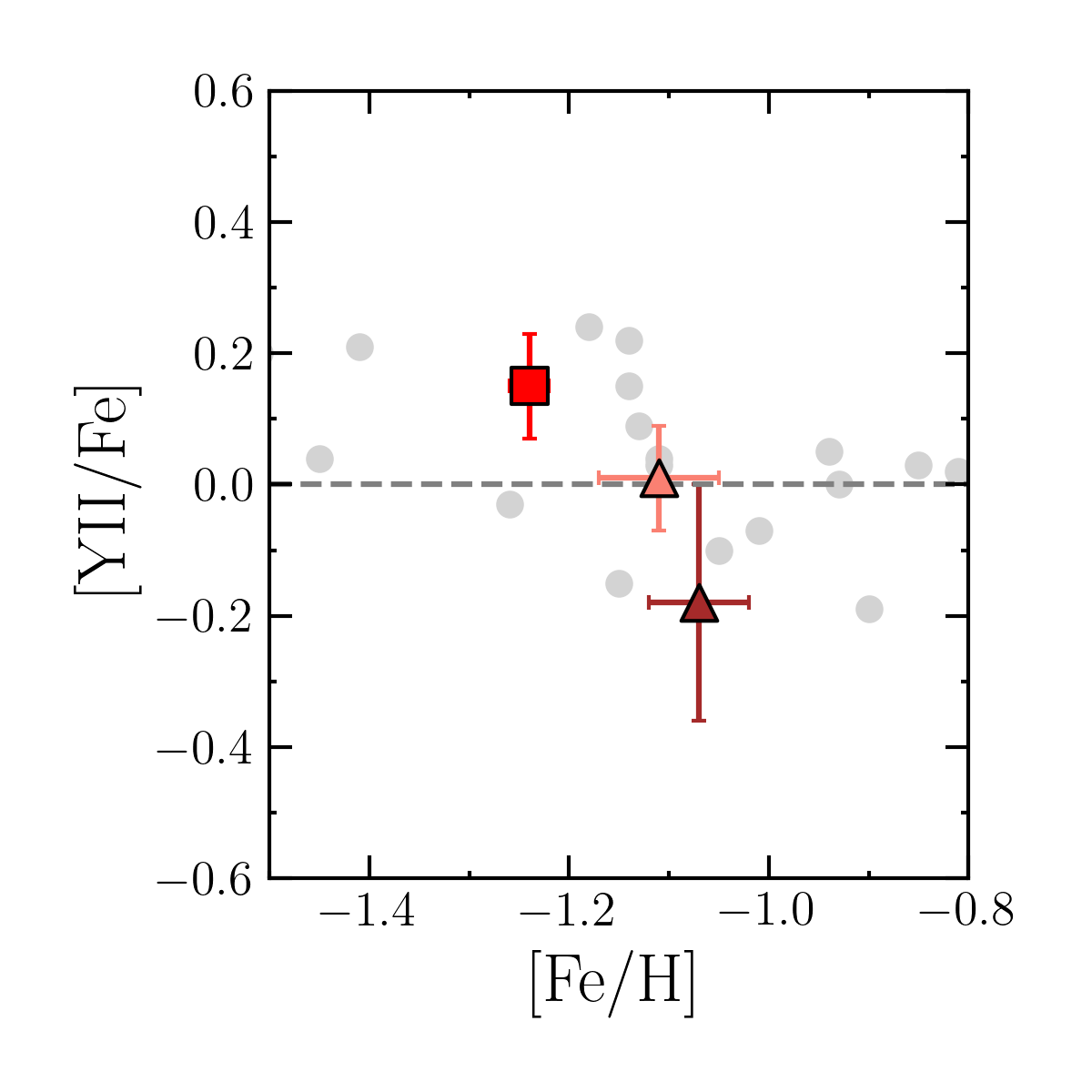}
   \end{minipage}
   \begin{minipage}{0.33\textwidth}
        \centering
        \includegraphics[width=1.0\textwidth]{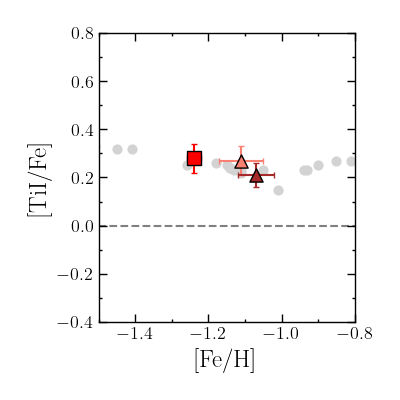}
        \includegraphics[width=1.0\textwidth]{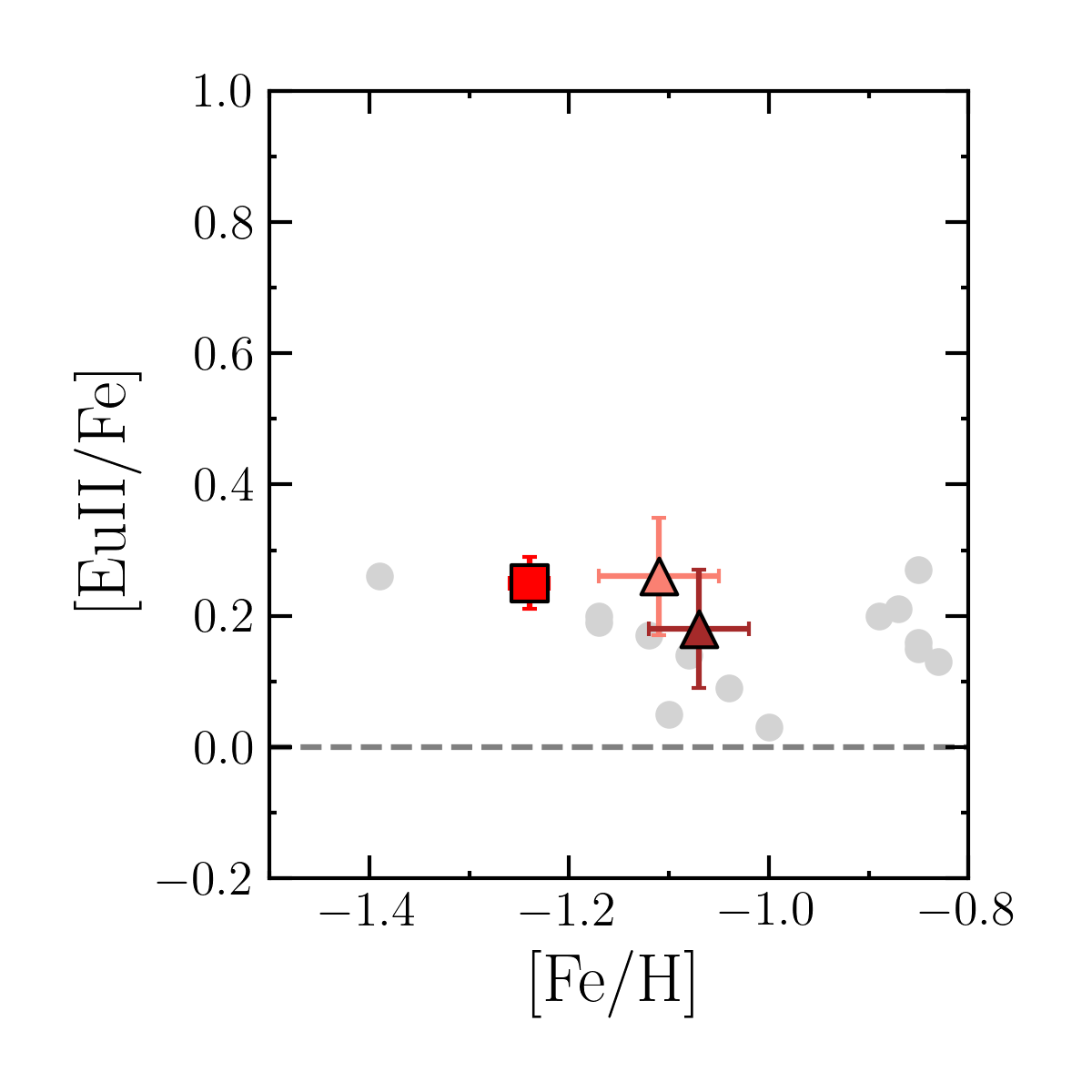}
   \end{minipage}    
   \caption{Mean abundances of NGC 6218, NGC 6522, and NGC 6626 derived in this work (the colour coding is the same as in Fig. \ref{FigAlpha}) compared with those derived for MW stars (grey filled points). [EuII/Fe] literature abundances are taken from \citet{fishlock2017}. Error bars show the standard deviation.} 
              \label{Fig:MW}%
    \end{figure*}    

   \begin{figure*}
   \centering
   \begin{minipage}{0.33\textwidth}
        \centering
        \includegraphics[width=1.0\textwidth]{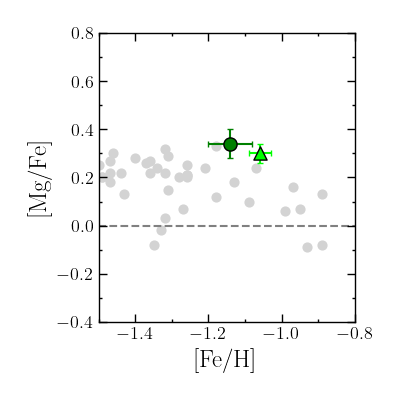} 
        \includegraphics[width=1.0\textwidth]{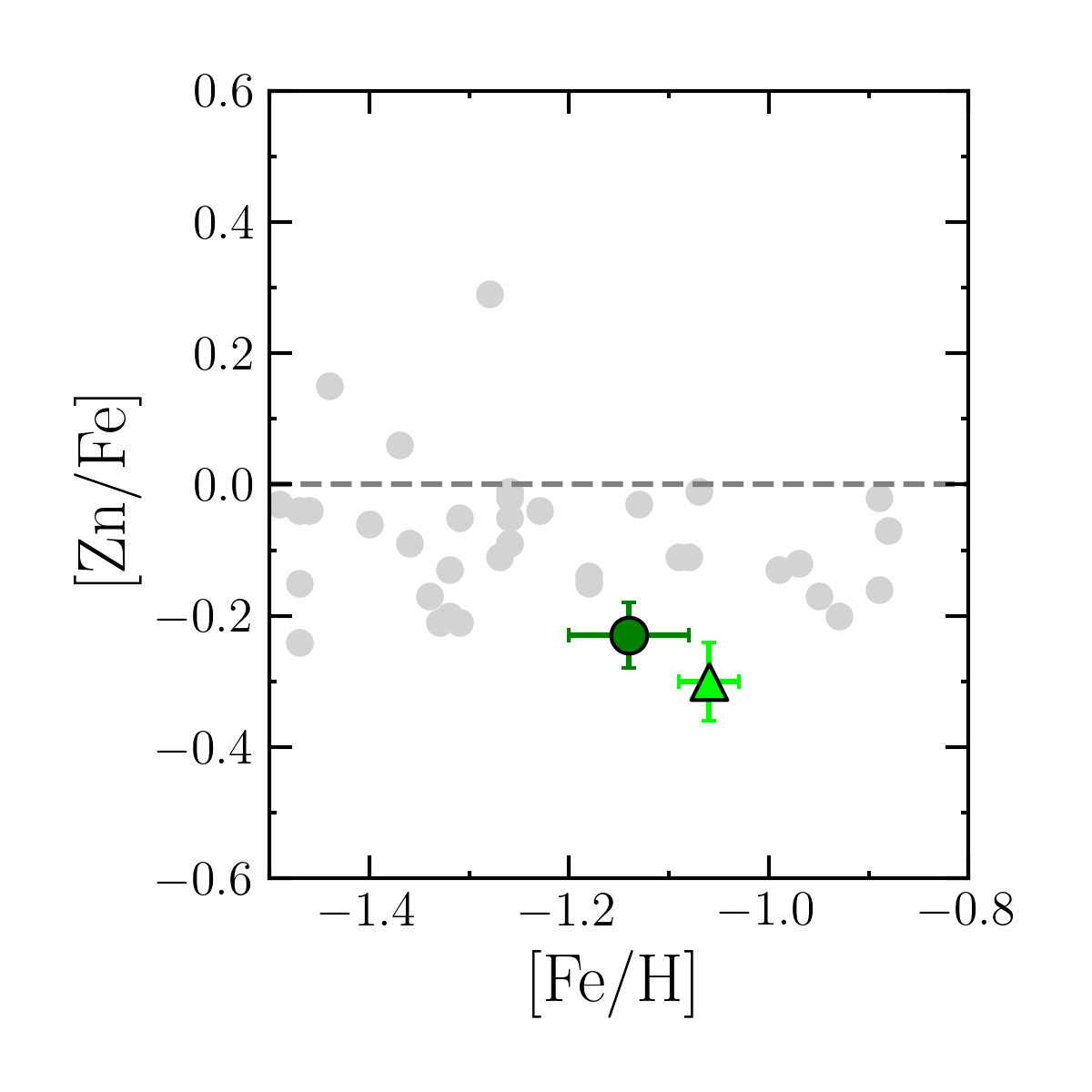}
   \end{minipage}\hfill
   \begin{minipage}{0.33\textwidth}
        \centering
        \includegraphics[width=1.0\textwidth]{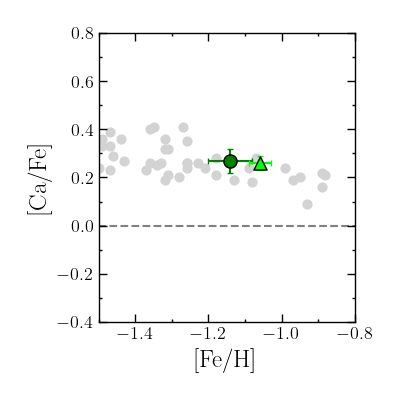}
        \includegraphics[width=1.0\textwidth]{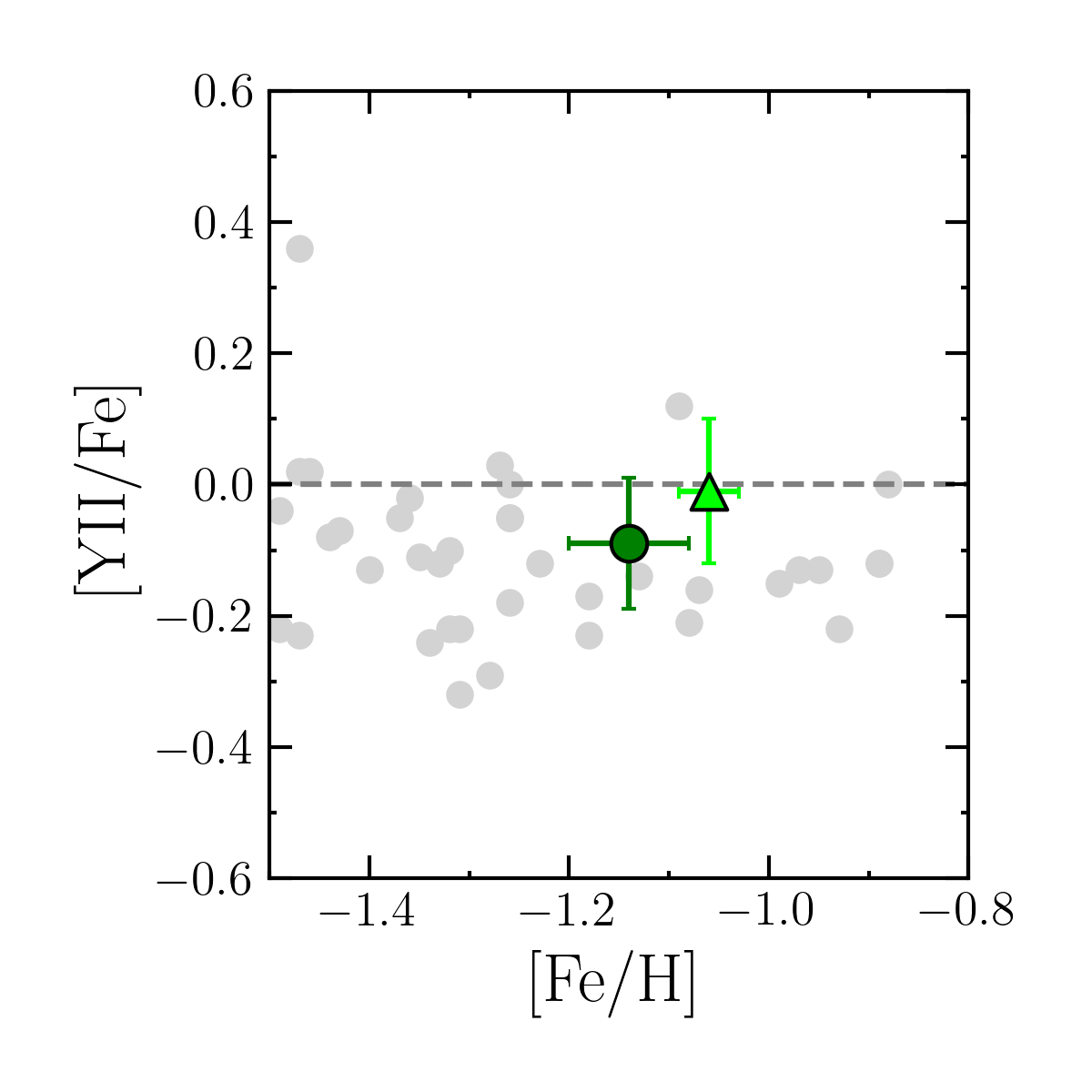}
   \end{minipage}
   \begin{minipage}{0.33\textwidth}
        \centering
        \includegraphics[width=1.0\textwidth]{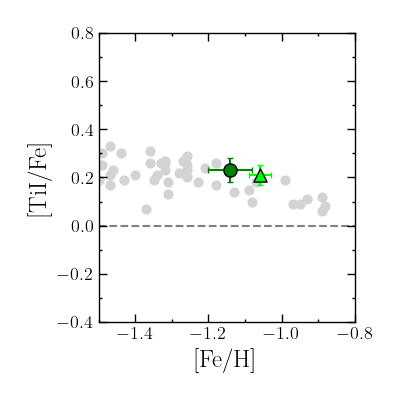}
        \includegraphics[width=1.0\textwidth]{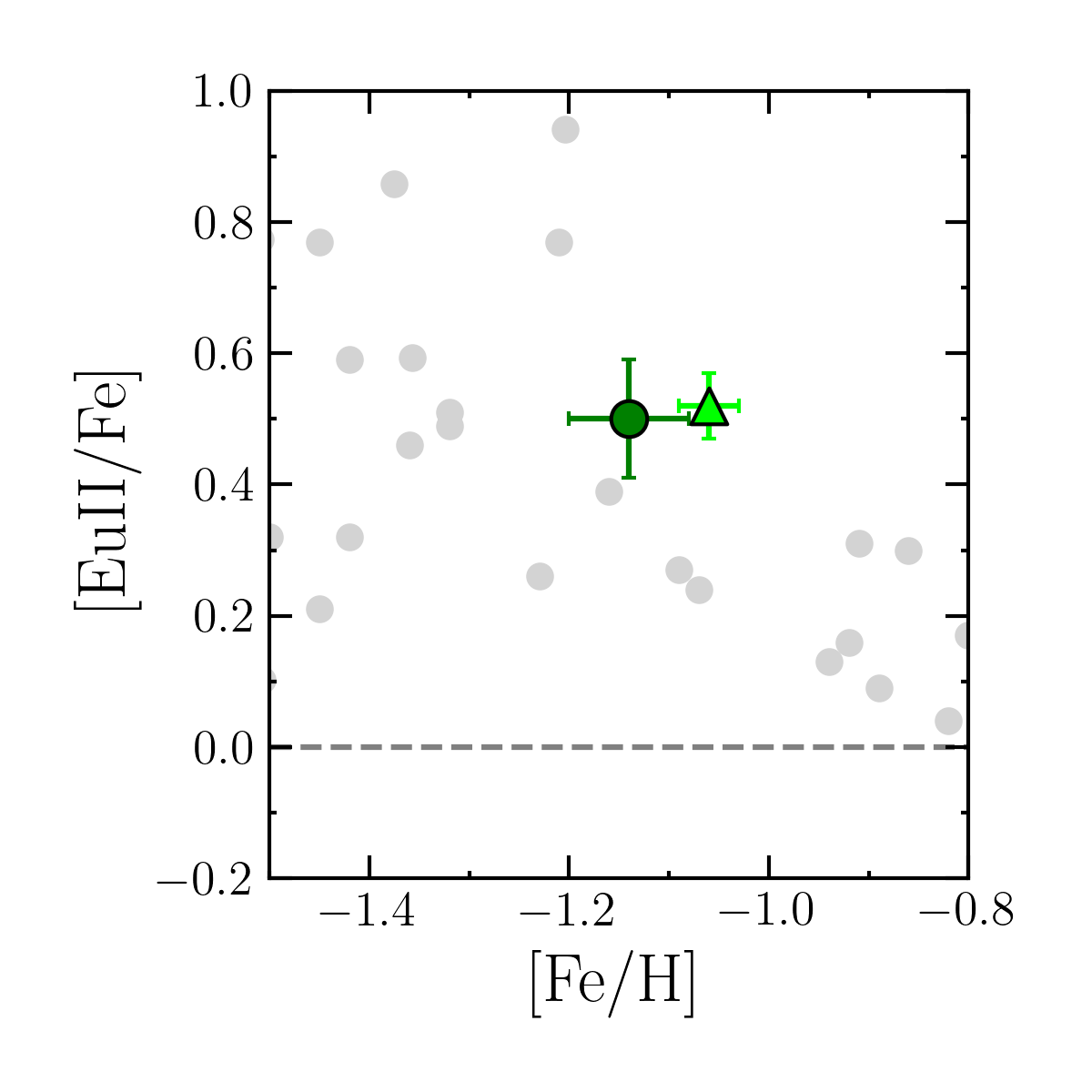}
   \end{minipage}    
   \caption{Mean abundances of NGC 362 and NGC 1261 derived in this work (the colour coding is the same as in Fig. \ref{FigAlpha}) compared with those of GSE stars (grey filled points, either derived in this work or by \citet{ceccarelli2024} for Mg, Ca, TiI, Zn, and YII). Literature abundances for [EuII/Fe] are taken from \citet{fishlock2017,aguado21,carrillo2022,giribaldi2023,francois2024}. Error bars stand for the standard deviation.} 
              \label{FigMW_GES}%
    \end{figure*}    


   \begin{figure*}
   \centering
   \begin{minipage}{0.33\textwidth}
        \centering
        \includegraphics[width=1.0\textwidth]{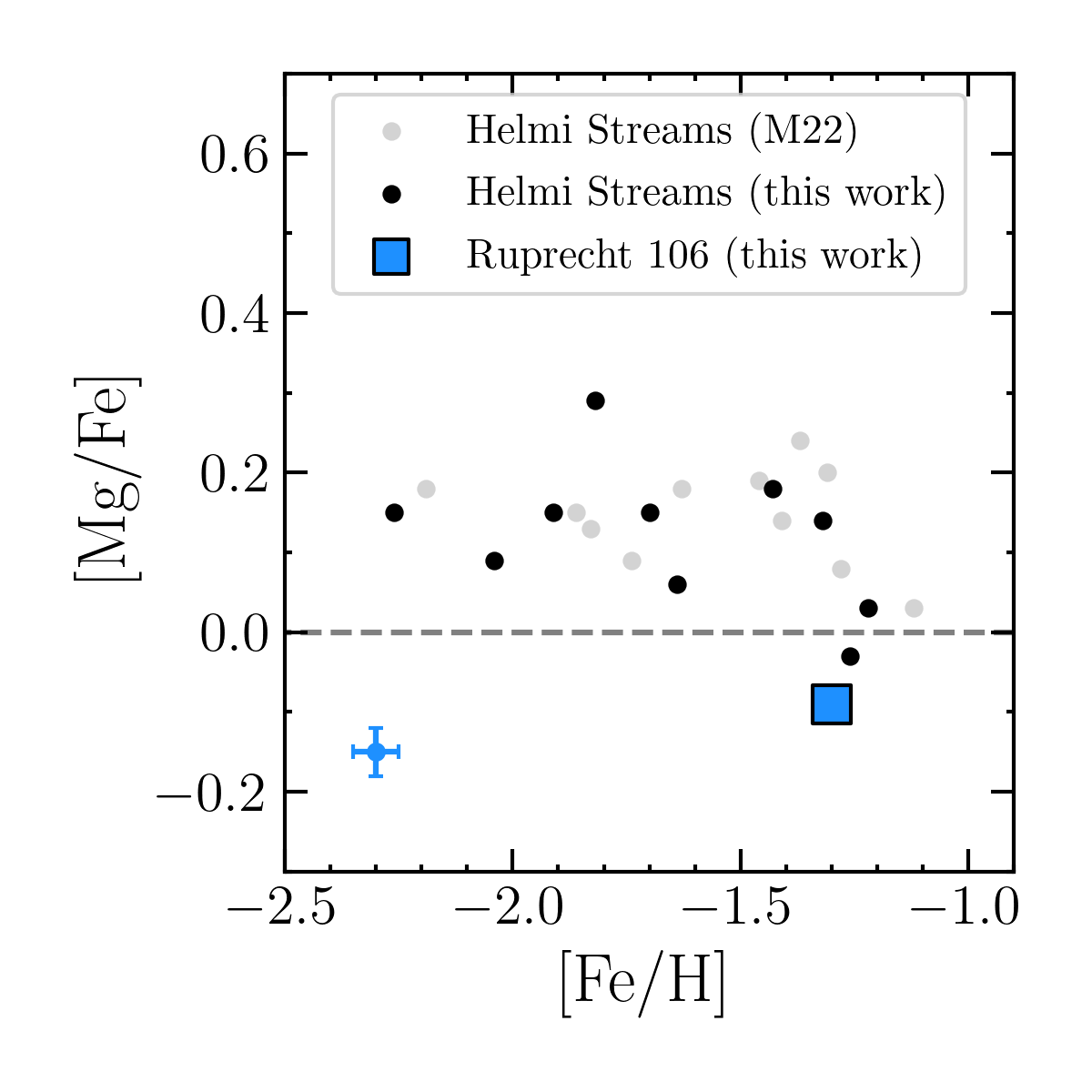} 
        \includegraphics[width=1.0\textwidth]{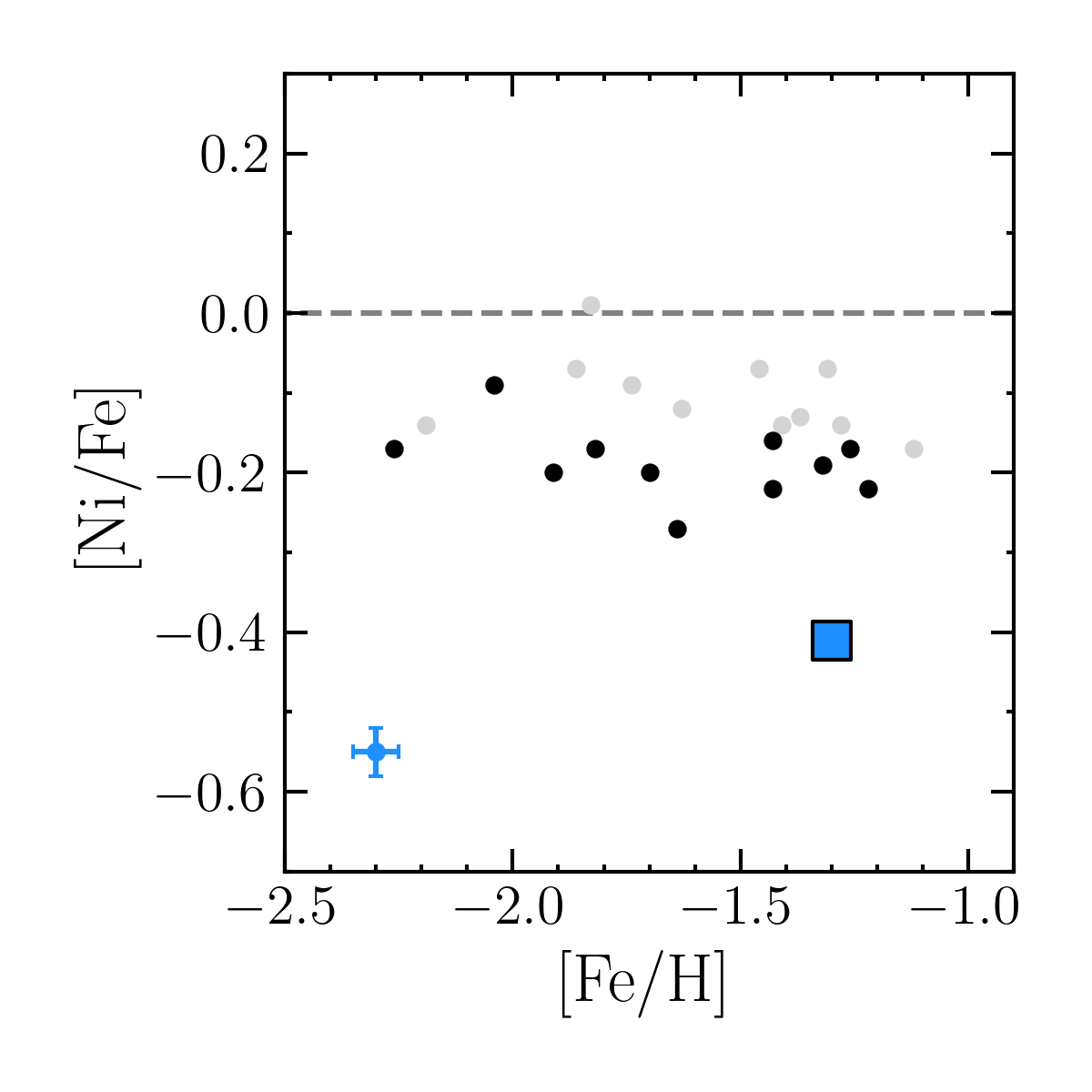}
   \end{minipage}\hfill
   \begin{minipage}{0.33\textwidth}
        \centering
        \includegraphics[width=1.0\textwidth]{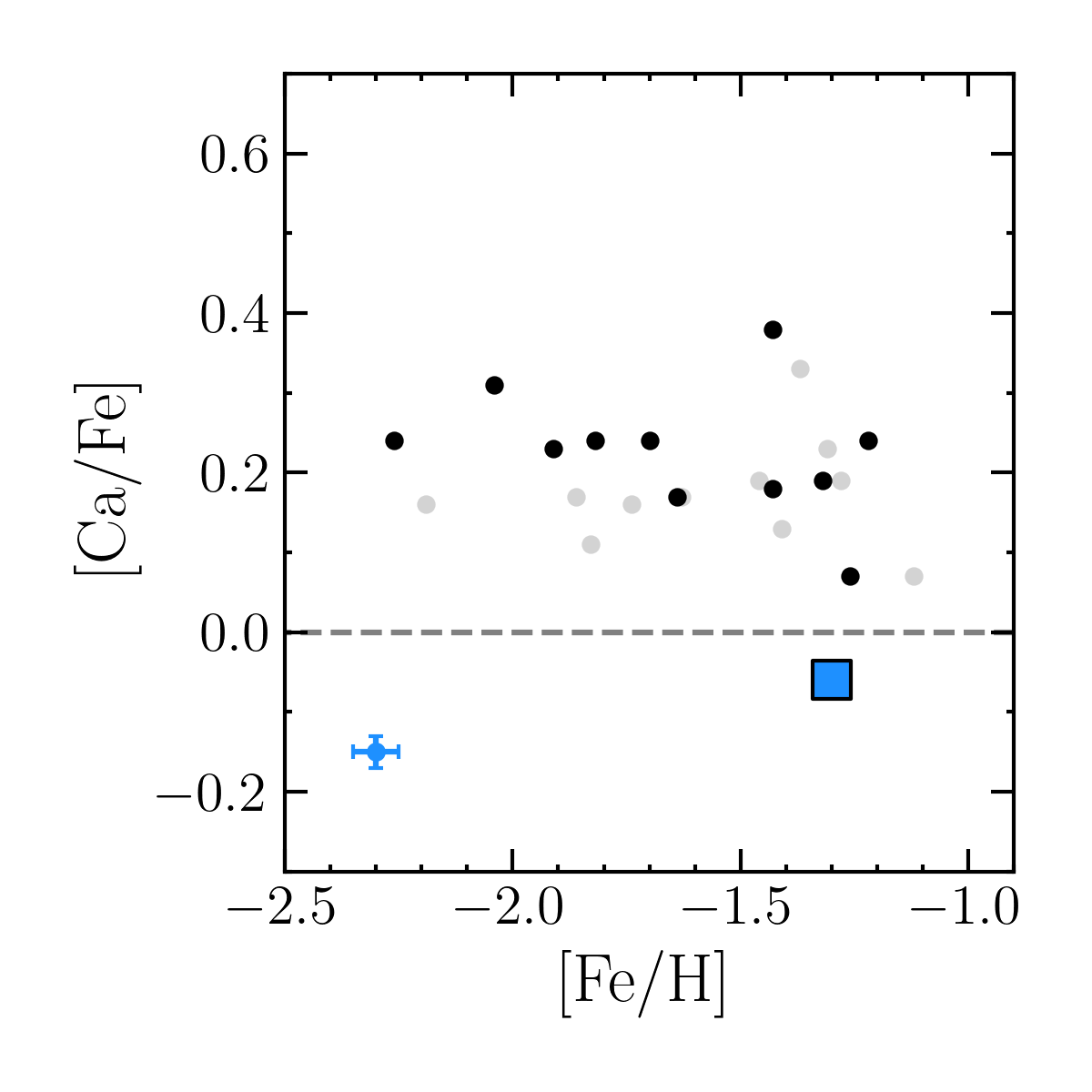}
        \includegraphics[width=1.0\textwidth]{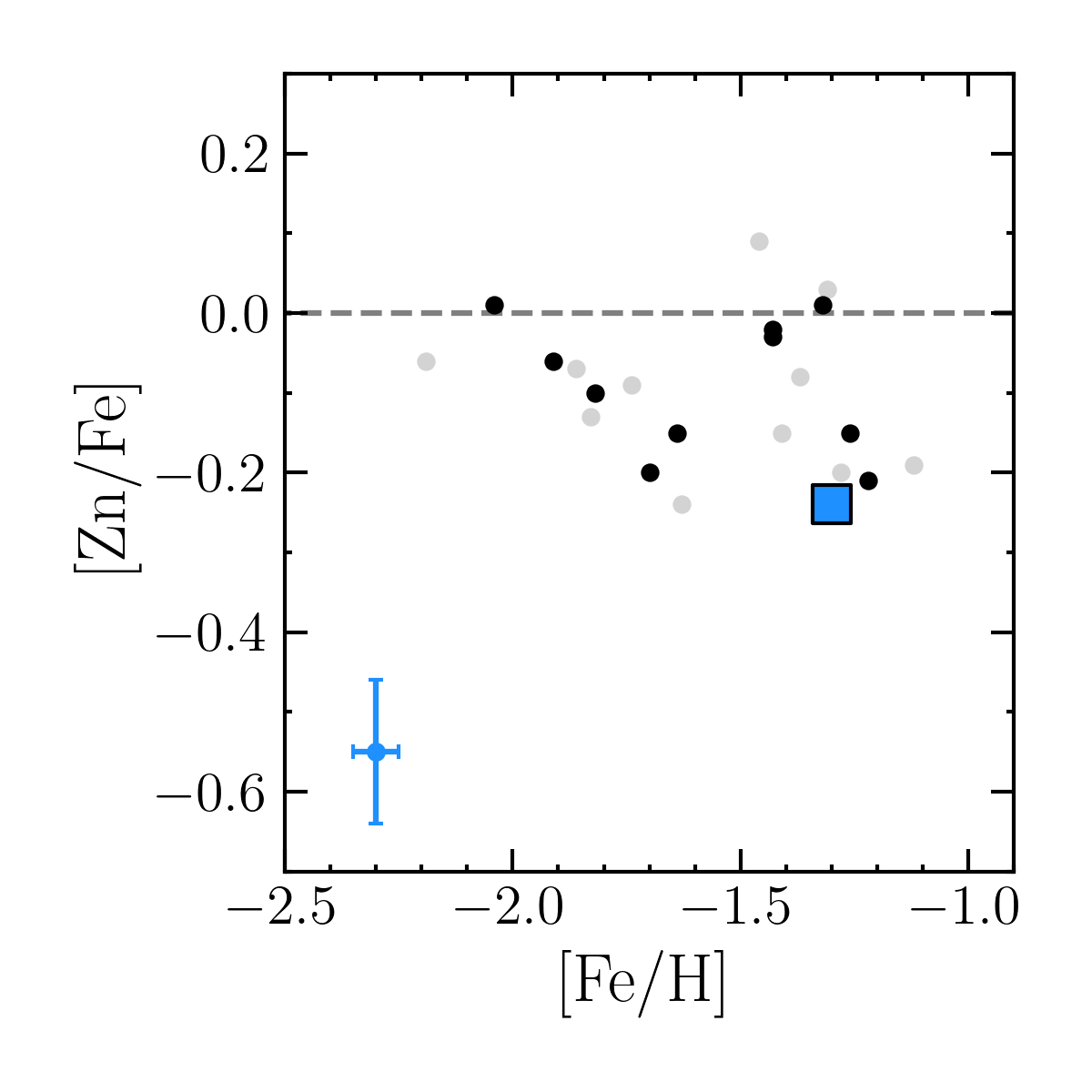}
   \end{minipage}
   \begin{minipage}{0.33\textwidth}
        \centering
        \includegraphics[width=1.0\textwidth]{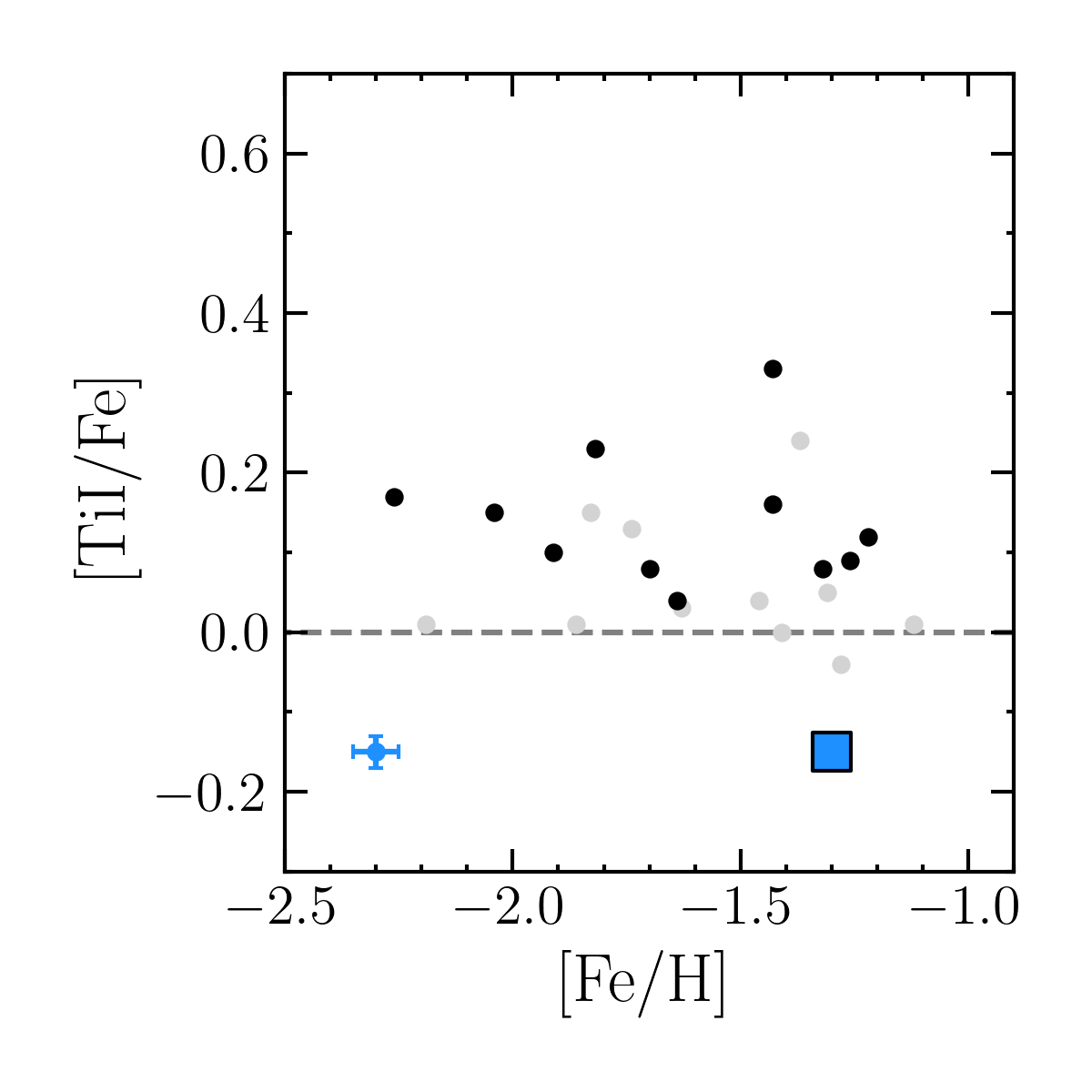}
        \includegraphics[width=1.0\textwidth]{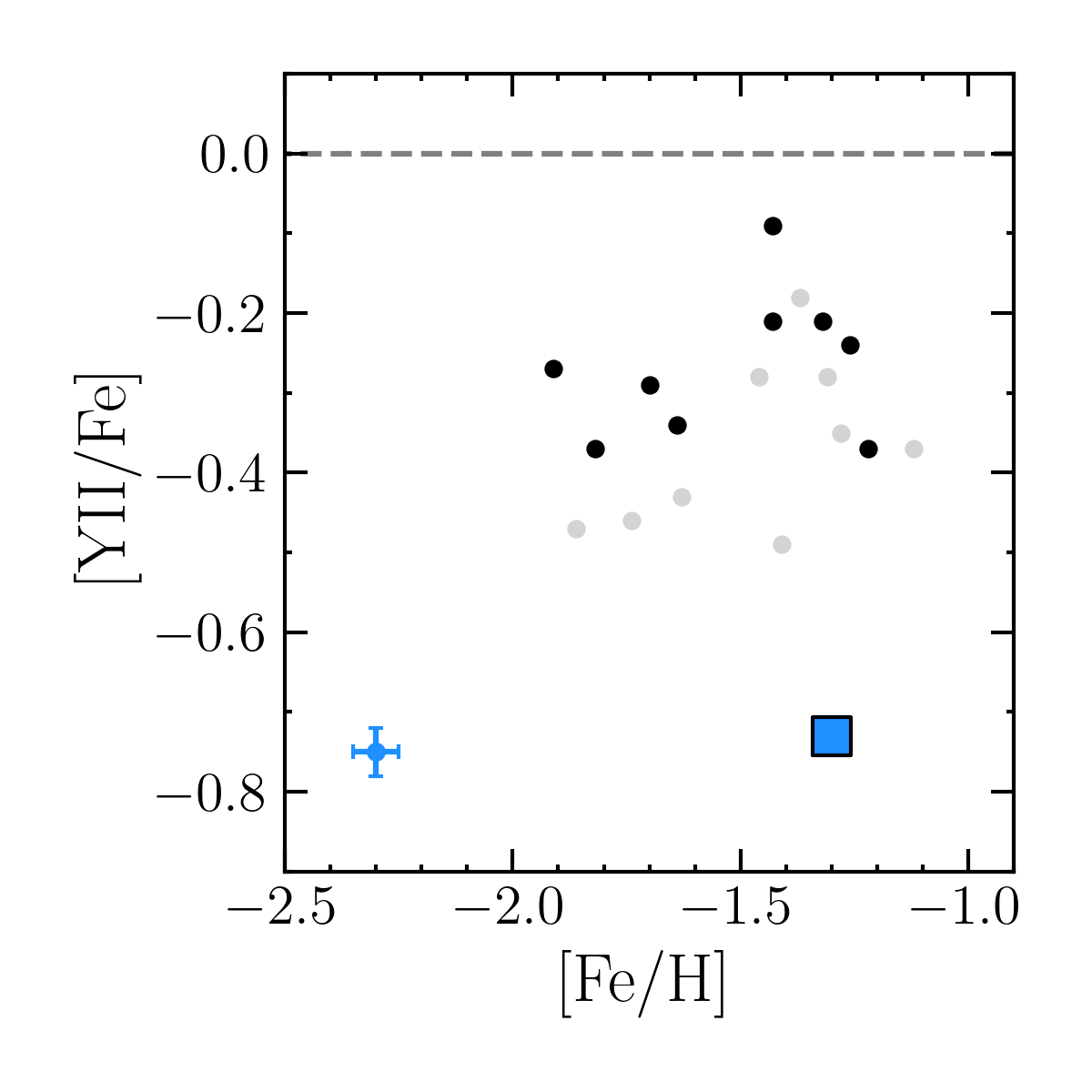}
   \end{minipage}    
   \caption{Mean abundances of Rup106 derived in this work (blue filled squares) compared with those derived 
   for field stars associated with the Helmi Streams (results from this work and from M22 are shown in black and grey, respectively). We plot the standard deviation in the bottom left corner.}
              \label{FigMatsuno}%
    \end{figure*}    
%


    In this Section we investigate the compatibility of the abundances derived for accreted GCs with those derived for the field stars associated with their respective progenitors, and in passing we note once again that homogeneity in this kind of comparison is crucial. Indeed, several factors may contribute to a zero-point offset in abundance between our results and those from the literature. Discrepancies in the assumed atmospheric parameters, model atmosphere, solar mixture, line list, and atomic data (e.g. log \textit{gf}) can potentially result in variations in the derived abundances.

\subsection{NGC 6218, NGC 6522, NGC 6626, and the Milky Way}\label{sec:MW}
    
    We use as a comparison the results obtained from high-resolution spectroscopy of MW field stars, focusing on the chemical elements that show significant differences between accreted and in situ GCs. To do so, given the extremely high quality of their stellar spectra, we selected the catalogue of \citet{nissen&schuster2010,nissen&schuster2011} as our benchmark comparison dataset, and in particular we focus on stars in the high-$\alpha$ sequence that have been observed with UVES at the VLT. To maintain homogeneity in the chemical analysis, we re-derived abundances from the spectra of these stars (see the detailed method in Sect. \ref{sec:abu}). The only exception in this homogeneous comparison is the element EuII, as (i) the number of available spectra in the literature is limited (and many of them were obtained with different instruments and S/N), and (ii) abundances are derived from different spectral lines due to the different evolutionary stages of the observed stars. We chose to compare our results with EuII abundances from the high-$\alpha$ sequence of \citet{fishlock2017}.

    Fig. \ref{Fig:MW} shows that the abundances of NGC 6218, NGC 6522, and NGC 6626 match those of the field stars (grey filled points) in all chemical spaces of interest (i.e. Ca, TiI, Zn, YII, and EuII) except for Mg, where the three in situ target GCs show enhancement in [Mg/Fe] compared to MW high-$\alpha$ sequence stars, which is found to be outside the $2\sigma$ of their distribution in the metallicity range of $-1.3 \le \feh \le -1.0 \dex$.

\subsection{NGC 362, NGC 1261, and \textit{Gaia}-Sausage-Enceladus}\label{sec:MW_GES}
    
   The reference catalogue used for this comparison is that of \citet{ceccarelli2024}, which was produced with the same methods as those described here. Additionally, we re-derived the abundances for the stars in the low-$\alpha$ sequence of \citet{nissen&schuster2010}. Finally, we compared our abundances with the results from the low-$\alpha$ stars in \citet{fishlock2017,aguado21,carrillo2022,giribaldi2023}, and \citet{francois2024} for EuII. 
    As displayed in Fig. \ref{FigMW_GES}, there is excellent agreement between the chemical composition of NGC 362 and NGC 1261 and the patterns shown by GSE stars (grey filled points) in the $\alpha$-elements Ca and TiI, while, as for in situ GCs, NGC 362 and NGC 1261 lay in the upper boundaries (at the 1.3$\sigma$ and 1.6$\sigma$ levels for NGC 362 and NGC 1261, respectively) of the distribution of GSE field stars in [Mg/Fe] in the metallicity range of interest. The value of [Zn/Fe] derived in NGC 1261 is consistent with those observed in the most Zn-poor stars in GSE, while NGC 362 seems to be slightly depleted ($\sim 0.1 \dex$) compared to them. Finally, the slightly subsolar values of [YII/Fe] found for NGC 362 and NGC 1261 are consistent with those observed in the literature \citep{aguado21,ceccarelli2024,francois2024}, as are the high [EuII/Fe] ratios \citep[see also,][]{aguado21,matsuno2021,francois2024,ou2024}.
   
    This comparison represents strong additional evidence that the two selected GCs were indeed born in the GSE dwarf galaxy, and provides important confirmation that GCs and stars trace the same chemical evolution when sharing the birth environment \citep[see also][]{monty2024}.

\subsection{Ruprecht 106 and the Helmi Streams}
  
    In this section, we investigate the compatibility of the chemical composition of Rup106 and that of stars formed in the galaxy that originated the Helmi Streams \citep{helmi99}, as a tentative dynamical connection between the two has been pointed out by several works \citep{massari19, forbes2020,callingham2022}. In Fig. \ref{FigMatsuno} we present the mean abundances of Rup106 derived in this work (blue filled squares) 
    compared to the high-resolution abundances of 11 stars linked to the Helmi Streams \citep[grey filled points;][M22 hereafter]{matsuno2022_helmi}. We note that we cannot compare EuII abundances as they were not published in M22, and we therefore show Ni as an additional comparison, representative of the iron-peak group. To ensure homogeneity in the comparison, we reanalysed the spectra of the Helmi Streams stars following the procedure described in Sect. \ref{sec:abu}. The results are shown in Fig. \ref{FigMatsuno} (black filled points). The average offset between our abundances and those of M22 is consistent with zero for all the elements we show in Fig. \ref{FigMatsuno}, except for TiI, Ni, and YII, for which we find mild offsets of $+0.08 \dex$ ($\sigma = 0.04$), $-0.08 \dex$ ($\sigma = 0.05$), and $+0.09 \dex$ ($\sigma = 0.07$), respectively. Regardless of the adopted data set (even though the most homogeneous comparison is the one with our reanalysed abundances), when comparing the mean abundance ratios  of Rup106 to the trend observed in the Helmi Streams, a noticeable difference emerges, particularly in the $\alpha$-elements. Our findings consistently indicate \afeh values approximately $0.2 \dex$ lower in comparison to the Helmi Streams, hinting at a lower star formation efficiency in the environment where Rup106 formed. Among the other elements studied, the most significant differences arise in Ni and YII with Rup106 being depleted by $\sim 0.2 \dex$ and by $\sim 0.4 \dex,$ respectively. 
    Therefore, in order to account for the observed differences shown in Fig. \ref{FigMatsuno}, the progenitor of the Helmi Streams would need to have been massive enough to be chemically inhomogeneous. Otherwise, this chemical incongruity is strong evidence in support of the interpretation that Rup106 was born in a different environment from the progenitor of the Helmi Streams.

\section{Possible origins of Ruprecht 106}\label{sec:Rup106}
   \begin{figure}[!th]
   \centering
   \includegraphics[width=.35\textwidth]{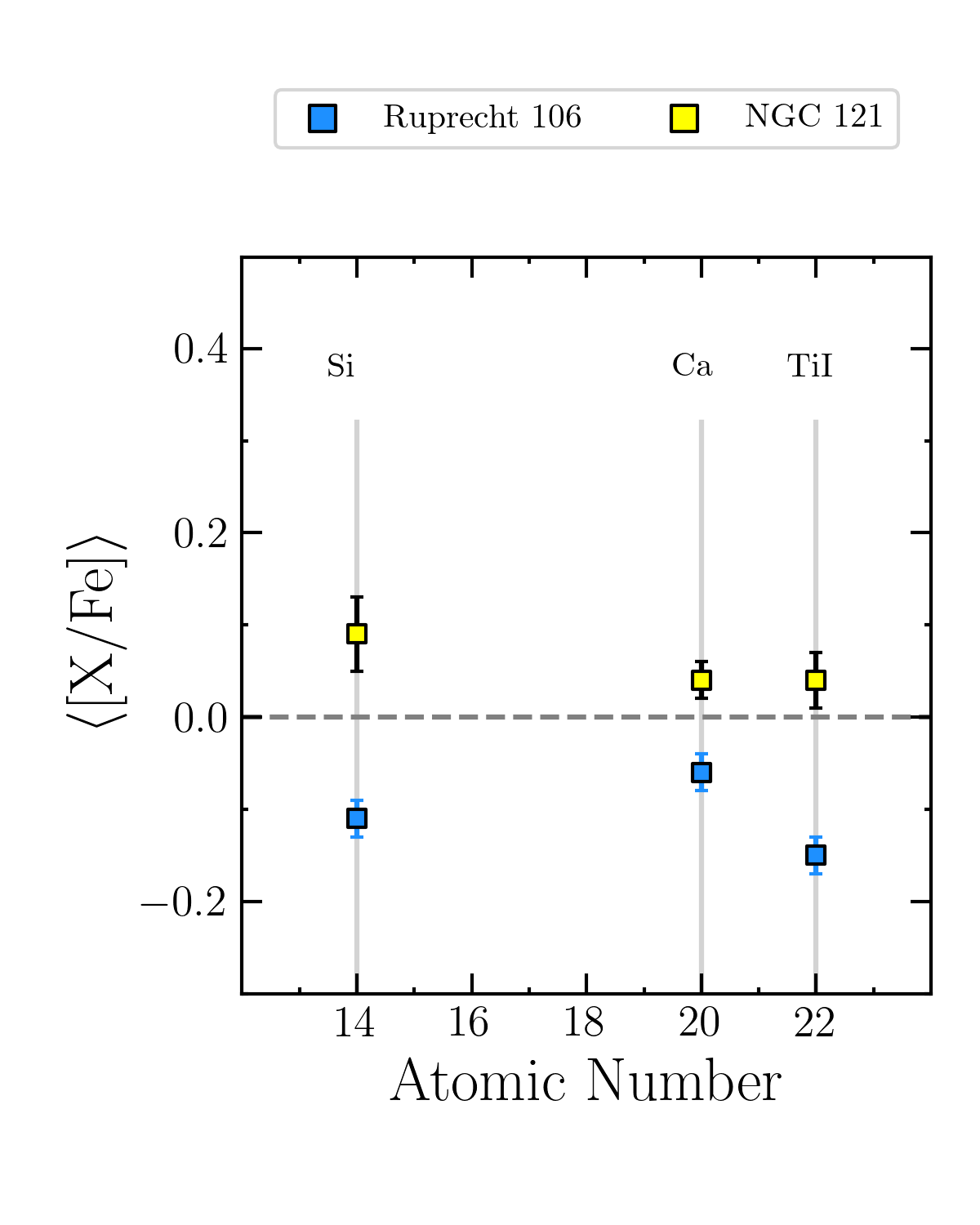} 
   \includegraphics[width=.35\textwidth]{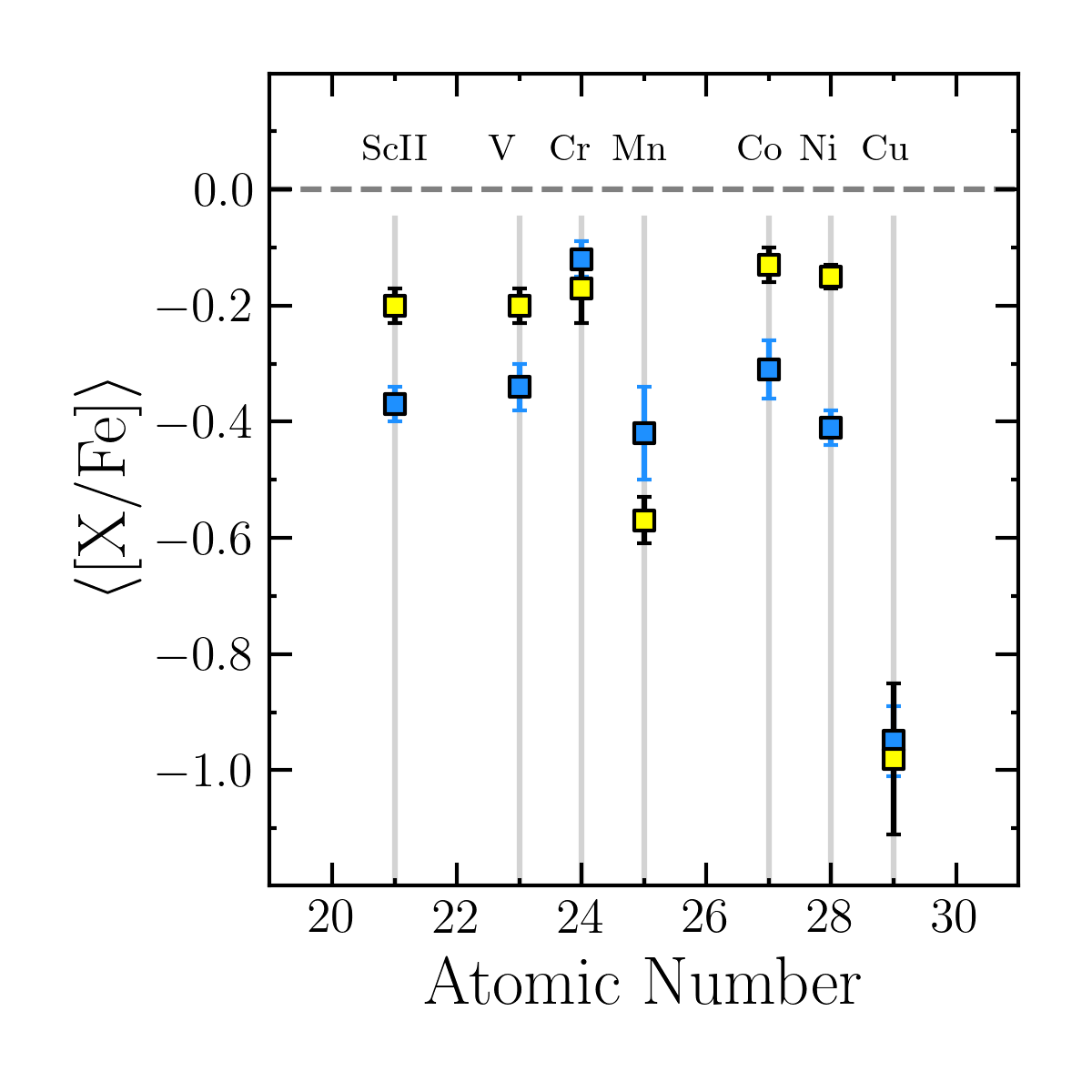}
   \includegraphics[width=.35\textwidth]{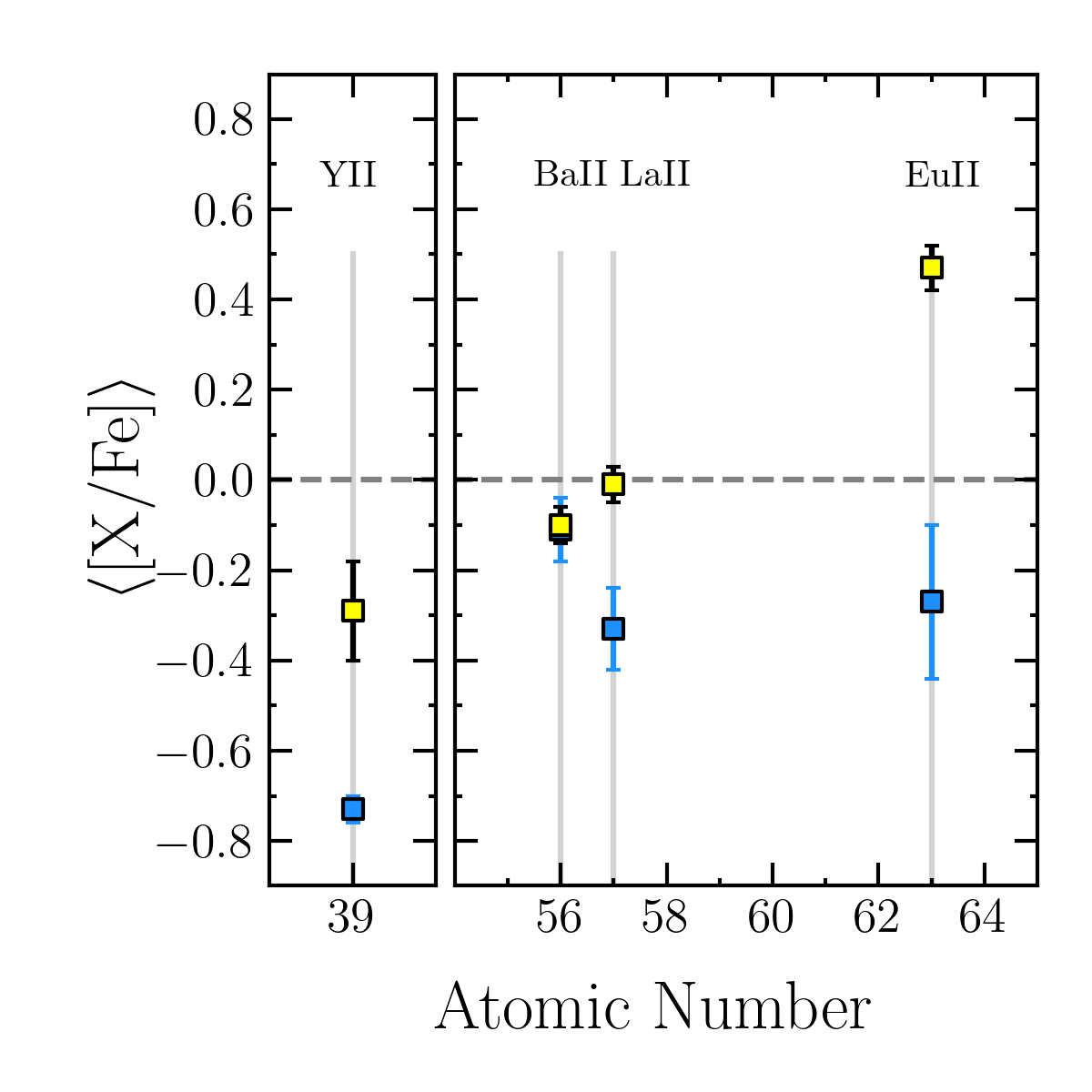}    
   \caption{Mean abundances of the $\alpha$- (top panel), iron-peak (middle panel), and neutron-capture (bottom panel) elements in Rup106 (blue filled squares) and NGC 121 (yellow filled triangles). Error bars indicate the standard deviation.}
              \label{FigNGC121}%
    \end{figure} 

    In light of the chemical difference between Rup106 and the Helmi Streams stars, and given its peculiar chemical composition compared to many other GCs \citep[e.g.][]{brown1997,carretta09,carretta13,villanova13,monaco2018,puls2018,crestani2019,masseron2019,horta2020,mucciarelli2021NatAs}, the birth environment of Rup106 likely underwent a unique chemical evolution. In the following, we discuss some possible interpretations.

\subsection{Comparison with NGC 121}

    Given its extremely peculiar chemical makeup, Rup106 has long been regarded as originating in an extragalactic environment. Initially, several works proposed that it was likely to have been accreted from the Magellanic Clouds \citep{lin1992, fusipecci1995}, and particularly from the Small Magellanic Cloud (SMC). Although recent post-\textit{Gaia} kinematics studies of the orbit in the MW of Rup106 (see Sect. \ref{introduction}) have excluded the possibility that it originated in the SMC, it is interesting to note that Rup106 is coeval with NGC 121, the oldest known SMC GC \citep[$\sim 10.5$ Gyr,][]{glatt2008, forbes&bridges2010, dotter2011} and shows very similar metallicity \citep{dalessandro2016,mucciarelli2023}. Therefore, this presents an excellent opportunity to explore the possibility that an SMC-like galaxy is the progenitor of Rup106. We directly compared our results with those from \citet{mucciarelli2023}, given the congruent analytical approaches. As shown in Fig. \ref{FigNGC121}, NGC 121 (yellow filled squares) has solar-scaled \afeh, hinting at an enrichment due to SN Ia already in place in the SMC $\sim 10.5$ Gyr ago. These values are, on average, $0.1$ - $0.2 \dex$ $\alpha$-enriched relative to what we find for Rup106, suggesting that Rup106 likely formed in an environment with even less efficient star formation than that of the SMC. Regarding the iron-peak elements, we find comparable values in the abundance ratios of Cu. The very low values of [Cu/Fe] found in both Rup106 and NGC 121 are consistent with those of the SMC field stars \citep{mucciarelli2023} and are 0.5 dex underabundant compared to MW GCs and stars, favouring a chemical-enrichment scenario in which the contribution by massive stars is extremely reduced, as also discussed in Sect. \ref{sec:iron-peak}. Additionally, Rup106 shows lower abundances ($\sim 0.2 \dex$) of other elements produced in massive stars (i.e. Sc, V, and Co) relative to NGC 121 and also in Ni. Interestingly, the only elements in which enhancement is detected in Rup106 are Cr and Mn. These elements are predominantly synthesised by SN Ia, with yields depending on the metallicity of the progenitor SN Ia. Unfortunately, \citet{mucciarelli2023} do not provide an estimate of [Zn/Fe] for NGC 121. When assessing $s$-process elements, we note similar Ba abundances between the two clusters, yet discern distinct values in Y and La, with Rup106 being, once again, depleted in these elements. Given that NGC 121 demonstrates pronounced deficiencies in these elements relative to the only other SMC GC of similar age, this discrepancy highlights the unique conditions under which Rup106 formed. Also, the efficiency of the production of $r$-process elements is extremely low, with Rup106 being $\sim 0.7 \dex$ underabundant in Eu relative to NGC 121, which conversely aligns with observations of GSE GCs and SMC stars. This comparison further supports the idea that Rup106 originated in an environment that was significantly different from the SMC of $10.5$ Gyr ago and was probably characterised by an exceptionally low star formation efficiency and likely featured a truncated initial mass function (IMF) with a minimal contribution from massive stars.  

\subsection{Comparison with surviving dwarf galaxies}

    A similar chemical composition to that of Rup106 has been observed in stars belonging to dwarf galaxies in the Local Group \citep{venn2004,tolstoy09}. Indeed, subsolar \afeh values at $\feh \le -1.5 \dex$ are observed in surviving dwarf galaxies such as Carina, Fornax, Sculptor, Sextans, and Ursa Minor \citep{venn2012,hendricks2014,hill2019,theler2020,fernandes2023,sestito2023}. The low values of Zn ([Zn/Fe] $\sim -0.3 \dex$) derived for Rup106 compared to those found in MW stars are also typical of these systems \citep[see Fig. 16 of][and reference therein]{skuladottir2017}. Finally, most of the dwarf galaxies show enhancement in [EuII/Fe] compared to MW stars due to either the very low star formation efficiency or the delayed contribution to the production of Eu by NSMs \citep{letarte2010,venn2012,lemasle2014}. However, even the extremely low [EuII/Fe] of Rup106 can be reproduced in some of these systems, such as Sculptor \citep{hill2019}. Indeed, these latter authors find several stars in Sculptor with subsolar values ($\sim -0.1 \dex$) of [EuII/BaII] at $\feh \sim -1.3 \dex$ , suggesting that the onset of $s$-process enrichment was already in place at this metallicity. It is interesting to note that the chemical composition of Rup106 resembles that found in iron-rich metal-poor stars (IRMP), despite these latter being unlikely to form in GCs, thus indicating a birth environment enriched by the (double) detonation of sub-Chandrasekhar-mass CO white dwarfs \citep{reggiani2023}. This further supports the idea that Rup106 formed in a dwarf galaxy, as these authors demonstrated that IRMP stars are easier to find in surviving dwarf galaxies and the Magellanic Clouds, rather than in the halo of the MW.
    
    It is important to note that the majority of the existing systems that reproduce the peculiar chemistry of Rup106 have a stellar mass \citep[$\sim 10^{6}$ $\mathrm{M_{\odot}}$,][]{mcconnachie2012} that is not compatible with that estimated for the progenitor of the Helmi Streams, with the latter being roughly 100 times more massive \citep{koppelman19}. Therefore, if the dynamical association of Rup106 is accurate, the possibility that it did not form directly in the progenitor of the Helmi Streams cannot be ruled out. We speculate that Rup106 could have formed in a dwarf galaxy with a mass comparable to that of Sculptor; the dynamical properties of  Rup106 suggest that this  could have been a satellite of the progenitor of the Helmi Streams or the Sagittarius dwarf \citep{davies2024}, similarly to the case of the GC NGC 2005 in the Large Magellanic Cloud \citep{mucciarelli2021NatAs}.
\section{Conclusion}\label{sec:conclusion}

    The primary aim of this work is to investigate the potential differences in elemental abundances between six GCs of very similar metallicity $( -1.3 \le \feh \le -1.0 \dex)$ that are expected to have formed in different environments, and to decipher which abundance ratios are most sensitive to their different origin.
 
        The outcome of this homogeneous chemical analysis reveals striking similarities in the chemical abundances of NGC 362 and NGC 1261, two GCs that were accreted during the merger event of the \textit{Gaia}-Sausage-Enceladus dwarf galaxy. When comparing the chemical abundances of these GCs with those of the in situ GCs NGC 6218, NGC 6522, and NGC 6626, a coherent and distinct overall pattern emerges in the $\alpha$-elements, especially in Mg, Si, and Ca, where we identify a depletion of $\sim 0.1 \dex$ in the GSE GCs with respect to in situ GCs. This finding agrees with observations of field stars of the Galactic halo in this metallicity range, where distinctions between chemical patterns of in situ and accreted stars are discernible, especially in the $\alpha$-elements, whereas trends become indistinguishable at lower metallicity \citep{horta23,ceccarelli2024}. This is consistent with the fact that it took 2 Gyr less for NGC 6218, NGC 6522, and NGC 6626 to reach the same metallicity as the GSE GCs \citep{dotter2010,dotter2011,vandenberg2013,villanova17,kerber2018}, indicating that the contribution of SN Ia to the chemical enrichment of the gas was not in place at the time when these systems formed. Moreover, statistically significant variations are observed in specific elements such as Zn and Eu, whose production sites are directly associated with massive stars (HNe for Zn and rCCSNe+NSMs for Eu). It is worth noting that that these events are inherently rare, and might exert a profound influence on the abundances within a given environment. These results reinforce the idea that chemical tagging is an effective tool for investigating the origin of GCs, even when comparing the chemical compositions of GCs formed in different progenitors, but at the same time demonstrate that not every element is equally sensitive in this sense.
    
    In this coherent picture, Rup106 stands out as a highly distinct GC. Indeed, Rup106 follows the AMR of accreted GCs \citep{forbes&bridges2010,dotter2010,dotter2011,vandenberg2013}, whereas MW in situ clusters of similar age clearly experienced a different chemical enrichment path, exhibiting significantly higher metallicity \citep[0.5 - 0.8 dex more metal rich;][]{massari2023}. Despite the congruence between Rup106 and the GSE GCs in the age--metallicity space, differences emerge when examining all their chemical elements. Particularly, Rup106 exhibits subsolar \afeh values typical of an environment already enriched by SNe Ia, in contrast to GSE clusters ($\afeh > 0.3 \dex$), where SNe Ia have only just started contributing to the enrichment of the gas. This comparison underlines distinct chemical evolution histories between GSE and the environment in which Rup106 formed, with the latter experiencing slower and less efficient star formation and likely an IMF truncated of most of the massive stars, indicating a lower mass for the host of Rup106 at $\sim$ 10.5 Gyr ago compared to GSE.
    
    In light of these results, this work demonstrates that the exceptionally high precision offered by a strictly homogeneous chemical analysis allows chemical tagging to play a crucial role in uncovering the true origin of GCs. Particularly, this method enables us not only to clearly discriminate  between GCs formed within the MW and those formed in other accreted galaxies, but also to distinguish the chemical signatures of GCs originating from independent accreted progenitors.
    
    Finally, we interpret the extreme peculiarity of the chemistry of Rup106 as due to the particular evolution of its progenitor, which could be the galaxy that gave rise to the Helmi Streams. However, the remarkably low values of the \afeh abundance ratios, as well as of some other heavier elements, compared to Helmi Streams stars lend weight to the possibility that Rup106 originated from an even smaller galaxy, with a stellar mass comparable to that of some surviving dwarf galaxies (e.g. Sculptor).

\section*{Data availability}

      Table \ref{tab:linelist_EW} with information on the line list used in this work is available on Zenodo at \url{https://doi.org/10.5281/zenodo.13907515}.

\begin{acknowledgements}

    Based on observations collected at the ESO-VLT under the programs 069.D-0227 (P.I. P. Fran\c{c}ois), 073.D-0211 (P.I. E. Carretta), 083.D-0208 (P.I. E. Carretta), 091.D-0535 (PI: C. Moni Bidin), 097.D-0175 (PI: B. Barbuy), 0101.D-0109 (P.I. A. Marino), 193.D-0232 (P.I. F. R. Ferraro), and 197.B-1074 (P.I. G. F. Gilmore).    

    This research is funded by the project \textit{LEGO – Reconstructing the building blocks of the Galaxy by chemical tagging} (P.I. A. Mucciarelli), granted by the Italian MUR through contract PRIN 2022LLP8TK\_001.

    MB, EC, AM and DM acknowledge the support to this study by the PRIN INAF 2023 grant ObFu \textit{CHAM - Chemodynamics of the Accreted Halo of the Milky Way} (P.I. M. Bellazzini).

    DM acknowledges financial support from PRIN-MIUR-22 ``CHRONOS: adjusting the clock(s) to unveil the CHRONO-chemo-dynamical Structure of the Galaxy” (PI: S. Cassisi) granted by the European Union - Next Generation EU.

    DM and MB acknowledge the support to activities related to the ESA/\textit{Gaia} mission by the Italian Space Agency (ASI) through contract 2018-24-HH.0 and its addendum 2018-24-HH.1-2022 to the National Institute for Astrophysics (INAF). 

    We thank the anonymous referee for the helpful comments that improved the quality of the paper.

    This work has made use of data from the European Space Agency (ESA) mission \textit{Gaia} \url{https://www.cosmos.esa.int/gaia}), processed by the \textit{Gaia} Data Processing and Analysis Consortium (DPAC, \url{https://www.cosmos.esa.int/web/gaia/dpac/consortium}). Funding for the DPAC has been provided by national institutions, in particular the institutions participating in the \textit{Gaia} Multilateral Agreement. 
    
    This work made use of SDSS-IV data. Funding for the Sloan Digital Sky Survey IV has been provided by the Alfred P. Sloan Foundation, the U.S. Department of Energy Office of Science, and the Participating Institutions. SDSS-IV acknowledges support and resources from the Center for High Performance Computing  at the University of Utah. The SDSS website is \url{www.sdss4.org}. SDSS-IV is managed by the Astrophysical Research Consortium for the Participating Institutions of the SDSS Collaboration including the Brazilian Participation Group, the Carnegie Institution for Science, Carnegie Mellon University, Center for Astrophysics | Harvard \& Smithsonian, the Chilean Participation Group, the French Participation Group, Instituto de Astrof\'isica de Canarias, The Johns Hopkins University, Kavli Institute for the Physics and Mathematics of the Universe (IPMU) / University of Tokyo, the Korean Participation Group, Lawrence Berkeley National Laboratory, Leibniz Institut f\"ur Astrophysik Potsdam (AIP),  Max-Planck-Institut f\"ur Astronomie (MPIA Heidelberg), Max-Planck-Institut f\"ur Astrophysik (MPA Garching), Max-Planck-Institut f\"ur Extraterrestrische Physik (MPE), National Astronomical Observatories of China, New Mexico State University, New York University, University of Notre Dame, Observat\'ario Nacional / MCTI, The Ohio State University, Pennsylvania State University, Shanghai Astronomical Observatory, United Kingdom Participation Group, Universidad Nacional Aut\'onoma de M\'exico, University of Arizona, University of Colorado Boulder, University of Oxford, University of Portsmouth, University of Utah, University of Virginia, University of Washington, University of Wisconsin, Vanderbilt University, and Yale University.
\end{acknowledgements}

\bibliographystyle{aa}
\bibliography{rup106.bib}

\begin{thebibliography}{130}
\expandafter\ifx\csname natexlab\endcsname\relax\def\natexlab#1{#1}\fi

\bibitem[{{Abdurro'uf} {et~al.}(2022){Abdurro'uf}, {Accetta}, {Aerts}, {Silva Aguirre}, {Ahumada}, {Ajgaonkar}, {Filiz Ak}, {Alam}, {Allende Prieto}, {Almeida}, {Anders}, {Anderson}, {Andrews}, {Anguiano}, {Aquino-Ort{\'\i}z}, {Arag{\'o}n-Salamanca}, {Argudo-Fern{\'a}ndez}, {Ata}, {Aubert}, {Avila-Reese}, {Badenes}, {Barb{\'a}}, {Barger}, {Barrera-Ballesteros}, {Beaton}, {Beers}, {Belfiore}, {Bender}, {Bernardi}, {Bershady}, {Beutler}, {Bidin}, {Bird}, {Bizyaev}, {Blanc}, {Blanton}, {Boardman}, {Bolton}, {Boquien}, {Borissova}, {Bovy}, {Brandt}, {Brown}, {Brownstein}, {Brusa}, {Buchner}, {Bundy}, {Burchett}, {Bureau}, {Burgasser}, {Cabang}, {Campbell}, {Cappellari}, {Carlberg}, {Wanderley}, {Carrera}, {Cash}, {Chen}, {Chen}, {Cherinka}, {Chiappini}, {Choi}, {Chojnowski}, {Chung}, {Clerc}, {Cohen}, {Comerford}, {Comparat}, {da Costa}, {Covey}, {Crane}, {Cruz-Gonzalez}, {Culhane}, {Cunha}, {Dai}, {Damke}, {Darling}, {Davidson}, {Davies}, {Dawson}, {De Lee}, {Diamond-Stanic}, {Cano-D{\'\i}az}, {S{\'a}nchez},
  {Donor}, {Duckworth}, {Dwelly}, {Eisenstein}, {Elsworth}, {Emsellem}, {Eracleous}, {Escoffier}, {Fan}, {Farr}, {Feng}, {Fern{\'a}ndez-Trincado}, {Feuillet}, {Filipp}, {Fillingham}, {Frinchaboy}, {Fromenteau}, {Galbany}, {Garc{\'\i}a}, {Garc{\'\i}a-Hern{\'a}ndez}, {Ge}, {Geisler}, {Gelfand}, {G{\'e}ron}, {Gibson}, {Goddy}, {Godoy-Rivera}, {Grabowski}, {Green}, {Greener}, {Grier}, {Griffith}, {Guo}, {Guy}, {Hadjara}, {Harding}, {Hasselquist}, {Hayes}, {Hearty}, {Hern{\'a}ndez}, {Hill}, {Hogg}, {Holtzman}, {Horta}, {Hsieh}, {Hsu}, {Hsu}, {Huber}, {Huertas-Company}, {Hutchinson}, {Hwang}, {Ibarra-Medel}, {Chitham}, {Ilha}, {Imig}, {Jaekle}, {Jayasinghe}, {Ji}, {Johnson}, {Jones}, {J{\"o}nsson}, {Katkov}, {Khalatyan}, {Kinemuchi}, {Kisku}, {Knapen}, {Kneib}, {Kollmeier}, {Kong}, {Kounkel}, {Kreckel}, {Krishnarao}, {Lacerna}, {Lane}, {Langgin}, {Lavender}, {Law}, {Lazarz}, {Leung}, {Leung}, {Lewis}, {Li}, {Li}, {Lian}, {Liang}, {Lin}, {Lin}, {Lin}, {Lintott}, {Long}, {Longa-Pe{\~n}a}, {L{\'o}pez-Cob{\'a}}, {Lu},
  {Lundgren}, {Luo}, {Mackereth}, {de la Macorra}, {Mahadevan}, {Majewski}, {Manchado}, {Mandeville}, {Maraston}, {Margalef-Bentabol}, {Masseron}, {Masters}, {Mathur}, {McDermid}, {Mckay}, {Merloni}, {Merrifield}, {Meszaros}, {Miglio}, {Di Mille}, {Minniti}, {Minsley}, {Monachesi}, {Moon}, {Mosser}, {Mulchaey}, {Muna}, {Mu{\~n}oz}, {Myers}, {Myers}, {Nadathur}, {Nair}, {Nandra}, {Neumann}, {Newman}, {Nidever}, {Nikakhtar}, {Nitschelm}, {O'Connell}, {Garma-Oehmichen}, {Luan Souza de Oliveira}, {Olney}, {Oravetz}, {Ortigoza-Urdaneta}, {Osorio}, {Otter}, {Pace}, {Padilla}, {Pan}, {Pan}, {Parikh}, {Parker}, {Peirani}, {Pe{\~n}a Ram{\'\i}rez}, {Penny}, {Percival}, {Perez-Fournon}, {Pinsonneault}, {Poidevin}, {Poovelil}, {Price-Whelan}, {B{\'a}rbara de Andrade Queiroz}, {Raddick}, {Ray}, {Rembold}, {Riddle}, {Riffel}, {Riffel}, {Rix}, {Robin}, {Rodr{\'\i}guez-Puebla}, {Roman-Lopes}, {Rom{\'a}n-Z{\'u}{\~n}iga}, {Rose}, {Ross}, {Rossi}, {Rubin}, {Salvato}, {S{\'a}nchez}, {S{\'a}nchez-Gallego}, {Sanderson}, {Santana
  Rojas}, {Sarceno}, {Sarmiento}, {Sayres}, {Sazonova}, {Schaefer}, {Schiavon}, {Schlegel}, {Schneider}, {Schultheis}, {Schwope}, {Serenelli}, {Serna}, {Shao}, {Shapiro}, {Sharma}, {Shen}, {Shetrone}, {Shu}, {Simon}, {Skrutskie}, {Smethurst}, {Smith}, {Sobeck}, {Spoo}, {Sprague}, {Stark}, {Stassun}, {Steinmetz}, {Stello}, {Stone-Martinez}, {Storchi-Bergmann}, {Stringfellow}, {Stutz}, {Su}, {Taghizadeh-Popp}, {Talbot}, {Tayar}, {Telles}, {Teske}, {Thakar}, {Theissen}, {Tkachenko}, {Thomas}, {Tojeiro}, {Hernandez Toledo}, {Troup}, {Trump}, {Trussler}, {Turner}, {Tuttle}, {Unda-Sanzana}, {V{\'a}zquez-Mata}, {Valentini}, {Valenzuela}, {Vargas-Gonz{\'a}lez}, {Vargas-Maga{\~n}a}, {Alfaro}, {Villanova}, {Vincenzo}, {Wake}, {Warfield}, {Washington}, {Weaver}, {Weijmans}, {Weinberg}, {Weiss}, {Westfall}, {Wild}, {Wilde}, {Wilson}, {Wilson}, {Wilson}, {Wolf}, {Wood-Vasey}, {Yan}, {Zamora}, {Zasowski}, {Zhang}, {Zhao}, {Zheng}, {Zheng}, \& {Zhu}}]{apogee22}
{Abdurro'uf}, {Accetta}, K., {Aerts}, C., {et~al.} 2022, \apjs, 259, 35

\bibitem[{{Aguado} {et~al.}(2021){Aguado}, {Belokurov}, {Myeong}, {Evans}, {Kobayashi}, {Sbordone}, {Chanam{\'e}}, {Navarrete}, \& {Koposov}}]{aguado21}
{Aguado}, D.~S., {Belokurov}, V., {Myeong}, G.~C., {et~al.} 2021, \apjl, 908, L8

\bibitem[{{Alvarez Garay} {et~al.}(2024){Alvarez Garay}, {Mucciarelli}, {Bellazzini}, {Lardo}, \& {Ventura}}]{deimer2024}
{Alvarez Garay}, D.~A., {Mucciarelli}, A., {Bellazzini}, M., {Lardo}, C., \& {Ventura}, P. 2024, \aap, 681, A54

\bibitem[{{Amarante} {et~al.}(2022){Amarante}, {Debattista}, {Beraldo e Silva}, {Laporte}, \& {Deg}}]{amarante2022}
{Amarante}, J. A.~S., {Debattista}, V.~P., {Beraldo e Silva}, L., {Laporte}, C. F.~P., \& {Deg}, N. 2022, \apj, 937, 12

\bibitem[{{Andrae} {et~al.}(2018){Andrae}, {Fouesneau}, {Creevey}, {Ordenovic}, {Mary}, {Burlacu}, {Chaoul}, {Jean-Antoine-Piccolo}, {Kordopatis}, {Korn}, {Lebreton}, {Panem}, {Pichon}, {Th{\'e}venin}, {Walmsley}, \& {Bailer-Jones}}]{andrae18}
{Andrae}, R., {Fouesneau}, M., {Creevey}, O., {et~al.} 2018, \aap, 616, A8

\bibitem[{{Barbuy} {et~al.}(2021){Barbuy}, {Cantelli}, {Muniz}, {Souza}, {Chiappini}, {Hirschi}, {Cescutti}, {Pignatari}, {Ortolani}, {Kerber}, {Maia}, {Bica}, \& {Depagne}}]{barbuy2021}
{Barbuy}, B., {Cantelli}, E., {Muniz}, L., {et~al.} 2021, \aap, 654, A29

\bibitem[{{Bastian} \& {Lardo}(2018)}]{bastian&lardo18}
{Bastian}, N. \& {Lardo}, C. 2018, \araa, 56, 83

\bibitem[{{Baumgardt} \& {Vasiliev}(2021)}]{baumgardt&vasiliev2021}
{Baumgardt}, H. \& {Vasiliev}, E. 2021, \mnras, 505, 5957

\bibitem[{{Bellazzini} {et~al.}(2020){Bellazzini}, {Ibata}, {Malhan}, {Martin}, {Famaey}, \& {Thomas}}]{bellazzini2020}
{Bellazzini}, M., {Ibata}, R., {Malhan}, K., {et~al.} 2020, \aap, 636, A107

\bibitem[{{Belokurov} {et~al.}(2018){Belokurov}, {Erkal}, {Evans}, {Koposov}, \& {Deason}}]{belokurov2018}
{Belokurov}, V., {Erkal}, D., {Evans}, N.~W., {Koposov}, S.~E., \& {Deason}, A.~J. 2018, \mnras, 478, 611

\bibitem[{{Belokurov} \& {Kravtsov}(2022)}]{belokurov22}
{Belokurov}, V. \& {Kravtsov}, A. 2022, \mnras, 514, 689

\bibitem[{{Belokurov} \& {Kravtsov}(2024)}]{belokurov&krastov2024}
{Belokurov}, V. \& {Kravtsov}, A. 2024, \mnras, 528, 3198

\bibitem[{{Belokurov} {et~al.}(2023){Belokurov}, {Vasiliev}, {Deason}, {Koposov}, {Fattahi}, {Dillamore}, {Davies}, \& {Grand}}]{belokurov23}
{Belokurov}, V., {Vasiliev}, E., {Deason}, A.~J., {et~al.} 2023, \mnras, 518, 6200

\bibitem[{{Brodie} \& {Strader}(2006)}]{brodie&starder2006}
{Brodie}, J.~P. \& {Strader}, J. 2006, \araa, 44, 193

\bibitem[{{Brown} {et~al.}(1997){Brown}, {Wallerstein}, \& {Zucker}}]{brown1997}
{Brown}, J.~A., {Wallerstein}, G., \& {Zucker}, D. 1997, \aj, 114, 180

\bibitem[{{Callingham} {et~al.}(2022){Callingham}, {Cautun}, {Deason}, {Frenk}, {Grand}, \& {Marinacci}}]{callingham2022}
{Callingham}, T.~M., {Cautun}, M., {Deason}, A.~J., {et~al.} 2022, \mnras, 513, 4107

\bibitem[{{Cardelli} {et~al.}(1989){Cardelli}, {Clayton}, \& {Mathis}}]{cardelli1989}
{Cardelli}, J.~A., {Clayton}, G.~C., \& {Mathis}, J.~S. 1989, \apj, 345, 245

\bibitem[{{Carretta} {et~al.}(2009){Carretta}, {Bragaglia}, {Gratton}, \& {Lucatello}}]{carretta09}
{Carretta}, E., {Bragaglia}, A., {Gratton}, R., \& {Lucatello}, S. 2009, \aap, 505, 139

\bibitem[{{Carretta} {et~al.}(2010){Carretta}, {Bragaglia}, {Gratton}, {Lucatello}, {Bellazzini}, \& {D'Orazi}}]{carretta2010}
{Carretta}, E., {Bragaglia}, A., {Gratton}, R., {et~al.} 2010, \apjl, 712, L21

\bibitem[{{Carretta} {et~al.}(2013){Carretta}, {Bragaglia}, {Gratton}, {Lucatello}, {D'Orazi}, {Bellazzini}, {Catanzaro}, {Leone}, {Momany}, \& {Sollima}}]{carretta13}
{Carretta}, E., {Bragaglia}, A., {Gratton}, R.~G., {et~al.} 2013, \aap, 557, A138

\bibitem[{{Carrillo} {et~al.}(2022){Carrillo}, {Hawkins}, {Jofr{\'e}}, {de Brito Silva}, {Das}, \& {Lucey}}]{carrillo2022}
{Carrillo}, A., {Hawkins}, K., {Jofr{\'e}}, P., {et~al.} 2022, \mnras, 513, 1557

\bibitem[{{Castelli} \& {Kurucz}(2003)}]{castelli&kurucz2003}
{Castelli}, F. \& {Kurucz}, R.~L. 2003, in Modelling of Stellar Atmospheres, ed. N.~{Piskunov}, W.~W. {Weiss}, \& D.~F. {Gray}, Vol. 210, A20

\bibitem[{{Ceccarelli} {et~al.}(2024){Ceccarelli}, {Massari}, {Mucciarelli}, {Bellazzini}, {Nunnari}, {Cusano}, {Lardo}, {Romano}, {Ilyin}, \& {Stokholm}}]{ceccarelli2024}
{Ceccarelli}, E., {Massari}, D., {Mucciarelli}, A., {et~al.} 2024, \aap, 684, A37

\bibitem[{{Cescutti} {et~al.}(2015){Cescutti}, {Romano}, {Matteucci}, {Chiappini}, \& {Hirschi}}]{cescutti2015}
{Cescutti}, G., {Romano}, D., {Matteucci}, F., {Chiappini}, C., \& {Hirschi}, R. 2015, \aap, 577, A139

\bibitem[{{Chen} \& {Gnedin}(2024)}]{chen&gnedin2024}
{Chen}, Y. \& {Gnedin}, O.~Y. 2024, The Open Journal of Astrophysics, 7, 23

\bibitem[{{Crestani} {et~al.}(2019){Crestani}, {Alves-Brito}, {Bono}, {Puls}, \& {Alonso-Garc{\'\i}a}}]{crestani2019}
{Crestani}, J., {Alves-Brito}, A., {Bono}, G., {Puls}, A.~A., \& {Alonso-Garc{\'\i}a}, J. 2019, \mnras, 487, 5463

\bibitem[{{Dalessandro} {et~al.}(2016){Dalessandro}, {Lapenna}, {Mucciarelli}, {Origlia}, {Ferraro}, \& {Lanzoni}}]{dalessandro2016}
{Dalessandro}, E., {Lapenna}, E., {Mucciarelli}, A., {et~al.} 2016, \apj, 829, 77

\bibitem[{{Davies} {et~al.}(2024){Davies}, {Monty}, {Belokurov}, \& {Dillamore}}]{davies2024}
{Davies}, E.~Y., {Monty}, S., {Belokurov}, V., \& {Dillamore}, A.~M. 2024, \mnras, 529, 772

\bibitem[{{Dekker} {et~al.}(2000){Dekker}, {D'Odorico}, {Kaufer}, {Delabre}, \& {Kotzlowski}}]{dekker2000}
{Dekker}, H., {D'Odorico}, S., {Kaufer}, A., {Delabre}, B., \& {Kotzlowski}, H. 2000, in Society of Photo-Optical Instrumentation Engineers (SPIE) Conference Series, Vol. 4008, Optical and IR Telescope Instrumentation and Detectors, ed. M.~{Iye} \& A.~F. {Moorwood}, 534--545

\bibitem[{{Dell'Agli} {et~al.}(2018){Dell'Agli}, {Garc{\'\i}a-Hern{\'a}ndez}, {Ventura}, {M{\'e}sz{\'a}ros}, {Masseron}, {Fern{\'a}ndez-Trincado}, {Tang}, {Shetrone}, {Zamora}, \& {Lucatello}}]{dellagli2018}
{Dell'Agli}, F., {Garc{\'\i}a-Hern{\'a}ndez}, D.~A., {Ventura}, P., {et~al.} 2018, \mnras, 475, 3098

\bibitem[{{Dotter} {et~al.}(2018){Dotter}, {Milone}, {Conroy}, {Marino}, \& {Sarajedini}}]{dotter18}
{Dotter}, A., {Milone}, A.~P., {Conroy}, C., {Marino}, A.~F., \& {Sarajedini}, A. 2018, \apjl, 865, L10

\bibitem[{{Dotter} {et~al.}(2011){Dotter}, {Sarajedini}, \& {Anderson}}]{dotter2011}
{Dotter}, A., {Sarajedini}, A., \& {Anderson}, J. 2011, \apj, 738, 74

\bibitem[{{Dotter} {et~al.}(2010){Dotter}, {Sarajedini}, {Anderson}, {Aparicio}, {Bedin}, {Chaboyer}, {Majewski}, {Mar{\'\i}n-Franch}, {Milone}, {Paust}, {Piotto}, {Reid}, {Rosenberg}, \& {Siegel}}]{dotter2010}
{Dotter}, A., {Sarajedini}, A., {Anderson}, J., {et~al.} 2010, \apj, 708, 698

\bibitem[{{Fernandes} {et~al.}(2023){Fernandes}, {Mason}, {Horta}, {Schiavon}, {Hayes}, {Hasselquist}, {Feuillet}, {Beaton}, {J{\"o}nsson}, {Kisku}, {Lacerna}, {Lian}, {Minniti}, \& {Villanova}}]{fernandes2023}
{Fernandes}, L., {Mason}, A.~C., {Horta}, D., {et~al.} 2023, \mnras, 519, 3611

\bibitem[{{Fishlock} {et~al.}(2017){Fishlock}, {Yong}, {Karakas}, {Alves-Brito}, {Mel{\'e}ndez}, {Nissen}, {Kobayashi}, \& {Casey}}]{fishlock2017}
{Fishlock}, C.~K., {Yong}, D., {Karakas}, A.~I., {et~al.} 2017, \mnras, 466, 4672

\bibitem[{{Forbes}(2020)}]{forbes2020}
{Forbes}, D.~A. 2020, \mnras, 493, 847

\bibitem[{{Forbes} \& {Bridges}(2010)}]{forbes&bridges2010}
{Forbes}, D.~A. \& {Bridges}, T. 2010, \mnras, 404, 1203

\bibitem[{{Fran{\c{c}}ois} {et~al.}(2024){Fran{\c{c}}ois}, {Cescutti}, {Bonifacio}, {Caffau}, {Monaco}, {Steffen}, {Puschnig}, {Calura}, {Cristallo}, {Di Marcantonio}, {Dobrovolskas}, {Franchini}, {Gallagher}, {Hansen}, {Korn}, {Ku{\v{c}}inskas}, {Lallement}, {Lombardo}, {Lucertini}, {Magrini}, {Matas Pinto}, {Matteucci}, {Mucciarelli}, {Sbordone}, {Spite}, {Spitoni}, \& {Valentini}}]{francois2024}
{Fran{\c{c}}ois}, P., {Cescutti}, G., {Bonifacio}, P., {et~al.} 2024, \aap, 686, A295

\bibitem[{{Frelijj} {et~al.}(2021){Frelijj}, {Villanova}, {Mu{\~n}oz}, \& {Fern{\'a}ndez-Trincado}}]{frelijj2021}
{Frelijj}, H., {Villanova}, S., {Mu{\~n}oz}, C., \& {Fern{\'a}ndez-Trincado}, J.~G. 2021, \mnras, 503, 867

\bibitem[{{Fusi Pecci} {et~al.}(1995){Fusi Pecci}, {Bellazzini}, {Cacciari}, \& {Ferraro}}]{fusipecci1995}
{Fusi Pecci}, F., {Bellazzini}, M., {Cacciari}, C., \& {Ferraro}, F.~R. 1995, \aj, 110, 1664

\bibitem[{{Gaia Collaboration} {et~al.}(2018){Gaia Collaboration}, {Babusiaux}, {van Leeuwen}, {Barstow}, {Jordi}, {Vallenari}, {Bossini}, {Bressan}, {Cantat-Gaudin}, {van Leeuwen}, {Brown}, {Prusti}, {de Bruijne}, {Bailer-Jones}, {Biermann}, {Evans}, {Eyer}, {Jansen}, {Klioner}, {Lammers}, {Lindegren}, {Luri}, {Mignard}, {Panem}, {Pourbaix}, {Randich}, {Sartoretti}, {Siddiqui}, {Soubiran}, {Walton}, {Arenou}, {Bastian}, {Cropper}, {Drimmel}, {Katz}, {Lattanzi}, {Bakker}, {Cacciari}, {Casta{\~n}eda}, {Chaoul}, {Cheek}, {De Angeli}, {Fabricius}, {Guerra}, {Holl}, {Masana}, {Messineo}, {Mowlavi}, {Nienartowicz}, {Panuzzo}, {Portell}, {Riello}, {Seabroke}, {Tanga}, {Th{\'e}venin}, {Gracia-Abril}, {Comoretto}, {Garcia-Reinaldos}, {Teyssier}, {Altmann}, {Andrae}, {Audard}, {Bellas-Velidis}, {Benson}, {Berthier}, {Blomme}, {Burgess}, {Busso}, {Carry}, {Cellino}, {Clementini}, {Clotet}, {Creevey}, {Davidson}, {De Ridder}, {Delchambre}, {Dell'Oro}, {Ducourant}, {Fern{\'a}ndez-Hern{\'a}ndez}, {Fouesneau},
  {Fr{\'e}mat}, {Galluccio}, {Garc{\'\i}a-Torres}, {Gonz{\'a}lez-N{\'u}{\~n}ez}, {Gonz{\'a}lez-Vidal}, {Gosset}, {Guy}, {Halbwachs}, {Hambly}, {Harrison}, {Hern{\'a}ndez}, {Hestroffer}, {Hodgkin}, {Hutton}, {Jasniewicz}, {Jean-Antoine-Piccolo}, {Jordan}, {Korn}, {Krone-Martins}, {Lanzafame}, {Lebzelter}, {L{\"o}ffler}, {Manteiga}, {Marrese}, {Mart{\'\i}n-Fleitas}, {Moitinho}, {Mora}, {Muinonen}, {Osinde}, {Pancino}, {Pauwels}, {Petit}, {Recio-Blanco}, {Richards}, {Rimoldini}, {Robin}, {Sarro}, {Siopis}, {Smith}, {Sozzetti}, {S{\"u}veges}, {Torra}, {van Reeven}, {Abbas}, {Abreu Aramburu}, {Accart}, {Aerts}, {Altavilla}, {{\'A}lvarez}, {Alvarez}, {Alves}, {Anderson}, {Andrei}, {Anglada Varela}, {Antiche}, {Antoja}, {Arcay}, {Astraatmadja}, {Bach}, {Baker}, {Balaguer-N{\'u}{\~n}ez}, {Balm}, {Barache}, {Barata}, {Barbato}, {Barblan}, {Barklem}, {Barrado}, {Barros}, {Bartholom{\'e} Mu{\~n}oz}, {Bassilana}, {Becciani}, {Bellazzini}, {Berihuete}, {Bertone}, {Bianchi}, {Bienaym{\'e}}, {Blanco-Cuaresma}, {Boch},
  {Boeche}, {Bombrun}, {Borrachero}, {Bouquillon}, {Bourda}, {Bragaglia}, {Bramante}, {Breddels}, {Brouillet}, {Br{\"u}semeister}, {Brugaletta}, {Bucciarelli}, {Burlacu}, {Busonero}, {Butkevich}, {Buzzi}, {Caffau}, {Cancelliere}, {Cannizzaro}, {Carballo}, {Carlucci}, {Carrasco}, {Casamiquela}, {Castellani}, {Castro-Ginard}, {Charlot}, {Chemin}, {Chiavassa}, {Cocozza}, {Costigan}, {Cowell}, {Crifo}, {Crosta}, {Crowley}, {Cuypers}, {Dafonte}, {Damerdji}, {Dapergolas}, {David}, {David}, {de Laverny}, {De Luise}, {De March}, {de Martino}, {de Souza}, {de Torres}, {Debosscher}, {del Pozo}, {Delbo}, {Delgado}, {Delgado}, {Diakite}, {Diener}, {Distefano}, {Dolding}, {Drazinos}, {Dur{\'a}n}, {Edvardsson}, {Enke}, {Eriksson}, {Esquej}, {Eynard Bontemps}, {Fabre}, {Fabrizio}, {Faigler}, {Falc{\~a}o}, {Farr{\`a}s Casas}, {Federici}, {Fedorets}, {Fernique}, {Figueras}, {Filippi}, {Findeisen}, {Fonti}, {Fraile}, {Fraser}, {Fr{\'e}zouls}, {Gai}, {Galleti}, {Garabato}, {Garc{\'\i}a-Sedano}, {Garofalo}, {Garralda}, {Gavel},
  {Gavras}, {Gerssen}, {Geyer}, {Giacobbe}, {Gilmore}, {Girona}, {Giuffrida}, {Glass}, {Gomes}, {Granvik}, {Gueguen}, {Guerrier}, {Guiraud}, {Guti{\'e}}, {Haigron}, {Hatzidimitriou}, {Hauser}, {Haywood}, {Heiter}, {Helmi}, {Heu}, {Hilger}, {Hobbs}, {Hofmann}, {Holland}, {Huckle}, {Hypki}, {Icardi}, {Jan{\ss}en}, {Jevardat de Fombelle}, {Jonker}, {Juh{\'a}sz}, {Julbe}, {Karampelas}, {Kewley}, {Klar}, {Kochoska}, {Kohley}, {Kolenberg}, {Kontizas}, {Kontizas}, {Koposov}, {Kordopatis}, {Kostrzewa-Rutkowska}, {Koubsky}, {Lambert}, {Lanza}, {Lasne}, {Lavigne}, {Le Fustec}, {Le Poncin-Lafitte}, {Lebreton}, {Leccia}, {Leclerc}, {Lecoeur-Taibi}, {Lenhardt}, {Leroux}, {Liao}, {Licata}, {Lindstr{\o}m}, {Lister}, {Livanou}, {Lobel}, {L{\'o}pez}, {Managau}, {Mann}, {Mantelet}, {Marchal}, {Marchant}, {Marconi}, {Marinoni}, {Marschalk{\'o}}, {Marshall}, {Martino}, {Marton}, {Mary}, {Massari}, {Matijevi{\v{c}}}, {Mazeh}, {McMillan}, {Messina}, {Michalik}, {Millar}, {Molina}, {Molinaro}, {Moln{\'a}r}, {Montegriffo}, {Mor},
  {Morbidelli}, {Morel}, {Morris}, {Mulone}, {Muraveva}, {Musella}, {Nelemans}, {Nicastro}, {Noval}, {O'Mullane}, {Ord{\'e}novic}, {Ord{\'o}{\~n}ez-Blanco}, {Osborne}, {Pagani}, {Pagano}, {Pailler}, {Palacin}, {Palaversa}, {Panahi}, {Pawlak}, {Piersimoni}, {Pineau}, {Plachy}, {Plum}, {Poggio}, {Poujoulet}, {Pr{\v{s}}a}, {Pulone}, {Racero}, {Ragaini}, {Rambaux}, {Ramos-Lerate}, {Regibo}, {Reyl{\'e}}, {Riclet}, {Ripepi}, {Riva}, {Rivard}, {Rixon}, {Roegiers}, {Roelens}, {Romero-G{\'o}mez}, {Rowell}, {Royer}, {Ruiz-Dern}, {Sadowski}, {Sagrist{\`a} Sell{\'e}s}, {Sahlmann}, {Salgado}, {Salguero}, {Sanna}, {Santana-Ros}, {Sarasso}, {Savietto}, {Schultheis}, {Sciacca}, {Segol}, {Segovia}, {S{\'e}gransan}, {Shih}, {Siltala}, {Silva}, {Smart}, {Smith}, {Solano}, {Solitro}, {Sordo}, {Soria Nieto}, {Souchay}, {Spagna}, {Spoto}, {Stampa}, {Steele}, {Steidelm{\"u}ller}, {Stephenson}, {Stoev}, {Suess}, {Surdej}, {Szabados}, {Szegedi-Elek}, {Tapiador}, {Taris}, {Tauran}, {Taylor}, {Teixeira}, {Terrett}, {Teyssandier},
  {Thuillot}, {Titarenko}, {Torra Clotet}, {Turon}, {Ulla}, {Utrilla}, {Uzzi}, {Vaillant}, {Valentini}, {Valette}, {van Elteren}, {Van Hemelryck}, {Vaschetto}, {Vecchiato}, {Veljanoski}, {Viala}, {Vicente}, {Vogt}, {von Essen}, {Voss}, {Votruba}, {Voutsinas}, {Walmsley}, {Weiler}, {Wertz}, {Wevers}, {Wyrzykowski}, {Yoldas}, {{\v{Z}}erjal}, {Ziaeepour}, {Zorec}, {Zschocke}, {Zucker}, {Zurbach}, \& {Zwitter}}]{GC18_extinction}
{Gaia Collaboration}, {Babusiaux}, C., {van Leeuwen}, F., {et~al.} 2018, \aap, 616, A10

\bibitem[{{Gaia Collaboration} {et~al.}(2021){Gaia Collaboration}, {Brown}, {Vallenari}, {Prusti}, {de Bruijne}, {Babusiaux}, {Biermann}, {Creevey}, {Evans}, {Eyer}, {Hutton}, {Jansen}, {Jordi}, {Klioner}, {Lammers}, {Lindegren}, {Luri}, {Mignard}, {Panem}, {Pourbaix}, {Randich}, {Sartoretti}, {Soubiran}, {Walton}, {Arenou}, {Bailer-Jones}, {Bastian}, {Cropper}, {Drimmel}, {Katz}, {Lattanzi}, {van Leeuwen}, {Bakker}, {Cacciari}, {Casta{\~n}eda}, {De Angeli}, {Ducourant}, {Fabricius}, {Fouesneau}, {Fr{\'e}mat}, {Guerra}, {Guerrier}, {Guiraud}, {Jean-Antoine Piccolo}, {Masana}, {Messineo}, {Mowlavi}, {Nicolas}, {Nienartowicz}, {Pailler}, {Panuzzo}, {Riclet}, {Roux}, {Seabroke}, {Sordo}, {Tanga}, {Th{\'e}venin}, {Gracia-Abril}, {Portell}, {Teyssier}, {Altmann}, {Andrae}, {Bellas-Velidis}, {Benson}, {Berthier}, {Blomme}, {Brugaletta}, {Burgess}, {Busso}, {Carry}, {Cellino}, {Cheek}, {Clementini}, {Damerdji}, {Davidson}, {Delchambre}, {Dell'Oro}, {Fern{\'a}ndez-Hern{\'a}ndez}, {Galluccio}, {Garc{\'\i}a-Lario},
  {Garcia-Reinaldos}, {Gonz{\'a}lez-N{\'u}{\~n}ez}, {Gosset}, {Haigron}, {Halbwachs}, {Hambly}, {Harrison}, {Hatzidimitriou}, {Heiter}, {Hern{\'a}ndez}, {Hestroffer}, {Hodgkin}, {Holl}, {Jan{\ss}en}, {Jevardat de Fombelle}, {Jordan}, {Krone-Martins}, {Lanzafame}, {L{\"o}ffler}, {Lorca}, {Manteiga}, {Marchal}, {Marrese}, {Moitinho}, {Mora}, {Muinonen}, {Osborne}, {Pancino}, {Pauwels}, {Petit}, {Recio-Blanco}, {Richards}, {Riello}, {Rimoldini}, {Robin}, {Roegiers}, {Rybizki}, {Sarro}, {Siopis}, {Smith}, {Sozzetti}, {Ulla}, {Utrilla}, {van Leeuwen}, {van Reeven}, {Abbas}, {Abreu Aramburu}, {Accart}, {Aerts}, {Aguado}, {Ajaj}, {Altavilla}, {{\'A}lvarez}, {{\'A}lvarez Cid-Fuentes}, {Alves}, {Anderson}, {Anglada Varela}, {Antoja}, {Audard}, {Baines}, {Baker}, {Balaguer-N{\'u}{\~n}ez}, {Balbinot}, {Balog}, {Barache}, {Barbato}, {Barros}, {Barstow}, {Bartolom{\'e}}, {Bassilana}, {Bauchet}, {Baudesson-Stella}, {Becciani}, {Bellazzini}, {Bernet}, {Bertone}, {Bianchi}, {Blanco-Cuaresma}, {Boch}, {Bombrun}, {Bossini},
  {Bouquillon}, {Bragaglia}, {Bramante}, {Breedt}, {Bressan}, {Brouillet}, {Bucciarelli}, {Burlacu}, {Busonero}, {Butkevich}, {Buzzi}, {Caffau}, {Cancelliere}, {C{\'a}novas}, {Cantat-Gaudin}, {Carballo}, {Carlucci}, {Carnerero}, {Carrasco}, {Casamiquela}, {Castellani}, {Castro-Ginard}, {Castro Sampol}, {Chaoul}, {Charlot}, {Chemin}, {Chiavassa}, {Cioni}, {Comoretto}, {Cooper}, {Cornez}, {Cowell}, {Crifo}, {Crosta}, {Crowley}, {Dafonte}, {Dapergolas}, {David}, {David}, {de Laverny}, {De Luise}, {De March}, {De Ridder}, {de Souza}, {de Teodoro}, {de Torres}, {del Peloso}, {del Pozo}, {Delbo}, {Delgado}, {Delgado}, {Delisle}, {Di Matteo}, {Diakite}, {Diener}, {Distefano}, {Dolding}, {Eappachen}, {Edvardsson}, {Enke}, {Esquej}, {Fabre}, {Fabrizio}, {Faigler}, {Fedorets}, {Fernique}, {Fienga}, {Figueras}, {Fouron}, {Fragkoudi}, {Fraile}, {Franke}, {Gai}, {Garabato}, {Garcia-Gutierrez}, {Garc{\'\i}a-Torres}, {Garofalo}, {Gavras}, {Gerlach}, {Geyer}, {Giacobbe}, {Gilmore}, {Girona}, {Giuffrida}, {Gomel}, {Gomez},
  {Gonzalez-Santamaria}, {Gonz{\'a}lez-Vidal}, {Granvik}, {Guti{\'e}rrez-S{\'a}nchez}, {Guy}, {Hauser}, {Haywood}, {Helmi}, {Hidalgo}, {Hilger}, {H{\l}adczuk}, {Hobbs}, {Holland}, {Huckle}, {Jasniewicz}, {Jonker}, {Juaristi Campillo}, {Julbe}, {Karbevska}, {Kervella}, {Khanna}, {Kochoska}, {Kontizas}, {Kordopatis}, {Korn}, {Kostrzewa-Rutkowska}, {Kruszy{\'n}ska}, {Lambert}, {Lanza}, {Lasne}, {Le Campion}, {Le Fustec}, {Lebreton}, {Lebzelter}, {Leccia}, {Leclerc}, {Lecoeur-Taibi}, {Liao}, {Licata}, {Lindstr{\o}m}, {Lister}, {Livanou}, {Lobel}, {Madrero Pardo}, {Managau}, {Mann}, {Marchant}, {Marconi}, {Marcos Santos}, {Marinoni}, {Marocco}, {Marshall}, {Martin Polo}, {Mart{\'\i}n-Fleitas}, {Masip}, {Massari}, {Mastrobuono-Battisti}, {Mazeh}, {McMillan}, {Messina}, {Michalik}, {Millar}, {Mints}, {Molina}, {Molinaro}, {Moln{\'a}r}, {Montegriffo}, {Mor}, {Morbidelli}, {Morel}, {Morris}, {Mulone}, {Munoz}, {Muraveva}, {Murphy}, {Musella}, {Noval}, {Ord{\'e}novic}, {Orr{\`u}}, {Osinde}, {Pagani}, {Pagano},
  {Palaversa}, {Palicio}, {Panahi}, {Pawlak}, {Pe{\~n}alosa Esteller}, {Penttil{\"a}}, {Piersimoni}, {Pineau}, {Plachy}, {Plum}, {Poggio}, {Poretti}, {Poujoulet}, {Pr{\v{s}}a}, {Pulone}, {Racero}, {Ragaini}, {Rainer}, {Raiteri}, {Rambaux}, {Ramos}, {Ramos-Lerate}, {Re Fiorentin}, {Regibo}, {Reyl{\'e}}, {Ripepi}, {Riva}, {Rixon}, {Robichon}, {Robin}, {Roelens}, {Rohrbasser}, {Romero-G{\'o}mez}, {Rowell}, {Royer}, {Rybicki}, {Sadowski}, {Sagrist{\`a} Sell{\'e}s}, {Sahlmann}, {Salgado}, {Salguero}, {Samaras}, {Sanchez Gimenez}, {Sanna}, {Santove{\~n}a}, {Sarasso}, {Schultheis}, {Sciacca}, {Segol}, {Segovia}, {S{\'e}gransan}, {Semeux}, {Shahaf}, {Siddiqui}, {Siebert}, {Siltala}, {Slezak}, {Smart}, {Solano}, {Solitro}, {Souami}, {Souchay}, {Spagna}, {Spoto}, {Steele}, {Steidelm{\"u}ller}, {Stephenson}, {S{\"u}veges}, {Szabados}, {Szegedi-Elek}, {Taris}, {Tauran}, {Taylor}, {Teixeira}, {Thuillot}, {Tonello}, {Torra}, {Torra}, {Turon}, {Unger}, {Vaillant}, {van Dillen}, {Vanel}, {Vecchiato}, {Viala}, {Vicente},
  {Voutsinas}, {Weiler}, {Wevers}, {Wyrzykowski}, {Yoldas}, {Yvard}, {Zhao}, {Zorec}, {Zucker}, {Zurbach}, \& {Zwitter}}]{GC21}
{Gaia Collaboration}, {Brown}, A.~G.~A., {Vallenari}, A., {et~al.} 2021, \aap, 649, A1

\bibitem[{{Gaia Collaboration} {et~al.}(2023){Gaia Collaboration}, {Vallenari}, {Brown}, {Prusti}, {de Bruijne}, {Arenou}, {Babusiaux}, {Biermann}, {Creevey}, {Ducourant}, {Evans}, {Eyer}, {Guerra}, {Hutton}, {Jordi}, {Klioner}, {Lammers}, {Lindegren}, {Luri}, {Mignard}, {Panem}, {Pourbaix}, {Randich}, {Sartoretti}, {Soubiran}, {Tanga}, {Walton}, {Bailer-Jones}, {Bastian}, {Drimmel}, {Jansen}, {Katz}, {Lattanzi}, {van Leeuwen}, {Bakker}, {Cacciari}, {Casta{\~n}eda}, {De Angeli}, {Fabricius}, {Fouesneau}, {Fr{\'e}mat}, {Galluccio}, {Guerrier}, {Heiter}, {Masana}, {Messineo}, {Mowlavi}, {Nicolas}, {Nienartowicz}, {Pailler}, {Panuzzo}, {Riclet}, {Roux}, {Seabroke}, {Sordo}, {Th{\'e}venin}, {Gracia-Abril}, {Portell}, {Teyssier}, {Altmann}, {Andrae}, {Audard}, {Bellas-Velidis}, {Benson}, {Berthier}, {Blomme}, {Burgess}, {Busonero}, {Busso}, {C{\'a}novas}, {Carry}, {Cellino}, {Cheek}, {Clementini}, {Damerdji}, {Davidson}, {de Teodoro}, {Nu{\~n}ez Campos}, {Delchambre}, {Dell'Oro}, {Esquej},
  {Fern{\'a}ndez-Hern{\'a}ndez}, {Fraile}, {Garabato}, {Garc{\'\i}a-Lario}, {Gosset}, {Haigron}, {Halbwachs}, {Hambly}, {Harrison}, {Hern{\'a}ndez}, {Hestroffer}, {Hodgkin}, {Holl}, {Jan{\ss}en}, {Jevardat de Fombelle}, {Jordan}, {Krone-Martins}, {Lanzafame}, {L{\"o}ffler}, {Marchal}, {Marrese}, {Moitinho}, {Muinonen}, {Osborne}, {Pancino}, {Pauwels}, {Recio-Blanco}, {Reyl{\'e}}, {Riello}, {Rimoldini}, {Roegiers}, {Rybizki}, {Sarro}, {Siopis}, {Smith}, {Sozzetti}, {Utrilla}, {van Leeuwen}, {Abbas}, {{\'A}brah{\'a}m}, {Abreu Aramburu}, {Aerts}, {Aguado}, {Ajaj}, {Aldea-Montero}, {Altavilla}, {{\'A}lvarez}, {Alves}, {Anders}, {Anderson}, {Anglada Varela}, {Antoja}, {Baines}, {Baker}, {Balaguer-N{\'u}{\~n}ez}, {Balbinot}, {Balog}, {Barache}, {Barbato}, {Barros}, {Barstow}, {Bartolom{\'e}}, {Bassilana}, {Bauchet}, {Becciani}, {Bellazzini}, {Berihuete}, {Bernet}, {Bertone}, {Bianchi}, {Binnenfeld}, {Blanco-Cuaresma}, {Blazere}, {Boch}, {Bombrun}, {Bossini}, {Bouquillon}, {Bragaglia}, {Bramante}, {Breedt},
  {Bressan}, {Brouillet}, {Brugaletta}, {Bucciarelli}, {Burlacu}, {Butkevich}, {Buzzi}, {Caffau}, {Cancelliere}, {Cantat-Gaudin}, {Carballo}, {Carlucci}, {Carnerero}, {Carrasco}, {Casamiquela}, {Castellani}, {Castro-Ginard}, {Chaoul}, {Charlot}, {Chemin}, {Chiaramida}, {Chiavassa}, {Chornay}, {Comoretto}, {Contursi}, {Cooper}, {Cornez}, {Cowell}, {Crifo}, {Cropper}, {Crosta}, {Crowley}, {Dafonte}, {Dapergolas}, {David}, {David}, {de Laverny}, {De Luise}, {De March}, {De Ridder}, {de Souza}, {de Torres}, {del Peloso}, {del Pozo}, {Delbo}, {Delgado}, {Delisle}, {Demouchy}, {Dharmawardena}, {Di Matteo}, {Diakite}, {Diener}, {Distefano}, {Dolding}, {Edvardsson}, {Enke}, {Fabre}, {Fabrizio}, {Faigler}, {Fedorets}, {Fernique}, {Fienga}, {Figueras}, {Fournier}, {Fouron}, {Fragkoudi}, {Gai}, {Garcia-Gutierrez}, {Garcia-Reinaldos}, {Garc{\'\i}a-Torres}, {Garofalo}, {Gavel}, {Gavras}, {Gerlach}, {Geyer}, {Giacobbe}, {Gilmore}, {Girona}, {Giuffrida}, {Gomel}, {Gomez}, {Gonz{\'a}lez-N{\'u}{\~n}ez},
  {Gonz{\'a}lez-Santamar{\'\i}a}, {Gonz{\'a}lez-Vidal}, {Granvik}, {Guillout}, {Guiraud}, {Guti{\'e}rrez-S{\'a}nchez}, {Guy}, {Hatzidimitriou}, {Hauser}, {Haywood}, {Helmer}, {Helmi}, {Sarmiento}, {Hidalgo}, {Hilger}, {H{\l}adczuk}, {Hobbs}, {Holland}, {Huckle}, {Jardine}, {Jasniewicz}, {Jean-Antoine Piccolo}, {Jim{\'e}nez-Arranz}, {Jorissen}, {Juaristi Campillo}, {Julbe}, {Karbevska}, {Kervella}, {Khanna}, {Kontizas}, {Kordopatis}, {Korn}, {K{\'o}sp{\'a}l}, {Kostrzewa-Rutkowska}, {Kruszy{\'n}ska}, {Kun}, {Laizeau}, {Lambert}, {Lanza}, {Lasne}, {Le Campion}, {Lebreton}, {Lebzelter}, {Leccia}, {Leclerc}, {Lecoeur-Taibi}, {Liao}, {Licata}, {Lindstr{\o}m}, {Lister}, {Livanou}, {Lobel}, {Lorca}, {Loup}, {Madrero Pardo}, {Magdaleno Romeo}, {Managau}, {Mann}, {Manteiga}, {Marchant}, {Marconi}, {Marcos}, {Marcos Santos}, {Mar{\'\i}n Pina}, {Marinoni}, {Marocco}, {Marshall}, {Martin Polo}, {Mart{\'\i}n-Fleitas}, {Marton}, {Mary}, {Masip}, {Massari}, {Mastrobuono-Battisti}, {Mazeh}, {McMillan}, {Messina}, {Michalik},
  {Millar}, {Mints}, {Molina}, {Molinaro}, {Moln{\'a}r}, {Monari}, {Mongui{\'o}}, {Montegriffo}, {Montero}, {Mor}, {Mora}, {Morbidelli}, {Morel}, {Morris}, {Muraveva}, {Murphy}, {Musella}, {Nagy}, {Noval}, {Oca{\~n}a}, {Ogden}, {Ordenovic}, {Osinde}, {Pagani}, {Pagano}, {Palaversa}, {Palicio}, {Pallas-Quintela}, {Panahi}, {Payne-Wardenaar}, {Pe{\~n}alosa Esteller}, {Penttil{\"a}}, {Pichon}, {Piersimoni}, {Pineau}, {Plachy}, {Plum}, {Poggio}, {Pr{\v{s}}a}, {Pulone}, {Racero}, {Ragaini}, {Rainer}, {Raiteri}, {Rambaux}, {Ramos}, {Ramos-Lerate}, {Re Fiorentin}, {Regibo}, {Richards}, {Rios Diaz}, {Ripepi}, {Riva}, {Rix}, {Rixon}, {Robichon}, {Robin}, {Robin}, {Roelens}, {Rogues}, {Rohrbasser}, {Romero-G{\'o}mez}, {Rowell}, {Royer}, {Ruz Mieres}, {Rybicki}, {Sadowski}, {S{\'a}ez N{\'u}{\~n}ez}, {Sagrist{\`a} Sell{\'e}s}, {Sahlmann}, {Salguero}, {Samaras}, {Sanchez Gimenez}, {Sanna}, {Santove{\~n}a}, {Sarasso}, {Schultheis}, {Sciacca}, {Segol}, {Segovia}, {S{\'e}gransan}, {Semeux}, {Shahaf}, {Siddiqui}, {Siebert},
  {Siltala}, {Silvelo}, {Slezak}, {Slezak}, {Smart}, {Snaith}, {Solano}, {Solitro}, {Souami}, {Souchay}, {Spagna}, {Spina}, {Spoto}, {Steele}, {Steidelm{\"u}ller}, {Stephenson}, {S{\"u}veges}, {Surdej}, {Szabados}, {Szegedi-Elek}, {Taris}, {Taylor}, {Teixeira}, {Tolomei}, {Tonello}, {Torra}, {Torra}, {Torralba Elipe}, {Trabucchi}, {Tsounis}, {Turon}, {Ulla}, {Unger}, {Vaillant}, {van Dillen}, {van Reeven}, {Vanel}, {Vecchiato}, {Viala}, {Vicente}, {Voutsinas}, {Weiler}, {Wevers}, {Wyrzykowski}, {Yoldas}, {Yvard}, {Zhao}, {Zorec}, {Zucker}, \& {Zwitter}}]{GC23}
{Gaia Collaboration}, {Vallenari}, A., {Brown}, A.~G.~A., {et~al.} 2023, \aap, 674, A1

\bibitem[{{Giribaldi} \& {Smiljanic}(2023)}]{giribaldi2023}
{Giribaldi}, R.~E. \& {Smiljanic}, R. 2023, \aap, 673, A18

\bibitem[{{Glatt} {et~al.}(2008){Glatt}, {Gallagher}, {Grebel}, {Nota}, {Sabbi}, {Sirianni}, {Clementini}, {Tosi}, {Harbeck}, {Koch}, \& {Cracraft}}]{glatt2008}
{Glatt}, K., {Gallagher}, John~S., I., {Grebel}, E.~K., {et~al.} 2008, \aj, 135, 1106

\bibitem[{{Grevesse} \& {Sauval}(1998)}]{grevesse1998}
{Grevesse}, N. \& {Sauval}, A.~J. 1998, \ssr, 85, 161

\bibitem[{{Harris}(2010)}]{harris2010}
{Harris}, W.~E. 2010, arXiv e-prints, arXiv:1012.3224

\bibitem[{{Helmi}(2020)}]{helmi2020}
{Helmi}, A. 2020, \araa, 58, 205

\bibitem[{{Helmi} {et~al.}(2018){Helmi}, {Babusiaux}, {Koppelman}, {Massari}, {Veljanoski}, \& {Brown}}]{helmi2018}
{Helmi}, A., {Babusiaux}, C., {Koppelman}, H.~H., {et~al.} 2018, \nat, 563, 85

\bibitem[{{Helmi} \& {de Zeeuw}(2000)}]{helmidezeeuw00}
{Helmi}, A. \& {de Zeeuw}, P.~T. 2000, \mnras, 319, 657

\bibitem[{{Helmi} {et~al.}(1999){Helmi}, {White}, {de Zeeuw}, \& {Zhao}}]{helmi99}
{Helmi}, A., {White}, S. D.~M., {de Zeeuw}, P.~T., \& {Zhao}, H. 1999, \nat, 402, 53

\bibitem[{{Hendricks} {et~al.}(2014){Hendricks}, {Koch}, {Walker}, {Johnson}, {Pe{\~n}arrubia}, \& {Gilmore}}]{hendricks2014}
{Hendricks}, B., {Koch}, A., {Walker}, M., {et~al.} 2014, \aap, 572, A82

\bibitem[{{Hill} {et~al.}(2019){Hill}, {Sk{\'u}lad{\'o}ttir}, {Tolstoy}, {Venn}, {Shetrone}, {Jablonka}, {Primas}, {Battaglia}, {de Boer}, {Fran{\c{c}}ois}, {Helmi}, {Kaufer}, {Letarte}, {Starkenburg}, \& {Spite}}]{hill2019}
{Hill}, V., {Sk{\'u}lad{\'o}ttir}, {\'A}., {Tolstoy}, E., {et~al.} 2019, \aap, 626, A15

\bibitem[{{Horta} {et~al.}(2020){Horta}, {Schiavon}, {Mackereth}, {Beers}, {Fern{\'a}ndez-Trincado}, {Frinchaboy}, {Garc{\'\i}a-Hern{\'a}ndez}, {Geisler}, {Hasselquist}, {J{\"o}nsson}, {Lane}, {Majewski}, {M{\'e}sz{\'a}ros}, {Bidin}, {Nataf}, {Roman-Lopes}, {Nitschelm}, {Vargas-Gonz{\'a}lez}, \& {Zasowski}}]{horta2020}
{Horta}, D., {Schiavon}, R.~P., {Mackereth}, J.~T., {et~al.} 2020, \mnras, 493, 3363

\bibitem[{{Horta} {et~al.}(2023){Horta}, {Schiavon}, {Mackereth}, {Weinberg}, {Hasselquist}, {Feuillet}, {O'Connell}, {Anguiano}, {Allende-Prieto}, {Beaton}, {Bizyaev}, {Cunha}, {Geisler}, {Garc{\'\i}a-Hern{\'a}ndez}, {Holtzman}, {J{\"o}nsson}, {Lane}, {Majewski}, {M{\'e}sz{\'a}ros}, {Minniti}, {Nitschelm}, {Shetrone}, {Smith}, \& {Zasowski}}]{horta23}
{Horta}, D., {Schiavon}, R.~P., {Mackereth}, J.~T., {et~al.} 2023, \mnras, 520, 5671

\bibitem[{{Ibata} {et~al.}(2024){Ibata}, {Malhan}, {Tenachi}, {Ardern-Arentsen}, {Bellazzini}, {Bianchini}, {Bonifacio}, {Caffau}, {Diakogiannis}, {Errani}, {Famaey}, {Ferrone}, {Martin}, {di Matteo}, {Monari}, {Renaud}, {Starkenburg}, {Thomas}, {Viswanathan}, \& {Yuan}}]{ibata2024}
{Ibata}, R., {Malhan}, K., {Tenachi}, W., {et~al.} 2024, \apj, 967, 89

\bibitem[{{Ibata} {et~al.}(1994){Ibata}, {Gilmore}, \& {Irwin}}]{ibata94}
{Ibata}, R.~A., {Gilmore}, G., \& {Irwin}, M.~J. 1994, \nat, 370, 194

\bibitem[{{Kerber} {et~al.}(2018){Kerber}, {Nardiello}, {Ortolani}, {Barbuy}, {Bica}, {Cassisi}, {Libralato}, \& {Vieira}}]{kerber2018}
{Kerber}, L.~O., {Nardiello}, D., {Ortolani}, S., {et~al.} 2018, \apj, 853, 15

\bibitem[{{Kobayashi} {et~al.}(2020){Kobayashi}, {Karakas}, \& {Lugaro}}]{kobayashi2020}
{Kobayashi}, C., {Karakas}, A.~I., \& {Lugaro}, M. 2020, \apj, 900, 179

\bibitem[{{Kobayashi} \& {Nomoto}(2009)}]{kobayashi2009}
{Kobayashi}, C. \& {Nomoto}, K. 2009, \apj, 707, 1466

\bibitem[{{Koch-Hansen} {et~al.}(2021){Koch-Hansen}, {Hansen}, \& {McWilliam}}]{koch-hansen2021}
{Koch-Hansen}, A.~J., {Hansen}, C.~J., \& {McWilliam}, A. 2021, \aap, 653, A2

\bibitem[{{Koppelman} {et~al.}(2019){Koppelman}, {Helmi}, {Massari}, {Price-Whelan}, \& {Starkenburg}}]{koppelman19}
{Koppelman}, H.~H., {Helmi}, A., {Massari}, D., {Price-Whelan}, A.~M., \& {Starkenburg}, T.~K. 2019, \aap, 631, L9

\bibitem[{{Kruijssen} {et~al.}(2019){Kruijssen}, {Pfeffer}, {Reina-Campos}, {Crain}, \& {Bastian}}]{kruijssen2019}
{Kruijssen}, J.~M.~D., {Pfeffer}, J.~L., {Reina-Campos}, M., {Crain}, R.~A., \& {Bastian}, N. 2019, \mnras, 486, 3180

\bibitem[{{Kurucz}(2005)}]{kurucz}
{Kurucz}, R.~L. 2005, Memorie della Societa Astronomica Italiana Supplementi, 8, 14

\bibitem[{{Lagioia} {et~al.}(2024){Lagioia}, {Milone}, {Legnardi}, {Cordoni}, {Dondoglio}, {Renzini}, {Tailo}, {Ziliotto}, {Carlos}, {Jang}, {Marino}, {Mohandasan}, {Qi}, {Rangwal}, {Bortolan}, \& {Muratore}}]{lagioia24}
{Lagioia}, E.~P., {Milone}, A.~P., {Legnardi}, M.~V., {et~al.} 2024, arXiv e-prints, arXiv:2406.16824

\bibitem[{{Lattimer} \& {Schramm}(1974)}]{lattimer1974}
{Lattimer}, J.~M. \& {Schramm}, D.~N. 1974, \apjl, 192, L145

\bibitem[{{Leaman} {et~al.}(2013){Leaman}, {VandenBerg}, \& {Mendel}}]{leaman2013}
{Leaman}, R., {VandenBerg}, D.~A., \& {Mendel}, J.~T. 2013, \mnras, 436, 122

\bibitem[{{Lemasle} {et~al.}(2014){Lemasle}, {de Boer}, {Hill}, {Tolstoy}, {Irwin}, {Jablonka}, {Venn}, {Battaglia}, {Starkenburg}, {Shetrone}, {Letarte}, {Fran{\c{c}}ois}, {Helmi}, {Primas}, {Kaufer}, \& {Szeifert}}]{lemasle2014}
{Lemasle}, B., {de Boer}, T.~J.~L., {Hill}, V., {et~al.} 2014, \aap, 572, A88

\bibitem[{{Letarte} {et~al.}(2010){Letarte}, {Hill}, {Tolstoy}, {Jablonka}, {Shetrone}, {Venn}, {Spite}, {Irwin}, {Battaglia}, {Helmi}, {Primas}, {Fran{\c{c}}ois}, {Kaufer}, {Szeifert}, {Arimoto}, \& {Sadakane}}]{letarte2010}
{Letarte}, B., {Hill}, V., {Tolstoy}, E., {et~al.} 2010, \aap, 523, A17

\bibitem[{{Limberg} {et~al.}(2022){Limberg}, {Souza}, {P{\'e}rez-Villegas}, {Rossi}, {Perottoni}, \& {Santucci}}]{limberg22}
{Limberg}, G., {Souza}, S.~O., {P{\'e}rez-Villegas}, A., {et~al.} 2022, \apj, 935, 109

\bibitem[{{Lin} \& {Richer}(1992)}]{lin1992}
{Lin}, D.~N.~C. \& {Richer}, H.~B. 1992, \apjl, 388, L57

\bibitem[{{Lucertini} {et~al.}(2023){Lucertini}, {Monaco}, {Caffau}, {Mucciarelli}, {Villanova}, {Bonifacio}, \& {Sbordone}}]{lucertini23}
{Lucertini}, F., {Monaco}, L., {Caffau}, E., {et~al.} 2023, \aap, 671, A137

\bibitem[{{Majewski} {et~al.}(2003){Majewski}, {Skrutskie}, {Weinberg}, \& {Ostheimer}}]{majewski2003}
{Majewski}, S.~R., {Skrutskie}, M.~F., {Weinberg}, M.~D., \& {Ostheimer}, J.~C. 2003, \apj, 599, 1082

\bibitem[{{Malhan} {et~al.}(2022){Malhan}, {Ibata}, {Sharma}, {Famaey}, {Bellazzini}, {Carlberg}, {D'Souza}, {Yuan}, {Martin}, \& {Thomas}}]{malhan2022}
{Malhan}, K., {Ibata}, R.~A., {Sharma}, S., {et~al.} 2022, \apj, 926, 107

\bibitem[{{Mar{\'\i}n-Franch} {et~al.}(2009){Mar{\'\i}n-Franch}, {Aparicio}, {Piotto}, {Rosenberg}, {Chaboyer}, {Sarajedini}, {Siegel}, {Anderson}, {Bedin}, {Dotter}, {Hempel}, {King}, {Majewski}, {Milone}, {Paust}, \& {Reid}}]{marin-franch2009}
{Mar{\'\i}n-Franch}, A., {Aparicio}, A., {Piotto}, G., {et~al.} 2009, \apj, 694, 1498

\bibitem[{{Marino} {et~al.}(2021){Marino}, {Milone}, {Renzini}, {Yong}, {Asplund}, {Da Costa}, {Jerjen}, {Cordoni}, {Carlos}, {Dondoglio}, {Lagioia}, {Jang}, \& {Tailo}}]{marino21}
{Marino}, A.~F., {Milone}, A.~P., {Renzini}, A., {et~al.} 2021, \apj, 923, 22

\bibitem[{{Marino} {et~al.}(2008){Marino}, {Villanova}, {Piotto}, {Milone}, {Momany}, {Bedin}, \& {Medling}}]{marino2008}
{Marino}, A.~F., {Villanova}, S., {Piotto}, G., {et~al.} 2008, \aap, 490, 625

\bibitem[{{Massari} {et~al.}(2023){Massari}, {Aguado-Agelet}, {Monelli}, {Cassisi}, {Pancino}, {Saracino}, {Gallart}, {Ruiz-Lara}, {Fern{\'a}ndez-Alvar}, {Surot}, {Stokholm}, {Salaris}, {Miglio}, \& {Ceccarelli}}]{massari2023}
{Massari}, D., {Aguado-Agelet}, F., {Monelli}, M., {et~al.} 2023, \aap, 680, A20

\bibitem[{{Massari} {et~al.}(2019){Massari}, {Koppelman}, \& {Helmi}}]{massari19}
{Massari}, D., {Koppelman}, H.~H., \& {Helmi}, A. 2019, \aap, 630, L4

\bibitem[{{Masseron} {et~al.}(2019){Masseron}, {Garc{\'\i}a-Hern{\'a}ndez}, {M{\'e}sz{\'a}ros}, {Zamora}, {Dell'Agli}, {Allende Prieto}, {Edvardsson}, {Shetrone}, {Plez}, {Fern{\'a}ndez-Trincado}, {Cunha}, {J{\"o}nsson}, {Geisler}, {Beers}, \& {Cohen}}]{masseron2019}
{Masseron}, T., {Garc{\'\i}a-Hern{\'a}ndez}, D.~A., {M{\'e}sz{\'a}ros}, S., {et~al.} 2019, \aap, 622, A191

\bibitem[{{Matsuno} {et~al.}(2022{\natexlab{a}}){Matsuno}, {Dodd}, {Koppelman}, {Helmi}, {Ishigaki}, {Aoki}, {Zhao}, {Yuan}, \& {Hattori}}]{matsuno2022_helmi}
{Matsuno}, T., {Dodd}, E., {Koppelman}, H.~H., {et~al.} 2022{\natexlab{a}}, \aap, 665, A46

\bibitem[{{Matsuno} {et~al.}(2021){Matsuno}, {Hirai}, {Tarumi}, {Hotokezaka}, {Tanaka}, \& {Helmi}}]{matsuno2021}
{Matsuno}, T., {Hirai}, Y., {Tarumi}, Y., {et~al.} 2021, \aap, 650, A110

\bibitem[{{Matsuno} {et~al.}(2022{\natexlab{b}}){Matsuno}, {Koppelman}, {Helmi}, {Aoki}, {Ishigaki}, {Suda}, {Yuan}, \& {Hattori}}]{matsuno22}
{Matsuno}, T., {Koppelman}, H.~H., {Helmi}, A., {et~al.} 2022{\natexlab{b}}, \aap, 661, A103

\bibitem[{{McConnachie}(2012)}]{mcconnachie2012}
{McConnachie}, A.~W. 2012, \aj, 144, 4

\bibitem[{{Milone} {et~al.}(2012){Milone}, {Piotto}, {Bedin}, {Aparicio}, {Anderson}, {Sarajedini}, {Marino}, {Moretti}, {Davies}, {Chaboyer}, {Dotter}, {Hempel}, {Mar{\'\i}n-Franch}, {Majewski}, {Paust}, {Reid}, {Rosenberg}, \& {Siegel}}]{milone2012}
{Milone}, A.~P., {Piotto}, G., {Bedin}, L.~R., {et~al.} 2012, \aap, 540, A16

\bibitem[{{Minelli} {et~al.}(2021){Minelli}, {Mucciarelli}, {Massari}, {Bellazzini}, {Romano}, \& {Ferraro}}]{minelli21}
{Minelli}, A., {Mucciarelli}, A., {Massari}, D., {et~al.} 2021, \apjl, 918, L32

\bibitem[{{Monaco} {et~al.}(2018){Monaco}, {Villanova}, {Carraro}, {Mucciarelli}, \& {Moni Bidin}}]{monaco2018}
{Monaco}, L., {Villanova}, S., {Carraro}, G., {Mucciarelli}, A., \& {Moni Bidin}, C. 2018, \aap, 616, A181

\bibitem[{{Monty} {et~al.}(2024){Monty}, {Belokurov}, {Sanders}, {Hansen}, {Sakari}, {McKenzie}, {Myeong}, {Davies}, {Ardern-Arentsen}, \& {Massari}}]{monty2024}
{Monty}, S., {Belokurov}, V., {Sanders}, J.~L., {et~al.} 2024, \mnras, 533, 2420

\bibitem[{{Monty} {et~al.}(2023{\natexlab{a}}){Monty}, {Yong}, {Marino}, {Karakas}, {McKenzie}, {Grundahl}, \& {Mura-Guzm{\'a}n}}]{monty23}
{Monty}, S., {Yong}, D., {Marino}, A.~F., {et~al.} 2023{\natexlab{a}}, \mnras, 518, 965

\bibitem[{{Monty} {et~al.}(2023{\natexlab{b}}){Monty}, {Yong}, {Massari}, {McKenzie}, {Myeong}, {Buder}, {Karakas}, {Freeman}, {Marino}, {Belokurov}, \& {Evans}}]{monty2023_2}
{Monty}, S., {Yong}, D., {Massari}, D., {et~al.} 2023{\natexlab{b}}, \mnras, 522, 4404

\bibitem[{{Moore} {et~al.}(1999){Moore}, {Ghigna}, {Governato}, {Lake}, {Quinn}, {Stadel}, \& {Tozzi}}]{moore1999}
{Moore}, B., {Ghigna}, S., {Governato}, F., {et~al.} 1999, \apjl, 524, L19

\bibitem[{{M{\"o}sta} {et~al.}(2018){M{\"o}sta}, {Roberts}, {Halevi}, {Ott}, {Lippuner}, {Haas}, \& {Schnetter}}]{mosta2018}
{M{\"o}sta}, P., {Roberts}, L.~F., {Halevi}, G., {et~al.} 2018, \apj, 864, 171

\bibitem[{{Mucciarelli}(2013)}]{4dao}
{Mucciarelli}, A. 2013, arXiv e-prints, arXiv:1311.1403

\bibitem[{{Mucciarelli} {et~al.}(2021{\natexlab{a}}){Mucciarelli}, {Bellazzini}, \& {Massari}}]{mucciarelli21}
{Mucciarelli}, A., {Bellazzini}, M., \& {Massari}, D. 2021{\natexlab{a}}, \aap, 653, A90

\bibitem[{{Mucciarelli} {et~al.}(2021{\natexlab{b}}){Mucciarelli}, {Massari}, {Minelli}, {Romano}, {Bellazzini}, {Ferraro}, {Matteucci}, \& {Origlia}}]{mucciarelli2021NatAs}
{Mucciarelli}, A., {Massari}, D., {Minelli}, A., {et~al.} 2021{\natexlab{b}}, Nature Astronomy, 5, 1247

\bibitem[{{Mucciarelli} {et~al.}(2023){Mucciarelli}, {Minelli}, {Lardo}, {Massari}, {Bellazzini}, {Romano}, {Origlia}, \& {Ferraro}}]{mucciarelli2023}
{Mucciarelli}, A., {Minelli}, A., {Lardo}, C., {et~al.} 2023, \aap, 677, A61

\bibitem[{{Mucciarelli} {et~al.}(2013){Mucciarelli}, {Pancino}, {Lovisi}, {Ferraro}, \& {Lapenna}}]{gala}
{Mucciarelli}, A., {Pancino}, E., {Lovisi}, L., {Ferraro}, F.~R., \& {Lapenna}, E. 2013, \apj, 766, 78

\bibitem[{{Myeong} {et~al.}(2018){Myeong}, {Evans}, {Belokurov}, {Sanders}, \& {Koposov}}]{myeong18}
{Myeong}, G.~C., {Evans}, N.~W., {Belokurov}, V., {Sanders}, J.~L., \& {Koposov}, S.~E. 2018, \apjl, 856, L26

\bibitem[{{Myeong} {et~al.}(2019){Myeong}, {Vasiliev}, {Iorio}, {Evans}, \& {Belokurov}}]{myeong19}
{Myeong}, G.~C., {Vasiliev}, E., {Iorio}, G., {Evans}, N.~W., \& {Belokurov}, V. 2019, \mnras, 488, 1235

\bibitem[{{Naidu} {et~al.}(2022{\natexlab{a}}){Naidu}, {Conroy}, {Bonaca}, {Zaritsky}, {Ting}, {Caldwell}, {Cargile}, {Speagle}, {Chandra}, {Johnson}, {Woody}, \& {Han}}]{naidu2022}
{Naidu}, R.~P., {Conroy}, C., {Bonaca}, A., {et~al.} 2022{\natexlab{a}}, arXiv e-prints, arXiv:2204.09057

\bibitem[{{Naidu} {et~al.}(2022{\natexlab{b}}){Naidu}, {Ji}, {Conroy}, {Bonaca}, {Ting}, {Zaritsky}, {van Son}, {Broekgaarden}, {Tacchella}, {Chandra}, {Caldwell}, {Cargile}, \& {Speagle}}]{naidu22}
{Naidu}, R.~P., {Ji}, A.~P., {Conroy}, C., {et~al.} 2022{\natexlab{b}}, \apjl, 926, L36

\bibitem[{{Newton} {et~al.}(2018){Newton}, {Cautun}, {Jenkins}, {Frenk}, \& {Helly}}]{newton18}
{Newton}, O., {Cautun}, M., {Jenkins}, A., {Frenk}, C.~S., \& {Helly}, J.~C. 2018, \mnras, 479, 2853

\bibitem[{{Nissen} \& {Schuster}(2010)}]{nissen&schuster2010}
{Nissen}, P.~E. \& {Schuster}, W.~J. 2010, \aap, 511, L10

\bibitem[{{Nissen} \& {Schuster}(2011)}]{nissen&schuster2011}
{Nissen}, P.~E. \& {Schuster}, W.~J. 2011, \aap, 530, A15

\bibitem[{{Ou} {et~al.}(2024){Ou}, {Ji}, {Frebel}, {Naidu}, \& {Limberg}}]{ou2024}
{Ou}, X., {Ji}, A.~P., {Frebel}, A., {Naidu}, R.~P., \& {Limberg}, G. 2024, arXiv e-prints, arXiv:2404.10067

\bibitem[{{Pagnini} {et~al.}(2023){Pagnini}, {Di Matteo}, {Khoperskov}, {Mastrobuono-Battisti}, {Haywood}, {Renaud}, \& {Combes}}]{pagnini2023}
{Pagnini}, G., {Di Matteo}, P., {Khoperskov}, S., {et~al.} 2023, \aap, 673, A86

\bibitem[{{Pasquini} {et~al.}(2002){Pasquini}, {Avila}, {Blecha}, {Cacciari}, {Cayatte}, {Colless}, {Damiani}, {de Propris}, {Dekker}, {di Marcantonio}, {Farrell}, {Gillingham}, {Guinouard}, {Hammer}, {Kaufer}, {Hill}, {Marteaud}, {Modigliani}, {Mulas}, {North}, {Popovic}, {Rossetti}, {Royer}, {Santin}, {Schmutzer}, {Simond}, {Vola}, {Waller}, \& {Zoccali}}]{pasquini02}
{Pasquini}, L., {Avila}, G., {Blecha}, A., {et~al.} 2002, The Messenger, 110, 1

\bibitem[{{Pe{\~n}arrubia} {et~al.}(2009){Pe{\~n}arrubia}, {Walker}, \& {Gilmore}}]{penarrubia2009}
{Pe{\~n}arrubia}, J., {Walker}, M.~G., \& {Gilmore}, G. 2009, \mnras, 399, 1275

\bibitem[{{Puls} {et~al.}(2018){Puls}, {Alves-Brito}, {Campos}, {Dias}, \& {Barbuy}}]{puls2018}
{Puls}, A.~A., {Alves-Brito}, A., {Campos}, F., {Dias}, B., \& {Barbuy}, B. 2018, \mnras, 476, 690

\bibitem[{{Recio-Blanco}(2018)}]{recioblanco2018}
{Recio-Blanco}, A. 2018, \aap, 620, A194

\bibitem[{{Reggiani} {et~al.}(2023){Reggiani}, {Schlaufman}, \& {Casey}}]{reggiani2023}
{Reggiani}, H., {Schlaufman}, K.~C., \& {Casey}, A.~R. 2023, \aj, 166, 128

\bibitem[{{Romano} {et~al.}(2010){Romano}, {Karakas}, {Tosi}, \& {Matteucci}}]{romano2010}
{Romano}, D., {Karakas}, A.~I., {Tosi}, M., \& {Matteucci}, F. 2010, \aap, 522, A32

\bibitem[{{Schiavon} {et~al.}(2024){Schiavon}, {Phillips}, {Myers}, {Horta}, {Minniti}, {Allende Prieto}, {Anguiano}, {Beaton}, {Beers}, {Brownstein}, {Cohen}, {Fern{\'a}ndez-Trincado}, {Frinchaboy}, {J{\"o}nsson}, {Kisku}, {Lane}, {Majewski}, {Mason}, {M{\'e}sz{\'a}ros}, \& {Stringfellow}}]{schiavon24}
{Schiavon}, R.~P., {Phillips}, S.~G., {Myers}, N., {et~al.} 2024, \mnras, 528, 1393

\bibitem[{{Sestito} {et~al.}(2023){Sestito}, {Zaremba}, {Venn}, {D'Aoust}, {Hayes}, {Jensen}, {Navarro}, {Jablonka}, {Fern{\'a}ndez-Alvar}, {Glover}, {McConnachie}, \& {Chen{\'e}}}]{sestito2023}
{Sestito}, F., {Zaremba}, D., {Venn}, K.~A., {et~al.} 2023, \mnras, 525, 2875

\bibitem[{{Siegel} {et~al.}(2019){Siegel}, {Barnes}, \& {Metzger}}]{siegel2019}
{Siegel}, D.~M., {Barnes}, J., \& {Metzger}, B.~D. 2019, \nat, 569, 241

\bibitem[{{Sk{\'u}lad{\'o}ttir} {et~al.}(2017){Sk{\'u}lad{\'o}ttir}, {Tolstoy}, {Salvadori}, {Hill}, \& {Pettini}}]{skuladottir2017}
{Sk{\'u}lad{\'o}ttir}, {\'A}., {Tolstoy}, E., {Salvadori}, S., {Hill}, V., \& {Pettini}, M. 2017, \aap, 606, A71

\bibitem[{{Stetson} \& {Pancino}(2008)}]{daospec}
{Stetson}, P.~B. \& {Pancino}, E. 2008, \pasp, 120, 1332

\bibitem[{{Theler} {et~al.}(2020){Theler}, {Jablonka}, {Lucchesi}, {Lardo}, {North}, {Irwin}, {Battaglia}, {Hill}, {Tolstoy}, {Venn}, {Helmi}, {Kaufer}, {Primas}, \& {Shetrone}}]{theler2020}
{Theler}, R., {Jablonka}, P., {Lucchesi}, R., {et~al.} 2020, \aap, 642, A176

\bibitem[{{Tolstoy} {et~al.}(2009){Tolstoy}, {Hill}, \& {Tosi}}]{tolstoy09}
{Tolstoy}, E., {Hill}, V., \& {Tosi}, M. 2009, \araa, 47, 371

\bibitem[{{Trujillo-Gomez} {et~al.}(2021){Trujillo-Gomez}, {Kruijssen}, {Reina-Campos}, {Pfeffer}, {Keller}, {Crain}, {Bastian}, \& {Hughes}}]{trujillo-gomez2021}
{Trujillo-Gomez}, S., {Kruijssen}, J.~M.~D., {Reina-Campos}, M., {et~al.} 2021, \mnras, 503, 31

\bibitem[{{Van der Swaelmen} {et~al.}(2013){Van der Swaelmen}, {Hill}, {Primas}, \& {Cole}}]{vanderswaelmen2013}
{Van der Swaelmen}, M., {Hill}, V., {Primas}, F., \& {Cole}, A.~A. 2013, \aap, 560, A44

\bibitem[{{VandenBerg} {et~al.}(2013){VandenBerg}, {Brogaard}, {Leaman}, \& {Casagrande}}]{vandenberg2013}
{VandenBerg}, D.~A., {Brogaard}, K., {Leaman}, R., \& {Casagrande}, L. 2013, \apj, 775, 134

\bibitem[{{Vasiliev} \& {Baumgardt}(2021)}]{vasiliev&baumgardt2021}
{Vasiliev}, E. \& {Baumgardt}, H. 2021, \mnras, 505, 5978

\bibitem[{{Venn} {et~al.}(2004){Venn}, {Irwin}, {Shetrone}, {Tout}, {Hill}, \& {Tolstoy}}]{venn2004}
{Venn}, K.~A., {Irwin}, M., {Shetrone}, M.~D., {et~al.} 2004, \aj, 128, 1177

\bibitem[{{Venn} {et~al.}(2012){Venn}, {Shetrone}, {Irwin}, {Hill}, {Jablonka}, {Tolstoy}, {Lemasle}, {Divell}, {Starkenburg}, {Letarte}, {Baldner}, {Battaglia}, {Helmi}, {Kaufer}, \& {Primas}}]{venn2012}
{Venn}, K.~A., {Shetrone}, M.~D., {Irwin}, M.~J., {et~al.} 2012, \apj, 751, 102

\bibitem[{{Ventura} {et~al.}(2013){Ventura}, {Di Criscienzo}, {Carini}, \& {D'Antona}}]{ventura2013}
{Ventura}, P., {Di Criscienzo}, M., {Carini}, R., \& {D'Antona}, F. 2013, \mnras, 431, 3642

\bibitem[{{Villanova} {et~al.}(2013){Villanova}, {Geisler}, {Carraro}, {Moni Bidin}, \& {Mu{\~n}oz}}]{villanova13}
{Villanova}, S., {Geisler}, D., {Carraro}, G., {Moni Bidin}, C., \& {Mu{\~n}oz}, C. 2013, \apj, 778, 186

\bibitem[{{Villanova} {et~al.}(2017){Villanova}, {Moni Bidin}, {Mauro}, {Munoz}, \& {Monaco}}]{villanova17}
{Villanova}, S., {Moni Bidin}, C., {Mauro}, F., {Munoz}, C., \& {Monaco}, L. 2017, \mnras, 464, 2730

\bibitem[{{White} \& {Frenk}(1991)}]{white&frenk1991}
{White}, S. D.~M. \& {Frenk}, C.~S. 1991, \apj, 379, 52

\bibitem[{{Yan} {et~al.}(2020){Yan}, {Jerabkova}, \& {Kroupa}}]{yan2020}
{Yan}, Z., {Jerabkova}, T., \& {Kroupa}, P. 2020, \aap, 637, A68

\end{thebibliography}

\begin{appendix} 
\section{Validation of the abundances} \label{app:A}
\subsection{GSE globular clusters}

      To validate our findings, we undertake a comparative analysis with two distinct independent studies conducted on two of the six target GCs. Specifically, we compare our results to those obtained by the seminal works by \citet{carretta09,carretta2010,carretta13} and to those provided by the latest data release of the APOGEE survey \citep{apogee22}. In particular, we use the APOGEE value-added catalogue of Galactic GC stars \citep{schiavon24}, limiting our comparison only to well-measured stars. To do so, we remove all the stars with the following flags\footnote{Flags definition available at \url{https://www.sdss4.org/dr17/irspec/apogee-bitmasks/}.} in the quality parameters:
    
    \begin{itemize}
        \item \texttt{ASPCAPFLAG} = 16, 17, 23;
        \item \texttt{STARFLAG} = 3, 4, 9, 12, 13, 16;
        \item \texttt{EXTRATARG} = 4;
        \item \texttt{VB\_PROB} $>$ 0.99.
    \end{itemize}
    
\begin{figure}[th!]
   \centering
   \includegraphics[width=.39\textwidth]{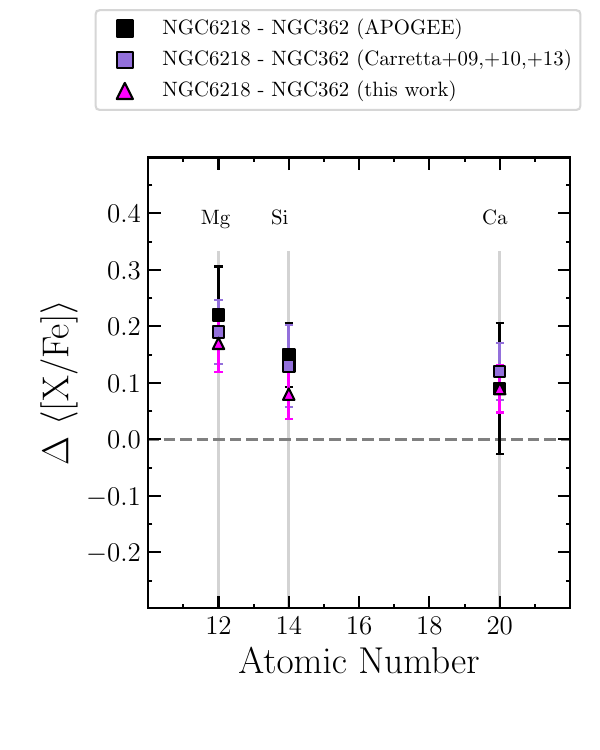} 
   \caption{Difference of mean abundance ratios of the $\alpha$-elements [Mg/Fe], [Si/Fe], and [Ca/Fe] between NGC 6218 and NGC 362 from this work (pink triangles), \citet{carretta09,carretta2010,carretta13} (purple squares), and APOGEE (black squares). Error bars indicate the standard deviation.}
              \label{FigAPOGEE}%
    \end{figure}

    The choice of using these two dataset was guided by their execution of identical chemical analyses on two out of the four target GCs, thereby ensuring consistency in their outcomes. 
    We limit our comparison with \citet{carretta09,carretta2010,carretta13} and APOGEE only to the chemical elements identified as sensitive to the origin of GCs available in these dataset, that are Mg, Si and Ca. As shown in Fig. \ref{FigAPOGEE}, the results from this work are in agreement within the uncertainties with those provided by \citet{carretta09,carretta2010,carretta13} and APOGEE, consistently yielding an enhancement of $0.1$ - $0.2 \dex$ in the $\alpha$-elements (Mg, Si, and Ca) of NGC 6218 with respect to NGC 362.


\subsection{Comparison with V13}\label{sec:V13}    

    A comparison with the results of V13, who analyzed the same Rup106 dataset, further confirms the peculiarity of this system. Indeed, their findings indicate that, within this cluster, the observed spread in all chemical elements is consistent with the uncertainty on the measurement, suggesting Rup106 to be a single population GC, as also confirmed photometrically in more recent works \citep{dotter18,lagioia24}. Moreover, their analysis reveals solar-like abundances in the $\alpha$-elements, alongside a notable underabundance in the iron peak elements relative to iron, ranging from $0.1$ up to $0.8$ dex. Firstly, when comparing the atmospheric parameters of Rup106 targets, we find a significant difference in the adopted $T_{\mathrm{eff}}$ scales, with an average discrepancy (this work – V13) of $+169$ K ($\sigma = 55$ K). The average difference for the surface gravity is $+0.65 \dex$ ($\sigma = 0.14 \dex$) and that for $v_{\mathrm{t}}$ is $-0.07$ km/s ($\sigma = 0.05$ km/s). These discrepancies arise from the fact that V13 constrained $T_{\mathrm{eff}}$ and log $g$ with a metodology similar to ours, deriving them from the photometry, but exploiting a different colour - $T_{\mathrm{eff}}$ relation. On the other hand, V13 assumed a $v_{\mathrm{t}}$ derived from the relation by \citet{marino2008}, while we constrained it spectroscopically (see Sect. \ref{sp}). To ensure a fair comparison between the results, we rescaled the abundance ratios reported in V13 on the \citet{grevesse1998} solar composition. Firstly, we note an average offset of $+0.17 \dex$ ($\sigma$ = $0.07 \dex$) in the star-to-star metallicity estimate of Rup106, which can be explained by the difference in the derived effective temperatures. 
    By comparing these two independent results performed on the same dataset, we find agreement with V13 at the $1\sigma$ level for Mg, Ti, ScII, Cr, Mn, Zn, YII, LaII, and EuII, while we note differences up to $\sim 0.2 \dex$ for Si, Ca, V, Co, Ni, Cu, and BaII. We attribute these inconsistencies to the different scale adopted for surface gravities and, most importantly, effective temperatures. This comparison proves that diverse assumptions in the procedure to derive chemical abundances make the results of two independent approaches not directly comparable and that the only reliable chemical analysis is within an homogeneous procedure. In general, notwithstanding differences stemming from varied assumptions in the chemical analysis, we reaffirm the distinct chemical characteristics of Rup106 within the MW GC system, alongside the minimal dispersion in chemical abundances among its constituent stars.     

\section{Line list}

    In Table \ref{tab:linelist_EW}, we list some useful information about the lines analyzed using the EW method, such as the wavelength, the log \textit{gf}, the excitation potential ($\chi$) and the EW measured with the code \texttt{DAOSPEC}.

\begin{table*}
  \caption{Lines analyzed in this work for each star (extract).}\label{tab:linelist_EW}
  \centering
  \begin{tabular}{cccccccc} 
   \hline             
Cluster & Star ID & Element & Wavelength & log \textit{gf} & $\chi$ & EW & $\sigma$(EW) \\ 
 & & & (\r{A}) & & (eV) & (m\r{A}) & (m\r{A}) \\ 
\hline 
NGC 362                  &                      1037     &                      Fe I      &        4804.517       &          -2.590       &           3.570        &           22.20       &            1.44        \\
NGC 362                  &                      1037     &                      Fe I      &        4807.708       &          -2.150       &           3.370        &           58.80       &            2.68        \\
NGC 362                  &                      1037     &                      Fe I      &        4808.148       &          -2.740       &           3.250        &           33.90       &            0.95        \\
NGC 362                  &                      1037     &                      Fe I      &        4813.113       &          -2.840       &           3.270        &           23.40       &            0.97        \\
NGC 362                  &                      1037     &                      Fe I      &        4867.529       &          -4.752       &           1.610        &           39.80       &            1.36        \\
NGC 362                  &                      1037     &                      Fe I      &        4869.463       &          -2.480       &           3.550        &           24.70       &            1.05        \\
NGC 362                  &                      1037     &                      Fe I      &        4882.143       &          -1.480       &           3.420        &           85.60       &            1.49        \\
NGC 362                  &                      1037     &                      Fe I      &        4918.012       &          -1.340       &           4.230        &           41.60       &            1.24        \\
NGC 362                  &                      1037     &                      Fe I      &        4927.863       &          -0.960       &           4.220        &           57.20       &            0.92        \\
NGC 362                  &                      1037     &                      Fe I      &        4950.105       &          -1.500       &           3.420        &           83.20       &            1.20        \\
... & ...                        & ...   & ...   & ...   & ...   & ...   & ... \\
   \hline
  \end{tabular}
\tablefoot{The entire table is available on Zenodo.}    
\end{table*}

\end{appendix}

\end{document}